\documentclass[number,preprint,3p]{elsarticle}

\usepackage{graphicx}
\usepackage{dcolumn}
\usepackage{bm}

\usepackage[utf8]{inputenc}
\usepackage[T1]{fontenc}
\usepackage{etoolbox}
\usepackage{amsmath}
\usepackage{amssymb}
\usepackage{gensymb}
\usepackage{natbib}
\usepackage{caption}
\usepackage{subcaption}
\usepackage{array}
\usepackage{tablefootnote}
\usepackage{multirow}
\usepackage{bigdelim}
\usepackage{makecell}
\usepackage{booktabs}
\usepackage{tabularray}
\UseTblrLibrary{booktabs}
\usepackage{adjustbox}
\usepackage{float}
\usepackage{siunitx}
\usepackage[version=4]{mhchem}

\usepackage[nonumberlist,nopostdot,shortcuts,style=alttree,section=section]{glossaries-extra}
\makeglossaries
\setabbreviationstyle[acronym]{long-short}
\newacronym{cmfx}{CMFX}{The Centrifugal Mirror Fusion Experiment}
\newacronym{mcf}{MCF}{magnetic confinement fusion}
\newacronym{icf}{ICF}{inertial confinement fusion}
\newacronym{dpa}{dpa}{displacements per atom}
\newacronym[shortplural={PFMs},longplural={plasma facing materials}]{pfm}{PFM}{plasma facing material}
\newacronym[shortplural={PFCs},longplural={plasma facing components}]{pfc}{PFC}{plasma facing component}
\newacronym{dbtt}{DBTT}{ductile-to-brittle transition temperature}
\newacronym{mhd}{MHD}{magnetohydrodynamic}
\newacronym{daq}{DAQ}{data acquisition system}
\newacronym{ni}{NI}{National Instruments}
\newacronym{umd}{UMD}{The University of Maryland}
\newacronym{pmi}{PMI}{plasma-material interactions}
\newacronym{hbn}{hBN}{hexagonal boron nitride}
\newacronym{bca}{BCA}{binary collision approximation}
\newacronym{ucsd}{UCSD}{The University of California, San Diego}
\newacronym{xrd}{XRD}{X-ray diffraction}
\newacronym{gui}{GUI}{graphical user interface}
\newacronym{rf}{RF}{radiofrequency}
\newacronym{cv}{CV}{coefficient of variation}
\newacronym{sem}{SEM}{scanning electron microscope}
\newacronym{tem}{TEM}{transmission electron microscope}
\newacronym{eds}{EDS}{energy-dispersive X-ray spectroscopy}
\newacronym{xps}{XPS}{X-ray photoelectron spectroscopy}
\newacronym{hv}{HV}{high voltage}
\newacronym{hvps}{HVPS}{high voltage power supply}
\newacronym{sca}{SCA}{single channel analyzer}
\newacronym{cad}{CAD}{computer aided design}
\newacronym{hdpe}{HDPE}{high density polyethylene}
\newacronym{nbi}{NBI}{neutral beam injection}
\newacronym{mcx}{MCX}{Maryland Centrifugal Experiment}
\newacronym{dc}{DC}{direct current}
\newacronym{cte}{CTE}{coefficient of thermal expansion}

\newcommand{\mctrans}{{\texttt{MCTrans++}}}

\providecommand{\eqref}[1]{Eq.\ (\ref{#1})}

\newcommand{\infrac}[2]{ \left.{#1}\middle/{#2}\right. }

\RequirePackage{amsmath}
\makeatletter
\newcommand*\rel@kern[1]{\kern#1\dimexpr\macc@kerna}
\newcommand*\widebar[1]{%
  \begingroup
  \def\mathaccent##1##2{%
    \rel@kern{0.8}%
    \overline{\rel@kern{-0.8}\macc@nucleus\rel@kern{0.2}}%
    \rel@kern{-0.2}%
  }%
  \macc@depth\@ne
  \let\math@bgroup\@empty \let\math@egroup\macc@set@skewchar
  \mathsurround\z@ \frozen@everymath{\mathgroup\macc@group\relax}%
  \macc@set@skewchar\relax
  \let\mathaccentV\macc@nested@a
  \macc@nested@a\relax111{#1}%
  \endgroup
}
\makeatother

\usepackage[capitalise, nameinlink]{cleveref}

\begin{document}
	
\begin{frontmatter}
	\title{Electrically insulating materials for centrifugal mirrors\tnoteref{t1}}
	\tnotetext[t1]{This work was supported by ARPA-E award no. DE-AR0001270 and US-DOE Cooperative Agreement No. DE-SC0022528.}
	
	\author[1]{Nick R. Schwartz}
	\author[2]{Carlos A. Romero-Talam\'as}
	\author[3]{Marlene I. Patino}
	\author[3]{Daisuke Nishijima}
	\author[3]{Matthew J. Baldwin}
	\author[3]{Russel P. Doerner}
	\author[2]{Artur Perevalov}
	\author[2]{Nathan Eschbach}
	\author[4]{Zachary D. Short}
	\author[1]{John Cumings}
	\author[5]{Ian G. Abel}
	\author[5]{Brian Beaudoin}
	\author[1]{Timothy W. Koeth\corref{cor1}}
	\ead{koeth@umd.edu}
	\cortext[cor1]{Corresponding author}
	\affiliation[1]{organization={Department of Materials Science and Engineering},
		addressline={University of Maryland},
		postcode={20742},
		city={College Park},
		state={MD},
		country={USA}}
	\affiliation[2]{organization={Department of Mechanical Engineering},
		addressline={University of Maryland, Baltimore County},
		postcode={21250},
		city={Baltimore},
		state={MD},
		country={USA}}
	\affiliation[3]{organization={Center for Energy Research},
		addressline={University of California San Diego},
		postcode={92093},
		city={La Jolla},
		state={CA},
		country={USA}}
	\affiliation[4]{organization={Department of Physics},
		addressline={University of Maryland},
		postcode={20742},
		city={College Park},
		state={MD},
		country={USA}}
	\affiliation[5]{organization={Institute for Research in Electronics and Applied Physics},
		addressline={University of Maryland},
		postcode={20742},
		city={College Park},
		state={MD},
		country={USA}}
	
	\begin{abstract}
		The centrifugal mirror confinement scheme incorporates supersonic rotation into a magnetic mirror device, which stabilizes and heats the plasma. This concept is under investigation in the Centrifugal Mirror Fusion Experiment (CMFX) at the University of Maryland. Plasma rotation is driven by an axial magnetic field and a radial electric field that lead to velocity drifts in the azimuthal direction. An electrically insulating material is required to prevent the applied voltage from shorting on the grounded chamber. Hexagonal boron nitride (hBN) is a promising candidate material for plasma-facing components in future centrifugal mirrors due to its exceptional thermal and electrical properties. However, its performance under intense particle and heat fluxes characteristic of the plasma edge in fusion devices remains largely unexplored. Computational modeling for ion- and neutron-material interactions was carried out with RustBCA and OpenMC, respectively, and predicts relatively good performance in comparison to other insulating materials. Material coupons were then exposed to plasma in PISCES-A at UCSD and CMFX. A load-locked sample feedthrough was constructed and installed on CMFX to test coupons. Two erosion mechanisms were identified -- sputtering and grain ejection -- both of which were more apparent in silicon carbide than hBN.
	\end{abstract}
	
\end{frontmatter}
\glsaddall
	
\section{Introduction}
\label{sec:Introduction}
The centrifugal mirror confinement scheme incorporates supersonic rotation of a plasma into a conventional axisymmetric magnetic mirror device. This rotation introduces a centrifugal force that complements the magnetic confinement, offering several advantages, including decreased parallel losses,  increased density and temperature, and shear-flow stabilization \citep{Schwartz2024}.

Indeed, supersonic rotation has been demonstrated to improve axial confinement \citep{Teodorescu2010} and stability \citep{Huang2001, Uzun2009} in a mirror configuration, thereby reducing both parallel and perpendicular losses. This concept has been studied in several past experiments, including those in \citep{Baker1961, Baker1961_2, Volosov2009, Ellis2005}, though none of these are currently in operation.

\Gls{cmfx} has been constructed at the Univeristy of Maryland and is currently in operation to demonstrate the viability of the centrifugal mirror as a fusion reactor.

\subsection{Centrifugal Mirrors}
\label{sec:Centrifugal Mirrors}
\begin{figure}
	\centering
	\includegraphics[width=0.8\textwidth]{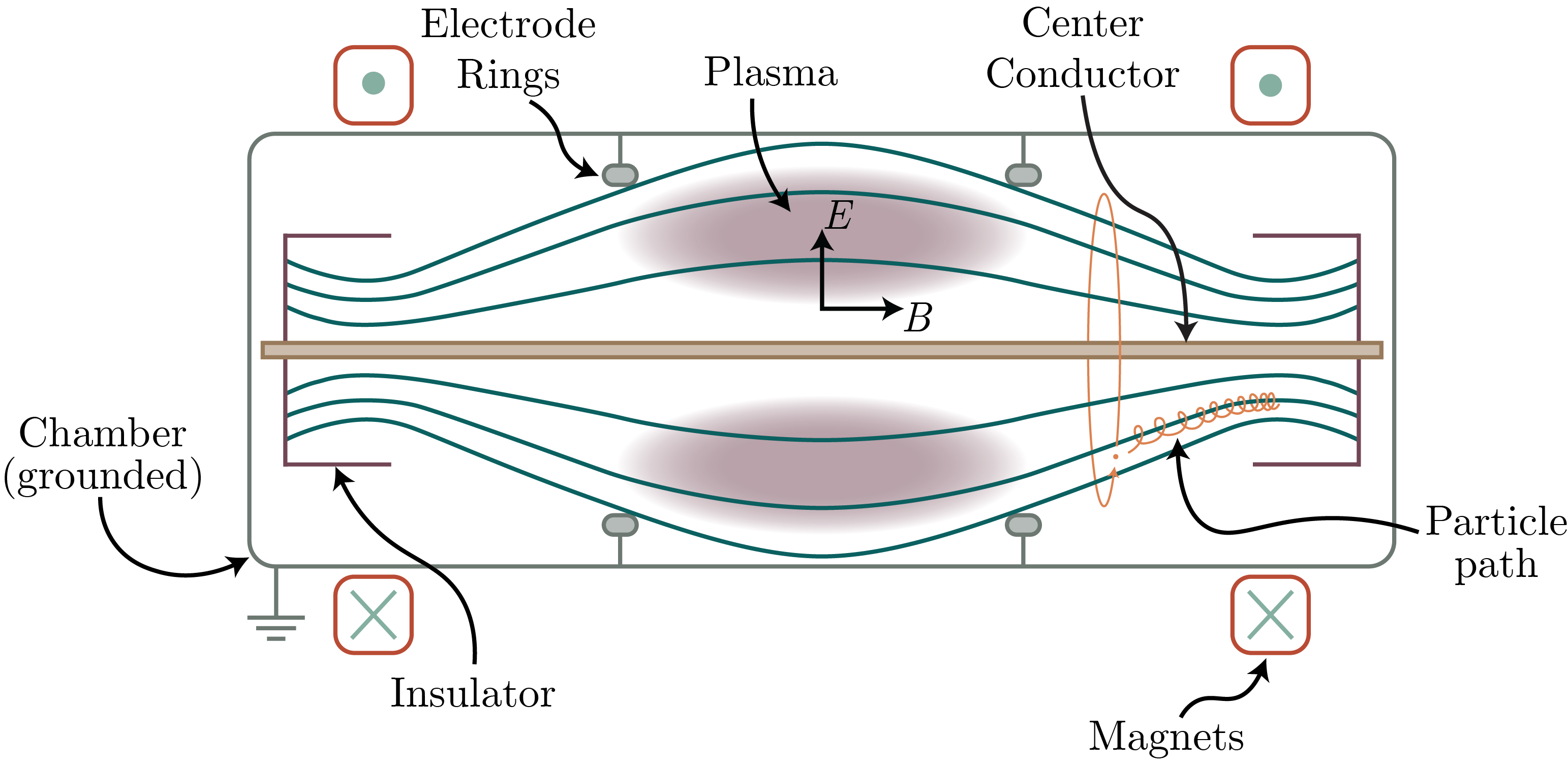}
	\caption{Diagram of a centrifugal mirror. High voltage is applied to a central conductor, generating a radial electric field. The combination of electric and magnetic fields produces an azimuthal $\bm{E} \times \bm{B}$ flow. Magnetic field lines terminate on insulating surfaces to avoid `shorting' the plasma. Electrode rings control the location of the outer flux surface.}
	\label{fig:cmfx_diagram}
\end{figure}

The centrifugal mirror introduces a transverse electric field $\bm{E}$ imposed through a center conductor to generate $\bm{E} \times \bm{B}$ azimuthal flows (\cref{fig:cmfx_diagram}). The centrifugal forces axially compress and radially expand the plasma, with most of the plasma rotating in a toroidal region around the center conductor.

Two superconducting magnets produce a magnetic field such that $B_\mathrm{max} = 3\ \si{\tesla}$, $B_\mathrm{min} = 0.34\ \si{\tesla}$. High voltage is applied to the central conductor, and an electrical current flowed across the magnetic field to electrode rings connected to the grounded chamber. The end sections of the chamber contain the conductor, supports, and large glass buckets that insulate the rotating plasma from the vacuum chamber.

In $\bm{E} \times \bm{B}$ rotating plasmas, each magnetic flux surface (i.e. field line, in the 2-D picture) is also a surface of constant voltage \citep{Lehnert1971}. Essentially, the electrons are very mobile along magnetic flux surfaces, so the magnetic field must terminate on an insulating surface to prevent the buildup of potential and breakdown across the flux surfaces, which would yield near-zero $\bm{E}$ field and no rotation. As can be seen in \cref{fig:cmfx_diagram}, the field lines terminate on insulators.

\subsection{Insulators in Centrifugal Mirrors}
\label{sec:Insulators in Centrifugal Mirrors}
While other fusion experiments have utilized electrically insulating materials, \gls{cmfx} and similar centrifugal mirror devices are distinct in that insulators are fundamental to their operation. Although \gls{cmfx} also faces conventional \gls{pfm} challenges with electrically conductive components (such as the central conductor, limiters, and vacuum chamber), here we focus primarily on the insulator-related challenges. Addressing these unique \gls{pmi} issues with insulators is crucial for the development of centrifugal mirrors.

\Gls{cmfx} has been designed, in part, as a test bed for evaluating electrically insulating materials under fusion-relevant conditions. In fact, the National Academies of Science report on fusion energy posits that one of the biggest challenges in achieving fusion is in designing and testing \glspl{pfm} \citep{NASReport}. Thus, \gls{cmfx} is a valuable resource to achieve this goal, providing a fusion-relevant environment for materials testing.

Currently, \gls{cmfx} uses glass insulators, but this is not a viable option for future power plants. \gls{cmfx} can use glass insulators due to the short plasma duration and relatively low total energy deposition. Future power plants will require active cooling of the insulators during continuous operation. This reality drives the focus of identifying insulating materials with specific key properties:

\begin{itemize}
	\item \textbf{High thermal conductivity} ($k$): Essential for efficient cooling and maintaining lower insulator temperatures.
	\item \textbf{High dielectric strength}: The surface must be able to standoff high electric fields to prevent flux surfaces from shorting.
	\item \textbf{Mechanical properties}: We desire a material with high strength but moderate to low modulus and hardness, allowing it to accommodate thermal strains due to a mismatch in the \gls{cte} with other components while maintaining structural integrity.
	\item \textbf{Low tritium retention}: The global supply of tritium is extremely limited, making inventory control and recycling essential for the economic viability of fusion energy \citep{Nishi2006,Ebenhoch2016,Ferry2023}.
	\item \textbf{Low plasma erosion rate}: Erosion of the insulators should be avoided to maintain a sufficient voltage stand-off over long operational periods.
	\item \textbf{Low implantation depth}: The depth of energetic ions or neutrons should be low to preserve the bulk properties.
	\item \textbf{Commercially available}: The insulating material should be widely available and relatively inexpensive.
\end{itemize}

Having established the unique insulator requirements for centrifugal mirrors, we conducted a systematic evaluation of candidate materials based on their thermal, electrical, and mechanical properties.

\section{Material Selection}
\label{sec:Material Selection}
\Glspl{pfm} must withstand high ion, neutron, and heat fluxes. As shown in \cref{tab:material_choices}, several materials were considered for the insulators, such as \gls{hbn}, silicon carbide, and diamond. Other ceramics, including \ce{Al2O3}, \ce{SiO2}, and \ce{TiC}, have been explored for use in fusion reactors but were ultimately ruled out due to various factors such as radiation swelling, high \gls{cte}, or poor toughness \citep{Hopkins1985, Bolt1993}. Cubic boron nitride (cBN) is also a promising insulating material due to its lower sputtering yield compared to \gls{hbn} \citep{Reinke1995}; however, it is significantly more difficult to produce cBN, so it was not considered in our present study.

\begin{table}
	\begin{center}
		\begin{tblr}{lcccc}
			\toprule
			\textbf{Property} & \ce{W} & \ce{SiC} & \gls{hbn} & Diamond \\ \midrule
			{Thermal conductivity\\(\si{\watt \per \centi \meter \per \kelvin})} & 1.8 \citep{Hu2017} & 4.9 \citep{Han2003} & {5.9 ($||$),\\0.045 ($\perp$) \citep{Yuan2019}} & 23 \citep{Coe2000} \\
			{Dielectric strength\\(\si{\mega \volt \per \centi \meter})} & - & 2.5 \citep{Han2003} & {3 ($||$),\\12 ($\perp$) \citep{Hattori2016}} & 2.4 \citep{Boettger1995} \\
			{Hydrogen diffusivity\\(\si{\meter \squared \per \second})\footnotemark[1]} & \num{4.5e-9} \citep{Liu2009,Liu2014} & \num{1.1e-18} \citep{CAUSEY1978} & \num{3.5e-15} \citep{Checchetto2000}\footnotemark[2] & \num{2.9e-18} \citep{Cherniak2018} \\
			Sputtering Yield\footnotemark[3] & \num{2.6e-3} \citep{Eckstein1991} & \num{1.4e-2} \citep{Balden2000} & - & - \\
			Effects on plasma & High-$Z$ \citep{Philipps2011,Wesson2004} & {Organic\\impurities} & {Beneficial\\\citep{Bortolon2019,Lunsford2019,Gilson2021,Lunsford2022}} & {Organic\\impurities} \\
			Elastic modulus (\si{\giga \pascal}) & 338 \citep{Yu2020} & 460 \citep{Snead2007} & 50 \citep{Duan2016} & 1100 \citep{Sussmann1994} \\ 
			{Fracture toughness\\(\si{\mega \pascal \meter \tothe{1/2}})} & 5.1 \citep{Gludovatz2010} & 3.0 \citep{Snead2007} & 2.8 \citep{Duan2016} & 6.0 \citep{Sussmann1994} \\ 
			Vicker's Hardness (\si{\giga \pascal}) & 5.6 \citep{Yu2020} & 27 \citep{Snead2007} & 0.8 \citep{Duan2016} & 95 \citep{Field2012} \\ 
			Cost (\$/g)\footnotemark[4] & 5.6 & 19 & 37 & 150\footnotemark[5] \\
			\bottomrule
		\end{tblr}
		\caption{Comparison of some \glspl{pfm} under consideration for the insulators. Tungsten is not a viable option for the insulators because it is electrically conductive, but it is included nonetheless because it is a widely used \gls{pfm} in fusion-relevant experiments \citep{Philipps2011}.}
		\label{tab:material_choices}
	\end{center}
\end{table}

After considering the material requirements outlined earlier, \gls{hbn} emerged as the most suitable candidate. Although \gls{hbn} is often described as "chalky" or brittle, it is a relatively inexpensive material that fulfills the most critical requirements for this application. Future work on composite materials may be necessary to improve upon the mechanical properties of \gls{hbn}.

It was not until 1990 that \citet{Buzhinskij1990, Buzhinskij1992} provided one of the first motivations for using \gls{hbn} in fusion devices, which was referred to as pyrolytic boron nitride at the time. Around the same time, \citet{Yamage1992} developed amorphous boron nitride coatings for fusion reactors. In 1995, \citet{Buzhinskij1995} irradiated \gls{hbn} with a 1 \si{\kilo \electronvolt} ($T_i + T_e$) plasma beam and observed "very small removal of material and no mechanical damage." More recently, it has become possible to regularly perform boronization using either boron powder or diborane (\ce{B2H6}). However, only a few experiments have been conducted with boron nitride powders \citep{Bortolon2019,Lunsford2019,Gilson2021,Lunsford2022}. While there is a significant body of work analyzing the effect of boron and nitrogen on plasma performance, there is currently very little research on how fusion plasmas affect bulk \gls{hbn}. A prior NASA report investigated boron nitride as a potential heat shield material, and exposed samples to heat fluxes that would be similar to those in fusion reactors, but this was from an atmospheric pressure plasma arc \cite{Okuno1966}.

To predict how these candidate materials would perform under fusion conditions, we employed computational modeling to simulate both ion and neutron damage mechanisms.

\section{Damage Modeling}
\label{sec:Damage Modeling}

\subsection{Ion Damage}
\label{sec:Ion Damage}
\label{sec:Plasma-Material Interaction Modeling}
\Gls{pmi} is primarily dominated by ion-material interactions. In this instance, we are most interested in reflection, sputtering, and implantation. There is a wide array of other effects present in \gls{pmi}, and the reader is directed to \citet{Naujoks2006} for a more detailed description.

\footnotetext[1]{At 1000 \si{\kelvin}, a relevant surface temperature for \glspl{pfm}.}
\footnotetext[2]{Extrapolated from the data in \citet{Checchetto2000}, which only goes up to 752 \si{\kelvin}.}
\footnotetext[3]{A beam of 100 \si{\electronvolt} \ce{D^+} ions perpendicular to the surface.}
\footnotetext[4]{Note that all the prices were retrieved from Millipore Sigma, a Sigma Aldrich company, in July of 2024 \citep{MilliporeSigma}. The form of each material is 10 mm diameter rods, except for the diamond, which is in powder form.}
\footnotetext[5]{The cost of diamond was for a powder form factor, so solid polycrystalline diamond is likely significantly more expensive.}

\subsubsection{Simulation Parameters}
\label{sec:Ion Parameters}

A predictive model has been developed to understand the \gls{pmi} performance of \gls{hbn} using the tool RustBCA \citep{Drobny2021}. RustBCA was chosen because it is open source, has been benchmarked against experimental data and molecular dynamics simulations for many species/materials at a wide range of input conditions, can be easily parallelized, and provides all the parameters of interest for this work. One drawback of RustBCA, and \gls{bca} codes in general, is that they model the target material amorphously, meaning it cannot capture crystalline effects like ion channeling or account for the anisotropy of \gls{hbn}.

Three materials were chosen for analysis -- hexagonal boron nitride, silicon carbide, and tungsten. We have chosen to model the 6H polymorph of \ce{SiC} because that is the one used in the experimental campaign discussed in \cref{sec:Exposure Results}. Additionally, while tungsten cannot be used as an insulating material in centrifugal mirrors, it has been included in the analysis because of its ubiquity in other fusion devices \citep{Philipps2011}.

The standard input parameters for all subsequent simulations are in \cref{tab:rustbca_settings}, unless otherwise noted. 

\begin{table}
	\begin{center}
		\begin{tblr}{colspec = {Q[r,m] Q[c,m]|Q[c,m]|Q[c,m]|Q[c,m]|},
				vlines = {2-16}{solid}}
			\cline{3-5}
			& & \gls{hbn} & \ce{SiC} & \ce{W} \\
			\hline
			\SetCell[r=5]{c} {Material\\Properties} & Number density (\si{\per \meter \cubed}) & \num{3.5e28} \citep{Lipp1989} & \num{4.8e28} \citep{Hvisdos2021} & \num{6.3e28} \\
			\hline
			& $Z$ & B: 5, N: 7 & Si: 14, C: 6 & 74 \\
			\hline
			& $m$ (\si{\gram \per \mole}) & {B: 10.81,\\N: 14.007} & {Si: 28.085,\\C: 12.011} & 183.84 \\
			\hline
			& Bulk binding, $E_b$ (\si{\electronvolt}) & 8.0 \citep{Ooi2005} & 6.34 \citep{Heera1996} & 8.9 \citep{kittel2005introduction} \\
			\hline
			& Surface binding, $E_s$ (\si{\electronvolt}) & 4.8 \citep{Smith2016} & 6.8 \citep{Heera1996} & 11.75 \citep{Yang2014} \\
			\hline
			\SetCell[r=4]{c} {Particle\\Properties} & Particle & \SetCell[c=3]{m} Deuterium, Tritium, Helium \\
			\hline
			& Particles, $N$ & \SetCell[c=3]{m} 300,000 \\
			\hline
			& Energy (\si{\kilo \electronvolt}) & \SetCell[c=3]{m} 1 \\
			\hline
			& Angle, $\theta$ (\si{\degree}) & \SetCell[c=3]{m} 0 \\
			\hline
			\SetCell[r=6]{c} {Simulation\\Options} & {Cutoff\\Energy (\si{\electronvolt})} & \SetCell[c=3]{m} 1 \\
			\hline
			& {Surface\\Binding Model} & \SetCell[c=3]{m} \texttt{TARGET} \\
			\hline
			& {Bulk Binding\\Model} & \SetCell[c=3]{m} \texttt{INDIVIDUAL} \\
			\hline
			& {Electronic\\Stopping Model} & \SetCell[c=3]{m} \texttt{LOW\_ENERGY\_EQUIPARTITION} \\
			\hline
			& {Interaction\\Potential} & \SetCell[c=3]{m} ZBL \\
			\hline
			& Quadrature & \SetCell[c=3]{m} Mendenhall-Weller \\
			\hline
		\end{tblr}
		\caption{Standard input parameters for all RustBCA simulations unless otherwise stated. Only deuterium was used for the convergence study, while the rest of the results simulate multiple particles.}
		\label{tab:rustbca_settings}
	\end{center}
\end{table}

A convergence study was performed to determine the appropriate number of particles, $N$, to simulate. A standard method for determining convergence is the \gls{cv}, a technique that estimates the statistical uncertainty in the mean value obtained from a simulation. In this approach, $N$ simulated particles are divided into $M$ batches, each containing $\lfloor \infrac{N}{M} \rfloor$ particles. The mean and standard deviation can then be computed among those batches and the \gls{cv} is the value of the standard deviation divided by the mean.

For the convergence study, $M$ was set to 10, and the simulation was considered sufficiently converged when the \gls{cv} was less than 10\% for both sputtering yield, $Y_s = \frac{\textrm{Ejected atoms}}{\textrm{Incident ions}}$, and reflection. Based on this criterion, $N$ was chosen to be 300,000 particles to ensure the desired level of convergence.

\subsubsection{Results}
\label{sec:Ion Results}
\begin{figure}
	\centering
	\begin{subfigure}{0.46\textwidth}
		\centering
		\captionsetup{justification=centering,font=small}
		\includegraphics[width=\textwidth]{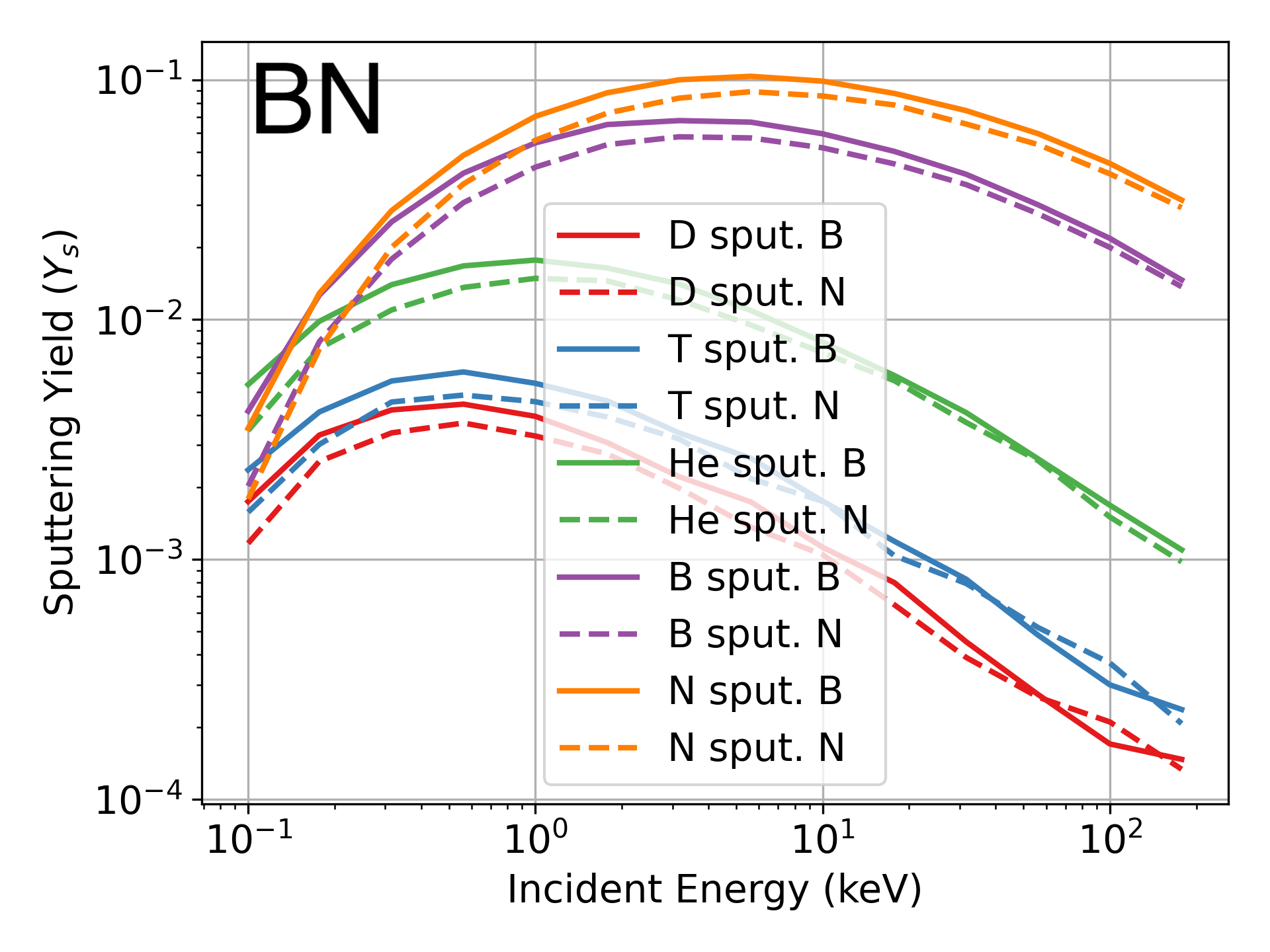}
		\caption{}
		\label{fig:Ys_energy_BN}
	\end{subfigure}
	\begin{subfigure}{0.46\textwidth}
		\centering
		\captionsetup{justification=centering,font=small}
		\includegraphics[width=\textwidth]{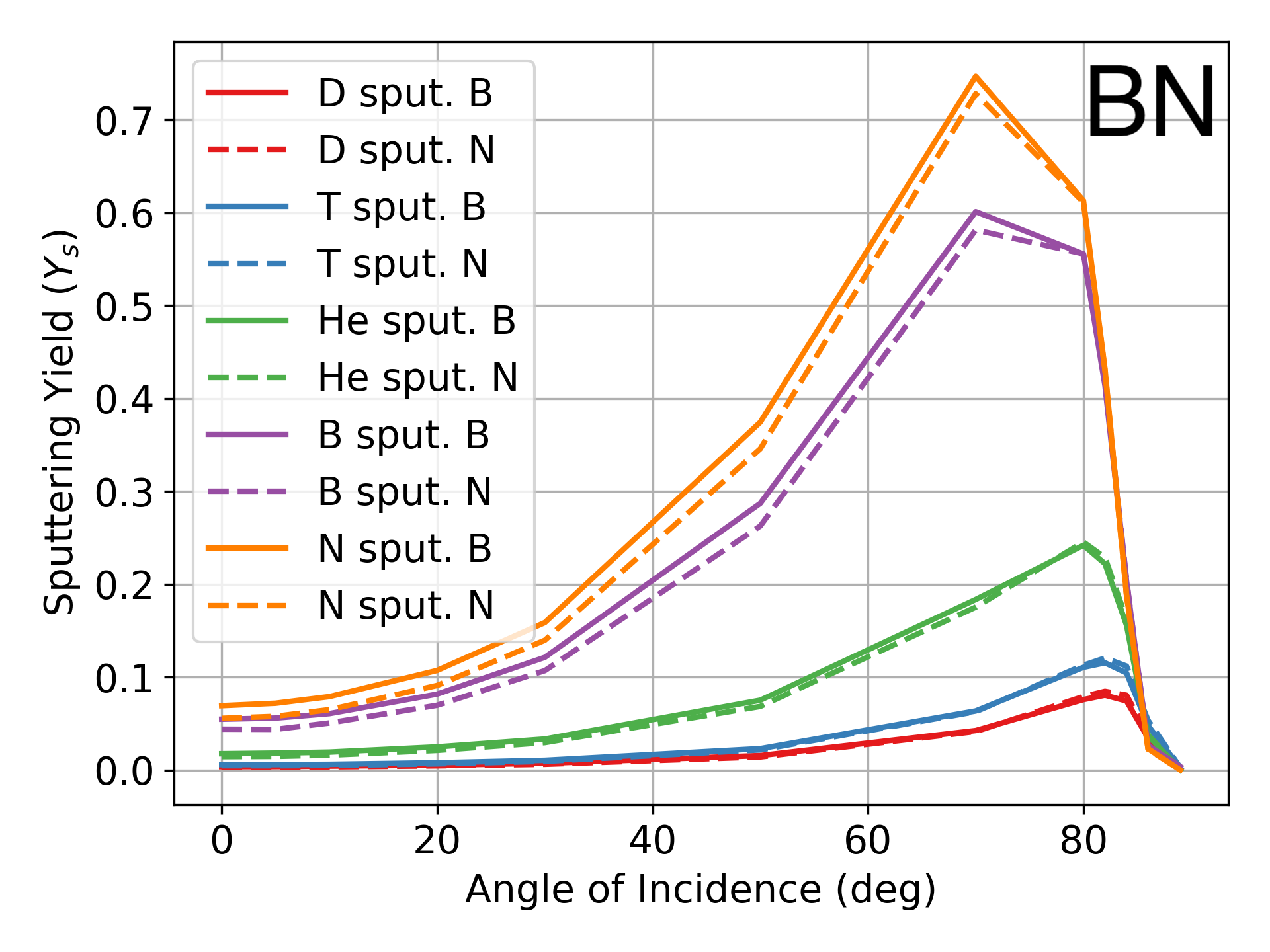}
		\caption{}
		\label{fig:Ys_theta_BN}
	\end{subfigure}
	\\
	\begin{subfigure}{0.46\textwidth}
		\centering
		\captionsetup{justification=centering,font=small}
		\includegraphics[width=\textwidth]{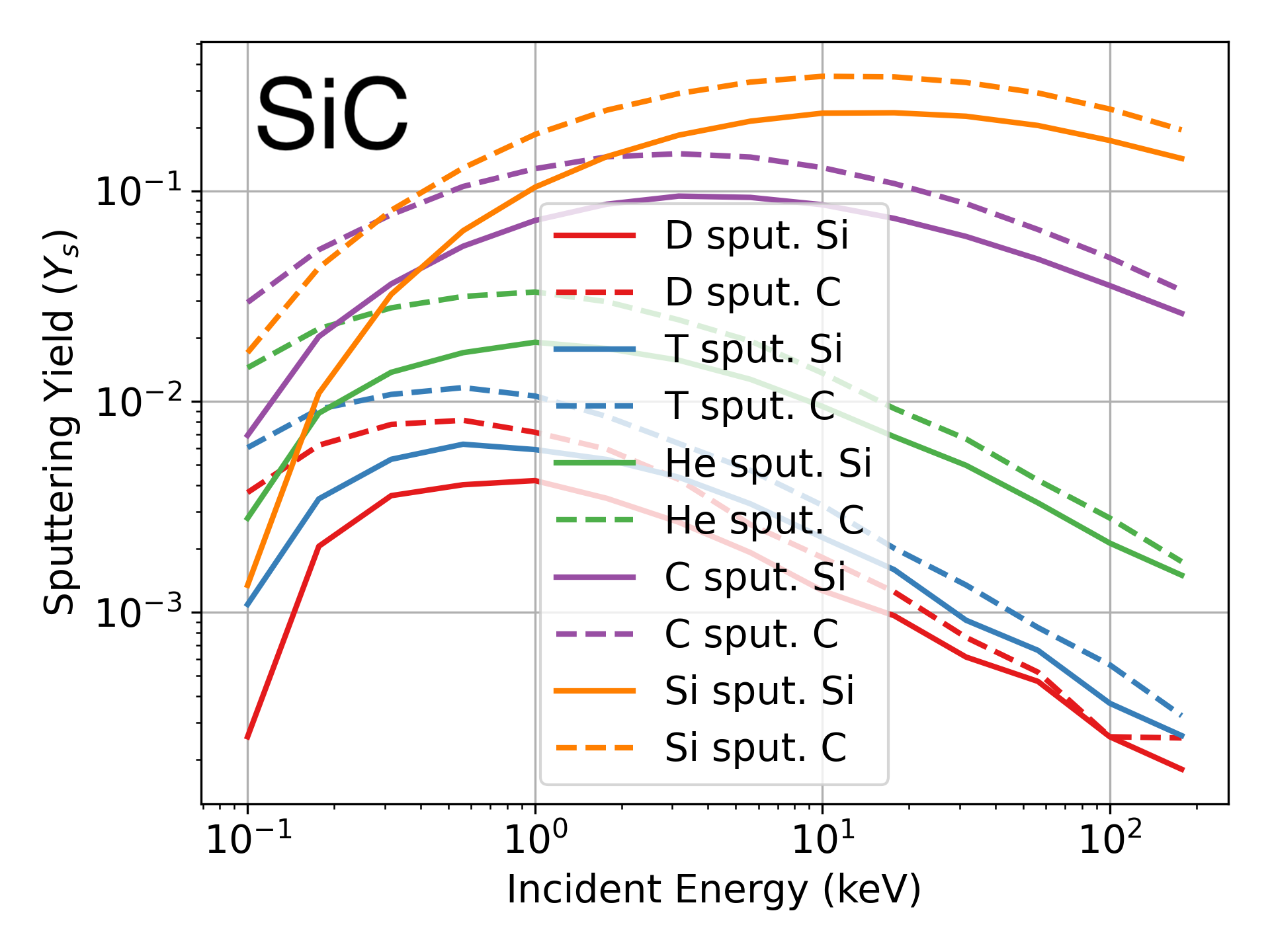}
		\caption{}
		\label{fig:Ys_energy_SiC}
	\end{subfigure}
	\begin{subfigure}{0.46\textwidth}
		\centering
		\captionsetup{justification=centering,font=small}
		\includegraphics[width=\textwidth]{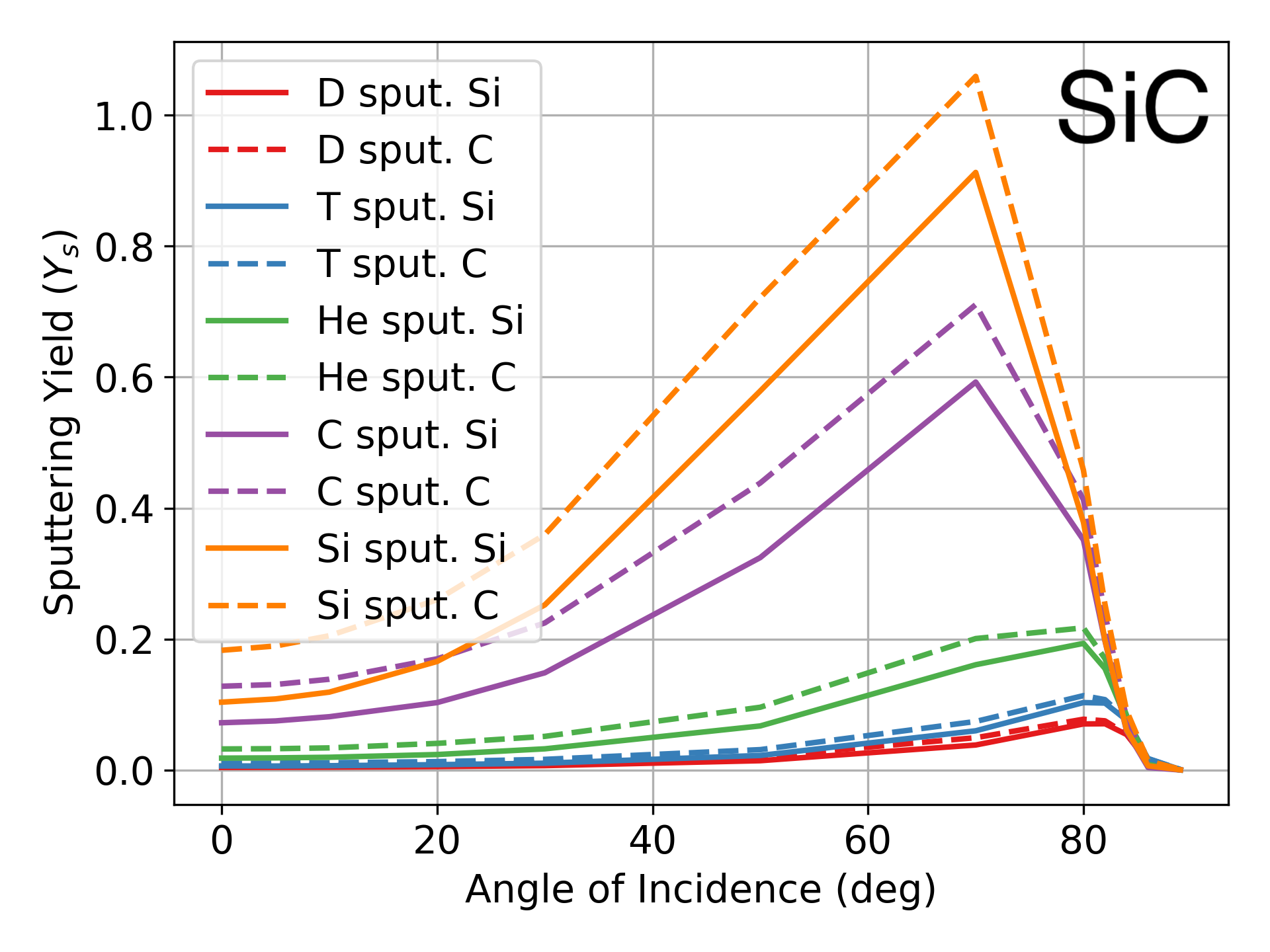}
		\caption{}
		\label{fig:Ys_theta_SiC}
	\end{subfigure}
	\\
	\begin{subfigure}{0.46\textwidth}
		\centering
		\captionsetup{justification=centering,font=small}
		\includegraphics[width=\textwidth]{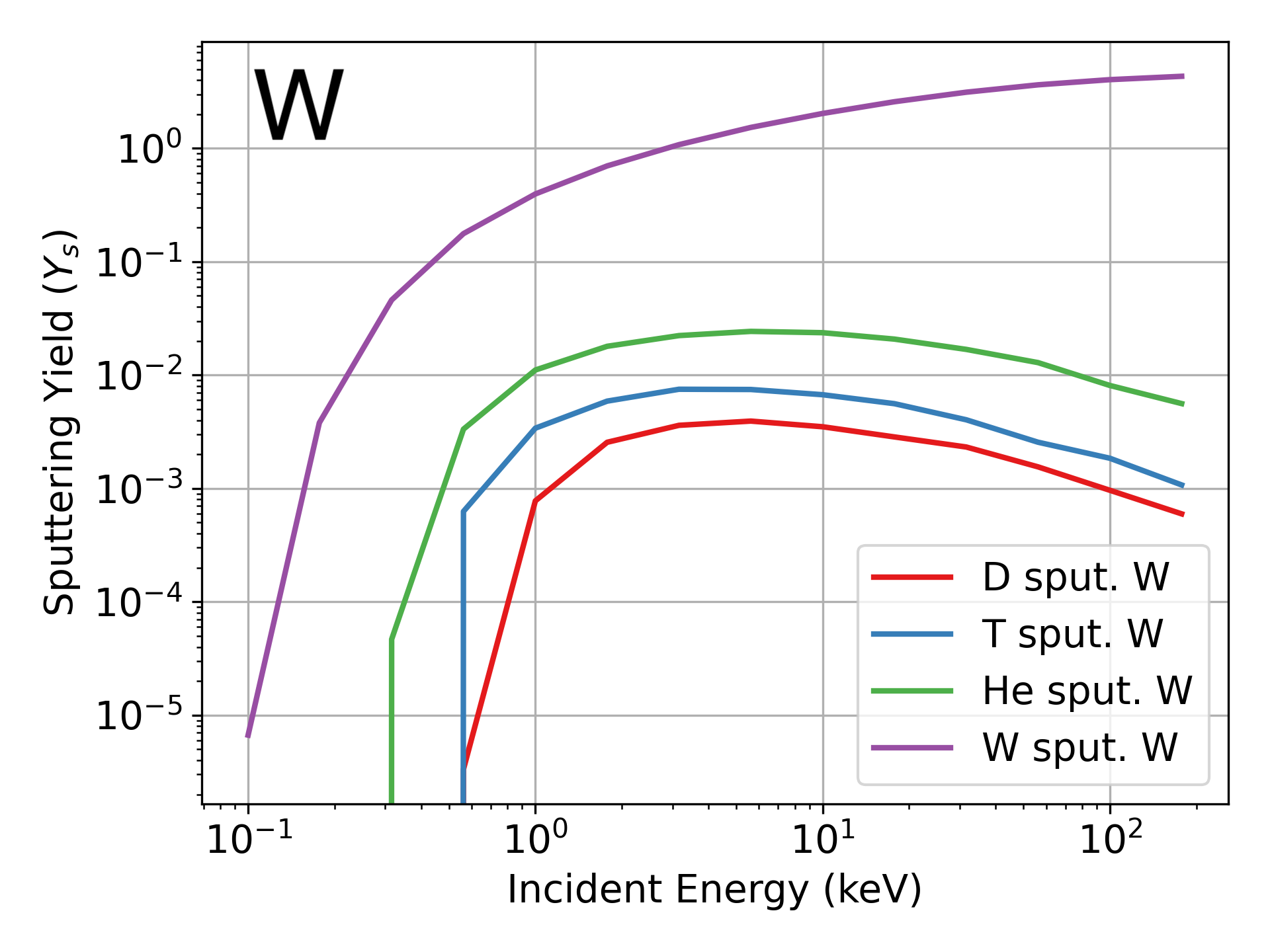}
		\caption{}
		\label{fig:Ys_energy_W}
	\end{subfigure}
	\begin{subfigure}{0.46\textwidth}
		\centering
		\captionsetup{justification=centering,font=small}
		\includegraphics[width=\textwidth]{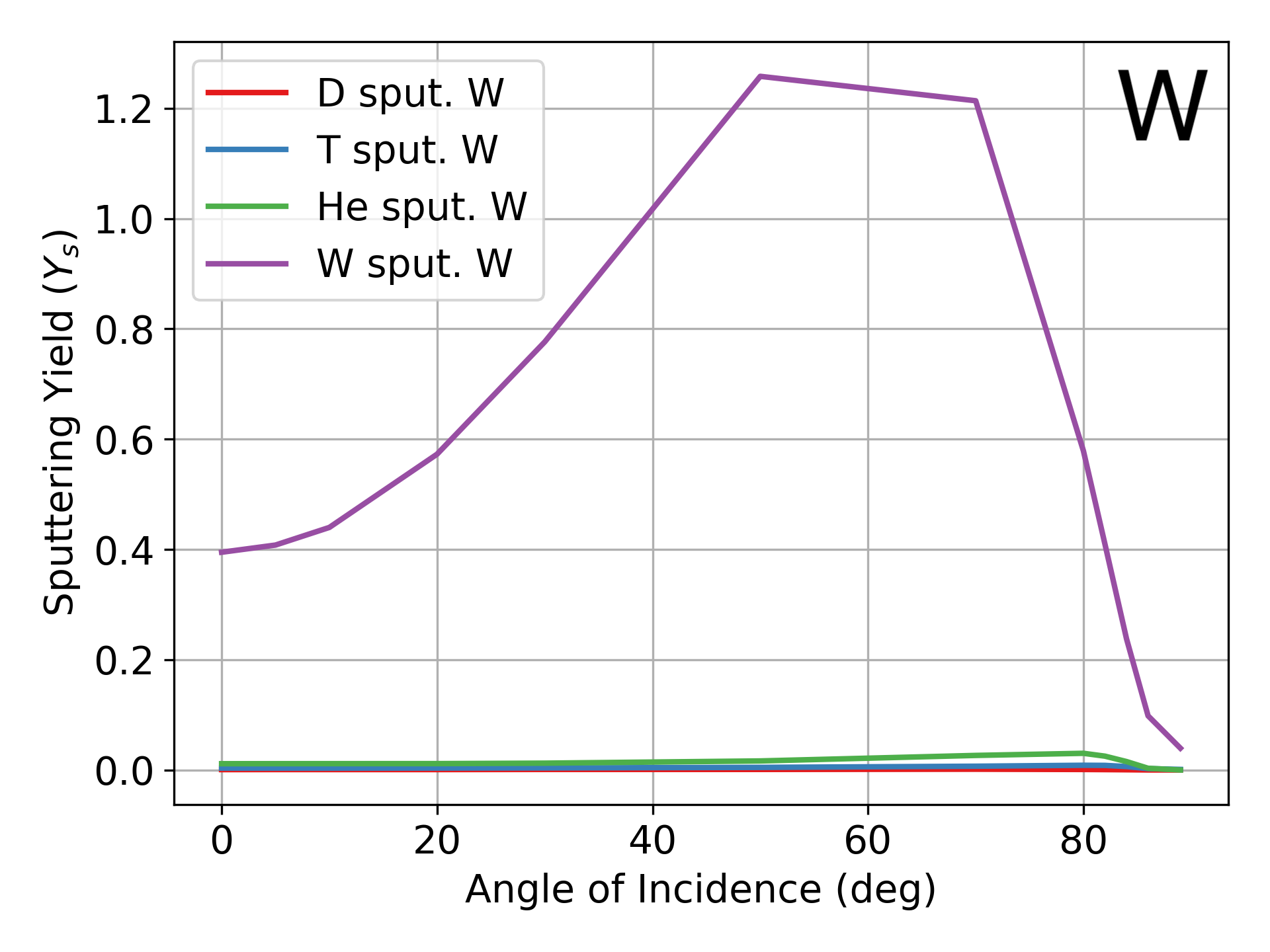}
		\caption{}
		\label{fig:Ys_theta_W}
	\end{subfigure}
	\caption{Results of sputtering yield study performed in RustBCA with the parameters in \cref{tab:rustbca_settings}. Common plasma ions and self-sputtering for (a) \gls{hbn}, (c) \ce{SiC}, and (e) \ce{W} at various ion energies and normal incidence. A similar study was performed for (b) \gls{hbn}, (d) \ce{SiC}, and (f) \ce{W} at various angles of incidence and 1 \si{\kilo \electronvolt} ion energy.}
	\label{fig:sputtering}
\end{figure}

A parameter sweep was performed to investigate the effects of incident energy and angle on sputtering yield, depth of penetration, and material damage. Sputtering yields for all primary ions (deuterium and tritium), $\alpha$ particles, and self-sputtering are shown in \cref{fig:sputtering}. The energy scan ranges from 100 \si{\electronvolt} to 180 \si{\kilo \electronvolt}, energies that might be expected in \gls{cmfx} and a future reactor. Additionally, $\alpha$ particles that have been confined and thermalized in the plasma are also included.

The sputtering yield ($Y_s$) remains below unity for both \ce{BN} and \ce{SiC} across all energies, angles of incidence, and ion species, with \ce{SiC} exhibiting higher overall sputtering rates. As ion energy increases, $Y_s$ exhibits local maxima (\cref{fig:Ys_energy_BN}), attributed to deeper ion implantation, which reduces energy transfer to surface atoms and consequently lowers sputtering rates. A notable decrease in sputtering occurs at very high angles of incidence for all materials. This reduction is due to insufficient perpendicular energy at glancing angles to overcome surface atom bonding. These findings suggest that future insulator designs could benefit from orienting targets either nearly parallel or with normal angles less than 20\degree to the magnetic field to minimize erosion.

\begin{figure}
	\centering
	\begin{subfigure}{0.46\textwidth}
		\centering
		\captionsetup{justification=centering,font=small}
		\includegraphics[width=\textwidth]{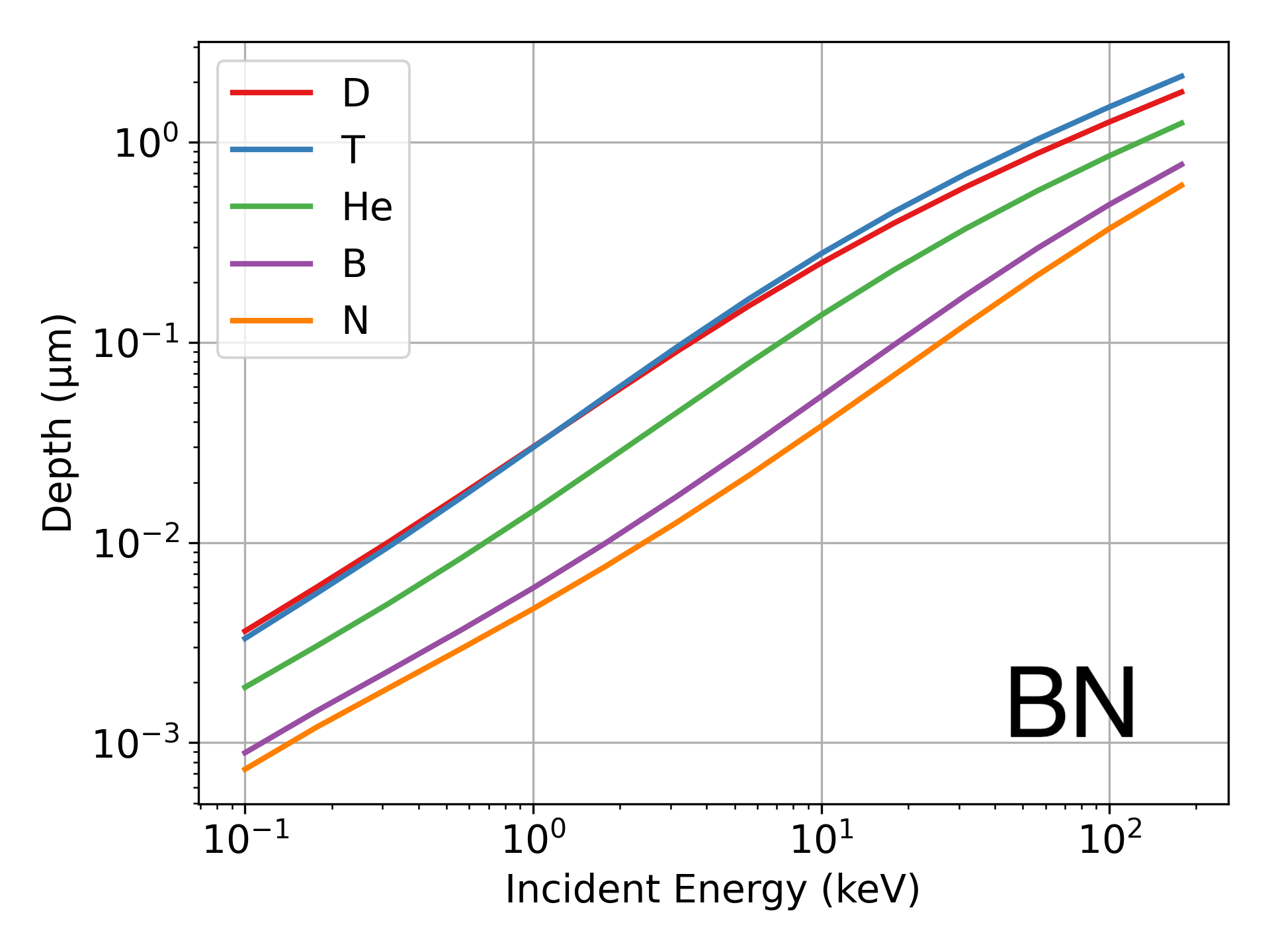}
		\caption{}
		\label{fig:d_energy_BN}
	\end{subfigure}
	\begin{subfigure}{0.46\textwidth}
		\centering
		\captionsetup{justification=centering,font=small}
		\includegraphics[width=\textwidth]{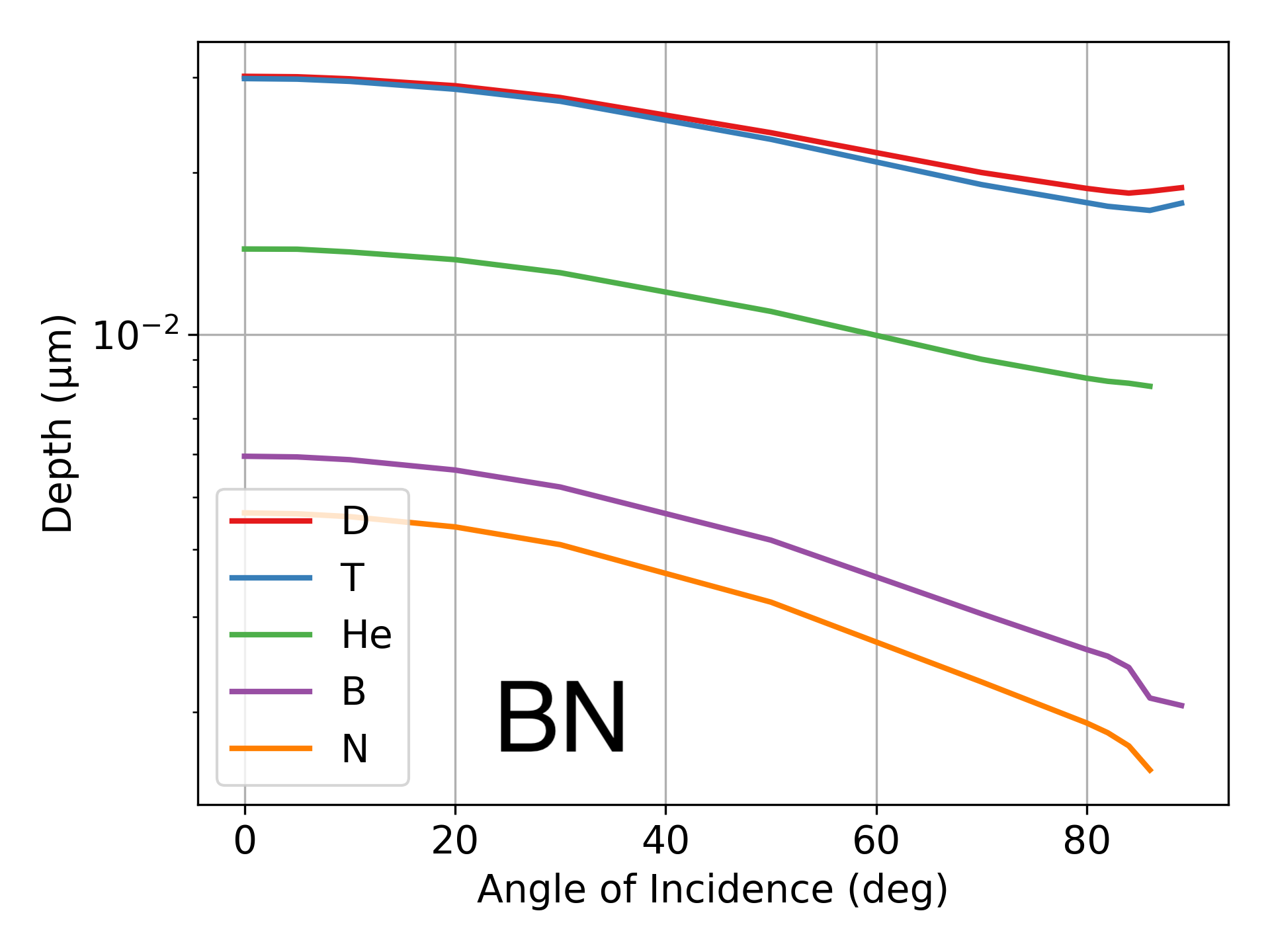}
		\caption{}
		\label{fig:d_theta_BN}
	\end{subfigure}
	\\
	\begin{subfigure}{0.46\textwidth}
		\centering
		\captionsetup{justification=centering,font=small}
		\includegraphics[width=\textwidth]{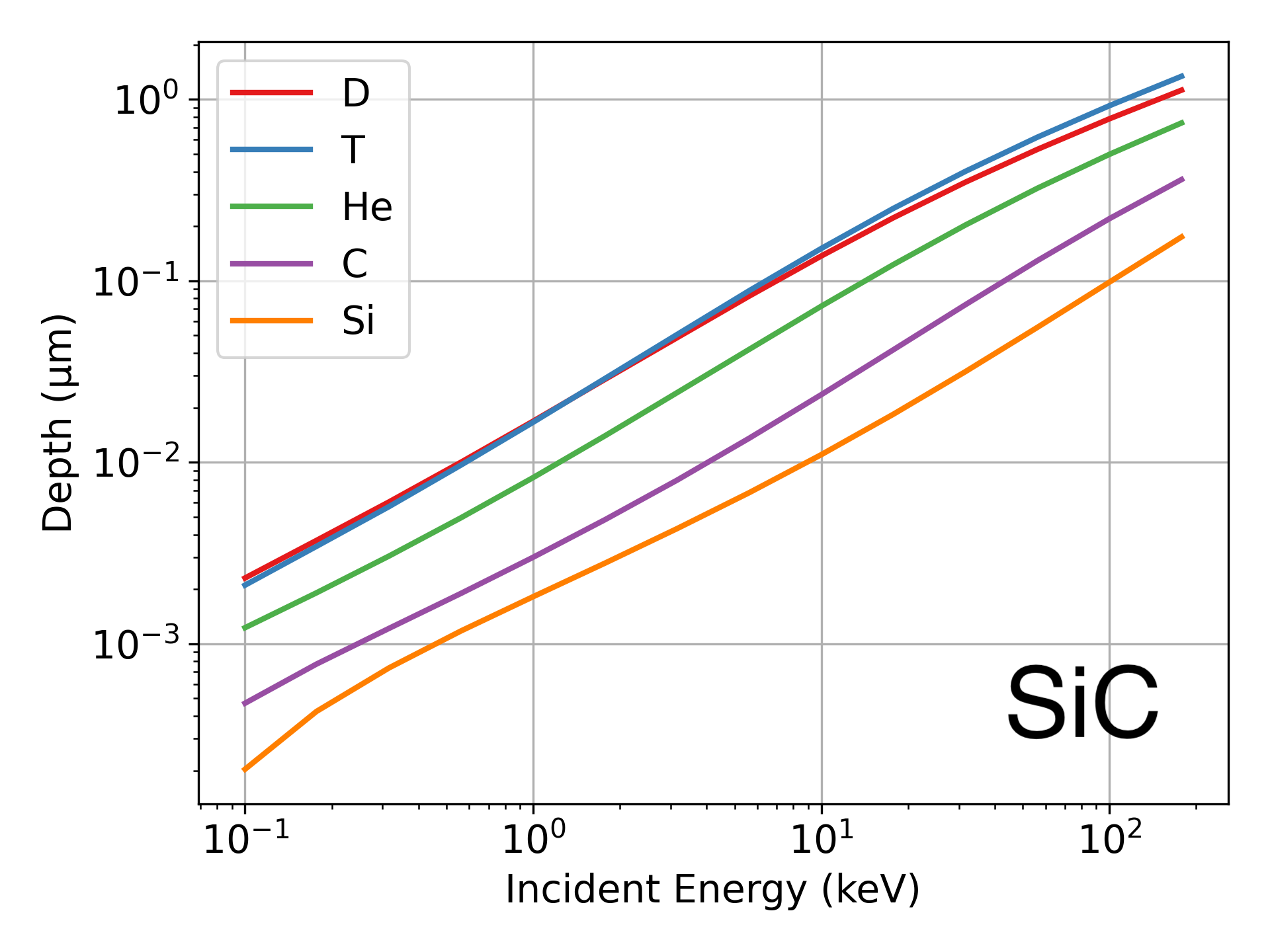}
		\caption{}
		\label{fig:d_energy_SiC}
	\end{subfigure}
	\begin{subfigure}{0.46\textwidth}
		\centering
		\captionsetup{justification=centering,font=small}
		\includegraphics[width=\textwidth]{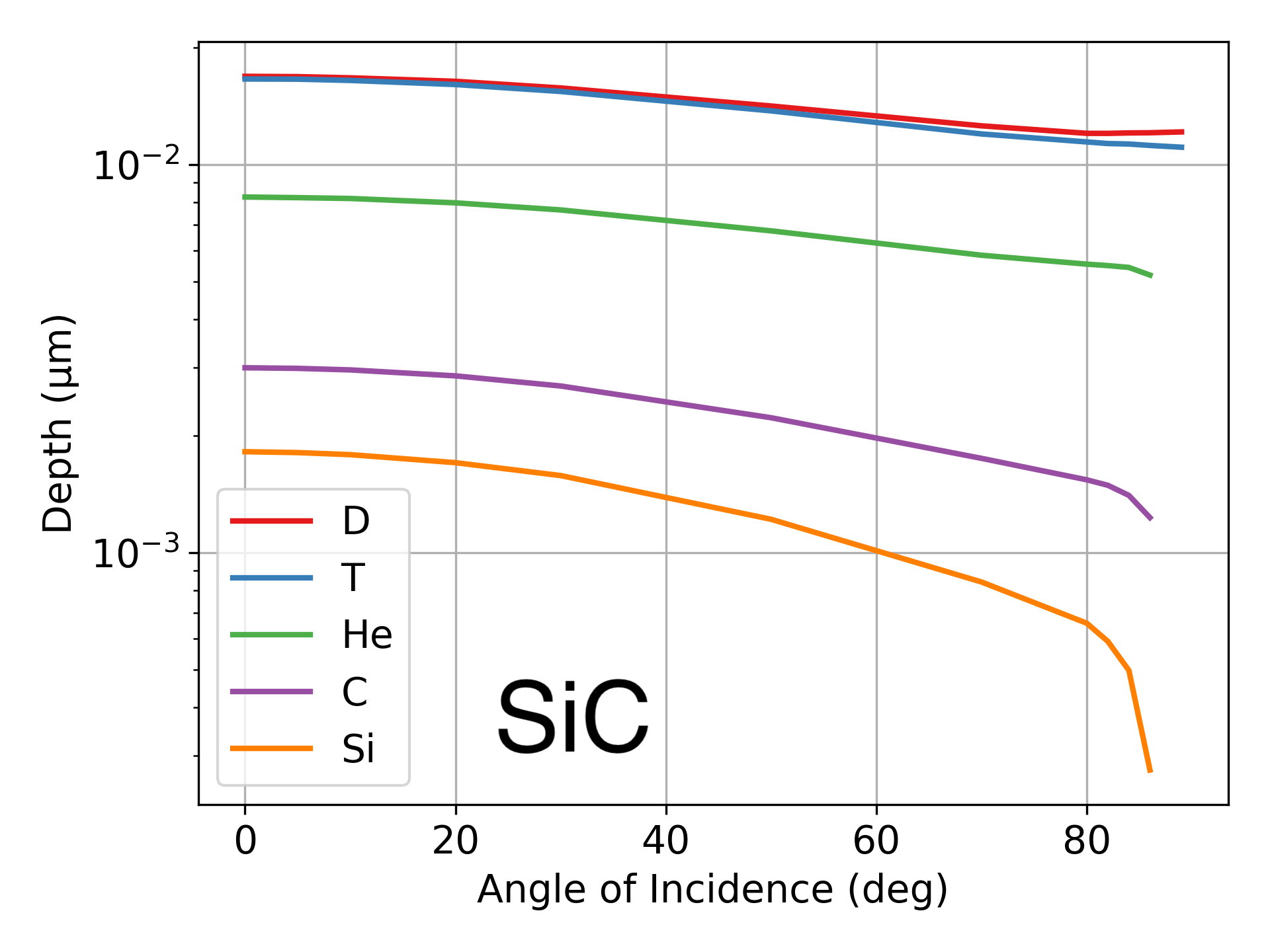}
		\caption{}
		\label{fig:d_theta_SiC}
	\end{subfigure}
	\\
	\begin{subfigure}{0.46\textwidth}
		\centering
		\captionsetup{justification=centering,font=small}
		\includegraphics[width=\textwidth]{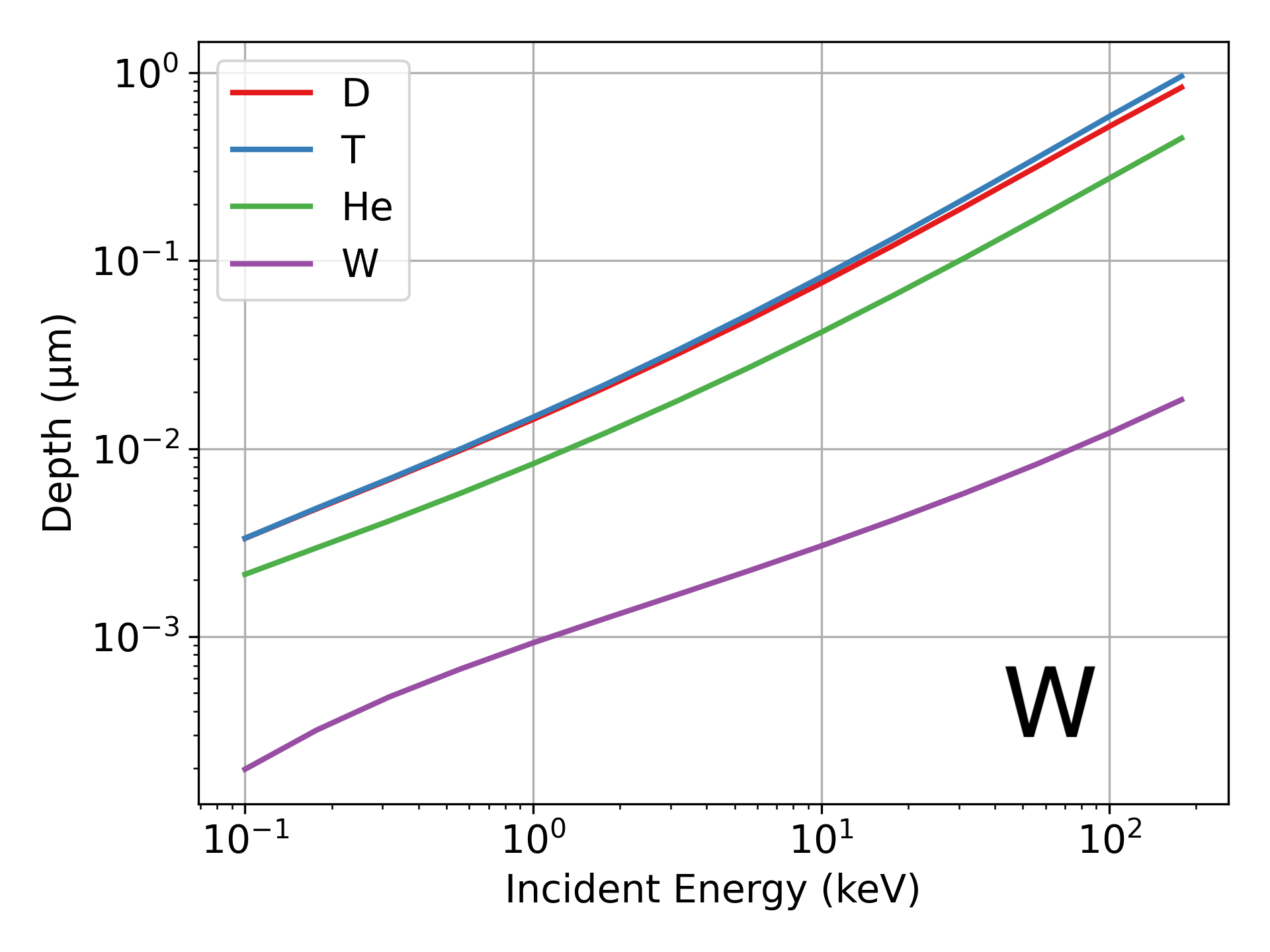}
		\caption{}
		\label{fig:d_energy_W}
	\end{subfigure}
	\begin{subfigure}{0.46\textwidth}
		\centering
		\captionsetup{justification=centering,font=small}
		\includegraphics[width=\textwidth]{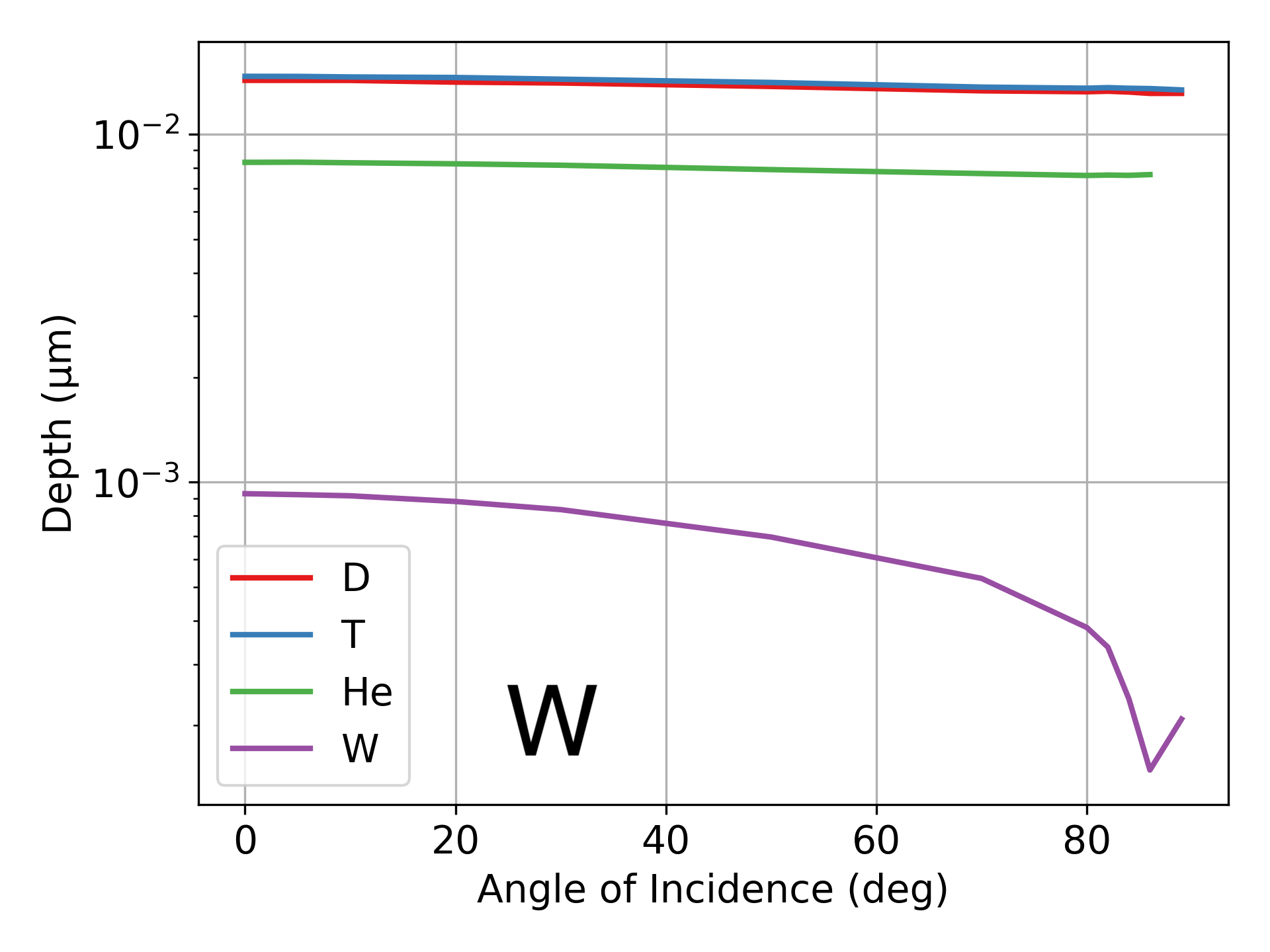}
		\caption{}
		\label{fig:d_theta_W}
	\end{subfigure}
	\caption{Results of average penetration depth study performed in RustBCA with the parameters in \cref{tab:rustbca_settings}. Common plasma ions and self-sputtering for (a) \gls{hbn}, (c) \ce{SiC}, and (e) \ce{W} at various ion energies and normal incidence. A similar study was performed for (b) \gls{hbn}, (d) \ce{SiC}, and (f) \ce{W} at various angles of incidence and 1 \si{\kilo \electronvolt} ion energy.}
	\label{fig:implantation}
\end{figure}

When ions implant in the material surface, they displace atoms in the existing lattice. Over time, this effect can transform a material from crystalline to amorphous. resulting in the degradation of electrical, thermal, and mechanical properties. We therefore assumed that average penetration depth of implanted ions is analogous to depth of amorphization (\cref{fig:implantation}). As long as this depth is near the surface of the material, it will not have a large effect on the bulk properties. As expected, deuterium and tritium can penetrate much deeper in all cases because they are light ions, and therefore have higher speeds for a given energy. However, the maximum depth is still $\lesssim$1 \si{\micro \meter}, which indicates that the depth of defect production is small compared to the thickness of \ce{BN} required to standoff the applied voltages ($\sim$ \si{\milli \meter}). Additionally, if the \glspl{pfc} are placed at glancing angles ($\gtrsim$82\si{\degree}), the average penetration depth and total count of implanted ions decreases greatly.

\begin{figure}
	\centering
	\begin{subfigure}{0.4\textwidth}
		\centering
		\captionsetup{justification=centering,font=small}
		\includegraphics[width=\textwidth]{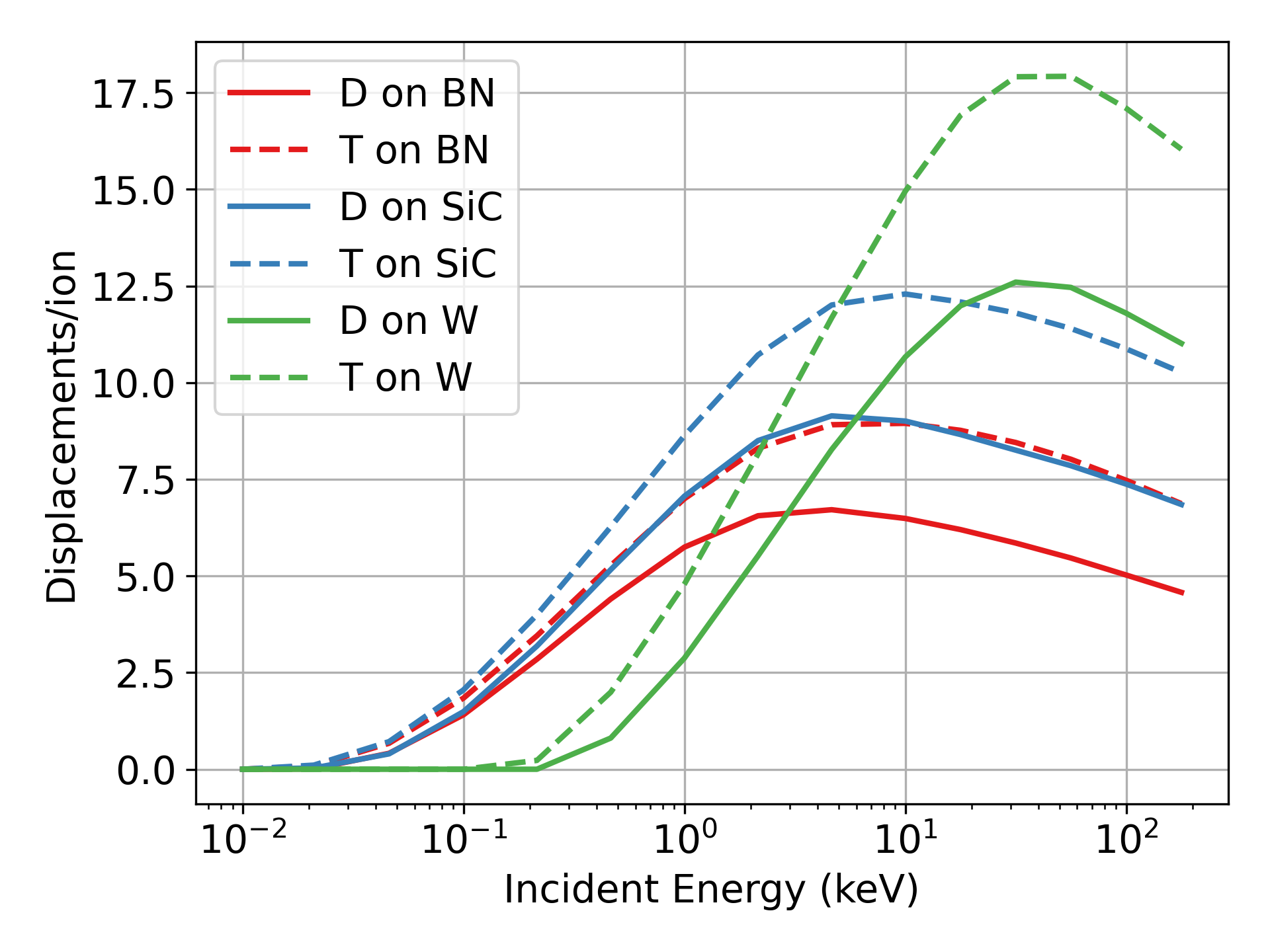}
		\caption{}
		\label{fig:dpa_energy}
	\end{subfigure}
	\begin{subfigure}{0.4\textwidth}
		\centering
		\captionsetup{justification=centering,font=small}
		\includegraphics[width=\textwidth]{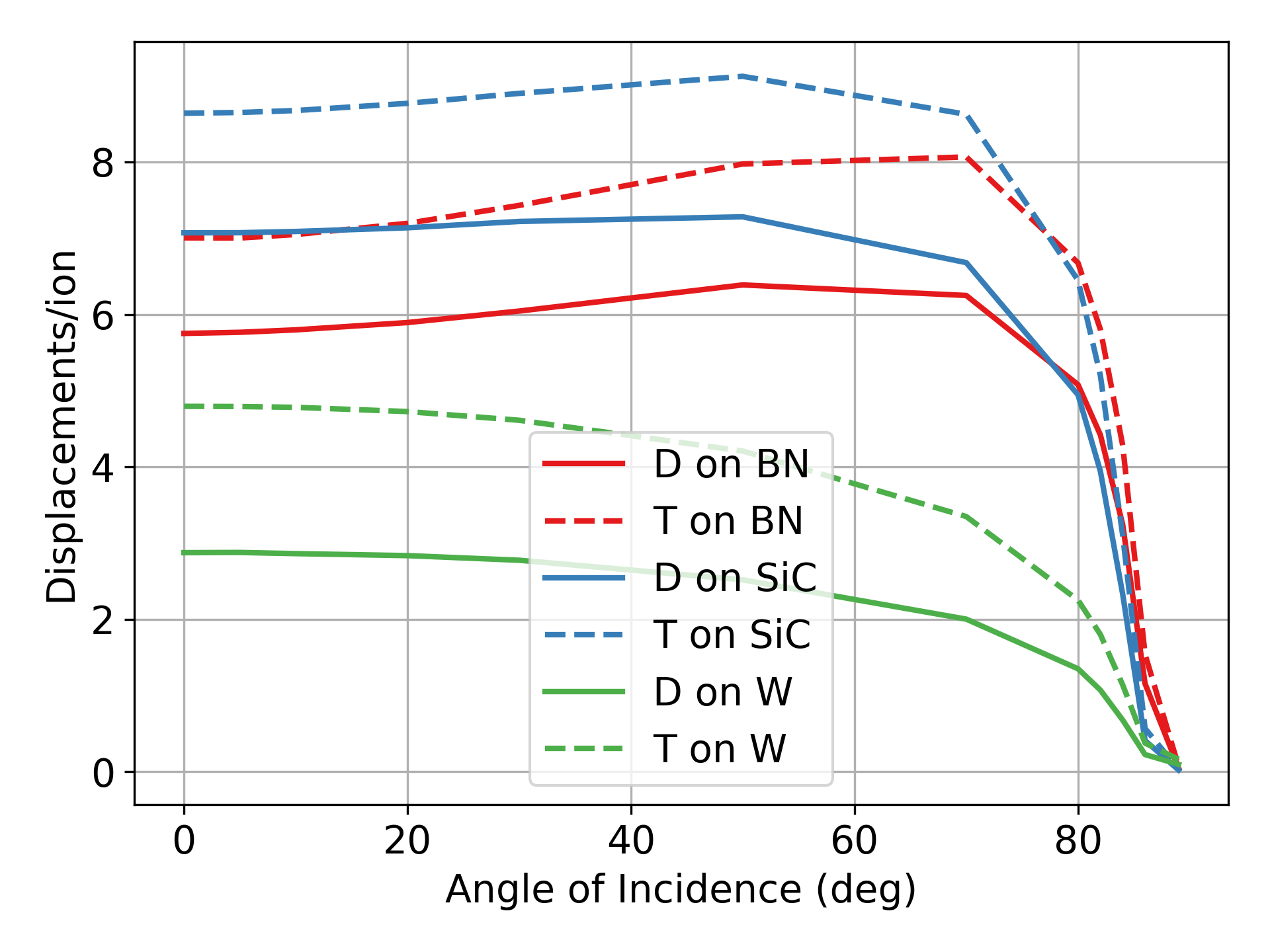}
		\caption{}
		\label{fig:dpa_theta}
	\end{subfigure}
	\caption{Results of \gls{dpa} study performed in RustBCA with the parameters in \cref{tab:rustbca_settings}. The curves are continous and not quantized to integer values because the total displacements were normalized by the total number of incoming ions. Results were found for a range of ion (a) energies at normal incidence and (b) angles of incidence with 1 \si{\kilo \electronvolt} ion energy.}
	\label{fig:dpa}
\end{figure}

Because absolute \gls{dpa} values are difficult to accurately predict without knowing the flux at the surface of a given component, results were computed normalized to the number of incident ions (\cref{fig:dpa}). Note that these values are not \gls{dpa} strictly, but simply how many atoms are displaced per incident particle. To predict \gls{dpa}, the total ion flux and exposure time would need to be known. Only primary plasma ion species, not impurities, were investigated.

The \gls{dpa} for \ce{BN} was less than that of \ce{SiC} in all cases. And, as expected, \gls{dpa} was higher for tritium than deuterium because of its larger mass. Especially if the \glspl{pfc} are at a glancing angle, \gls{dpa} drops significantly. Past a threshold in energy, \gls{dpa} decreases because the particles penetrate more deeply, losing their energy through small angle collisions (electronic stopping) that do not displace the target atoms (nuclear stopping).

\begin{figure}
	\centering
	\includegraphics[width=0.6\textwidth]{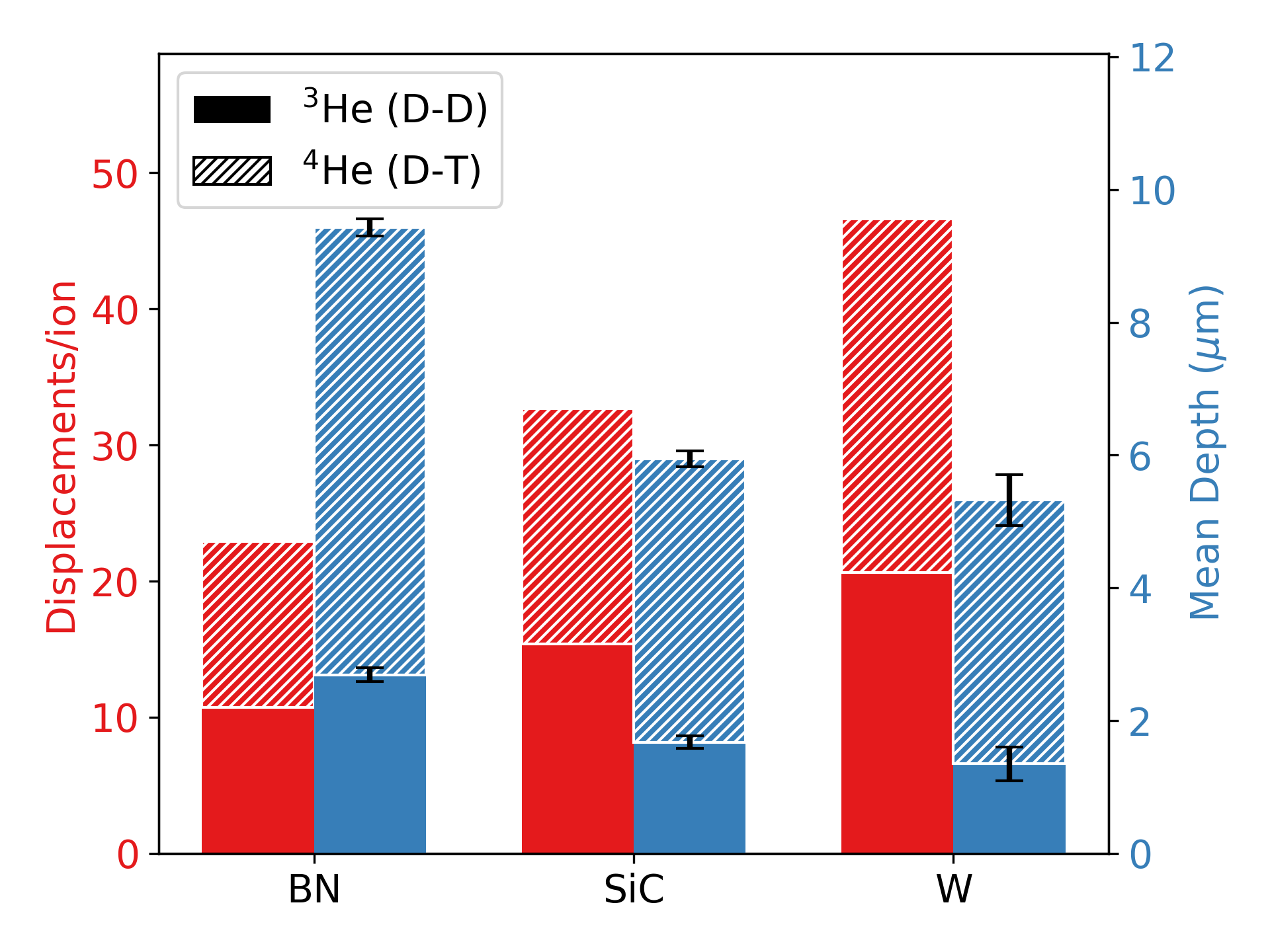}
	\caption{A RustBCA simulation was performed with fusion \ce{^3He} and $\alpha$ particles from the D-D (solid) and D-T (hashed) reactions, respectively. The number of displacements per ion (red) and the depth of penetration (blue) were compared between \gls{hbn}, \ce{SiC} and \ce{W}. Note values in each bar segment are stacked progressively, such that the values are cumulative.}
	\label{fig:dpa_alpha}
\end{figure}

The results thus far have only discussed $\alpha$ particles that have been confined in the plasma and thermalized to $T_i$. However, fusion helium particles are born at much higher energies from either the \ce{D + T -> ^{4}He\ (3.52\ MeV) + n\ (14.1\ MeV)} or \ce{D + D -> ^{3}He\ (0.82\ MeV) + n\ (2.45\ MeV)} reactions. While these $\alpha$ particles are not primarily responsible for \gls{pfc} degradation through sputtering \citep{Bauer_1979}, they are the cause of sub-surface damage that can be much more detrimental to the structural properties. Additionally, they will penetrate into the material significantly further than the primary plasma ions, which means that the depth of amorphization can be larger. Both the displacements and the depth of penetration for fusion $\alpha$ and \ce{^3He} particles are shown in \cref{fig:dpa_alpha}.

As the mass of the target species increases, the number of displacements increases, while the depth of penetration decreases. The helium particles can impart more energy to heavy targets during elastic collisions, and because the bulk binding energy is roughly the same for all targets, it is more likely that a heavier atom will be displaced than a lighter one. Indeed, for a similar reason, the depth of penetration has an inverted behavior. Because the helium particles lose less energy during a collision with a light atom than a heavy one, they can travel farther in the light material. It would seem that a lighter target material would be more suitable for exposure to fusion $\alpha$ particles because there is cumulatively less material damage, and it is spread over a larger distance. That being said, there is also a downside to larger depth of penetration in that a larger portion of the surface may become amorphized and lose its bulk properties.

There are some differences between D-D \ce{^3He} and D-T $\alpha$ particles as well. The penetration depth is significantly larger for D-T because the $\alpha$ energy (3.52 \si{\mega \electronvolt}) is larger than the \ce{^3He} energy (0.82 \si{\mega \electronvolt}), though the increase is not linear with energy. Similarly, the displacements generated by D-T $\alpha$ particles are slightly larger than those of D-D \ce{^3He} particles. This behavior can be attributed to the higher energy and mass of $\alpha$ particles.

While ion-material interactions dominate surface effects, neutron damage presents distinct challenges for bulk material properties that require separate consideration.

\subsection{Neutron Damage}
\label{sec:Neutron Damage}
Fusion neutrons can also cause significant damage through displacement and transmutation. Luckily, there are no long-lived radioactive isotopes that are produced from boron transmutation \citep{Jung2024}, though helium is produced through the \ce{^{10}B(n, $\alpha$)^7Li} reaction. Note that naturally occurring boron is $\sim$80\% \ce{^{11}B}, but the neutron capture cross-section for \ce{^{11}B} is much smaller than that of \ce{^{10}B} \citep{Brown2018}. This helium could accumulate and produce ``bubbles'' as has been observed in other nitrides \citep{Kozlovskiy2021}. Note that ``fuzz'' is unlikely to occur, because it has only been observed in metallics to date \citep{Hammond2017}.

For nitrogen, on the other hand, several absorption reactions could be important \citep{Brown2018}. At thermal energies, the \ce{^{14}N(n, p)^{14}C} reaction is dominant, which does produce activated carbon-14 with a half-life of 5,730 years. At higher energies ($\gtrsim$1 \si{\mega \electronvolt}), there is a resonance for both the \ce{^{14}N(n, $\alpha$)^{11}B} and \ce{^{14}N(n, 2$\alpha$)^{7}Li} reactions, neither of which produce radioactive products. However, these reactions may also contribute to the buildup of helium in the \glspl{pfc}. Moreover, it is possible that the \gls{hbn} layer is sufficiently thin such that the neutrons do not thermalize and therefore do not produce radioactive \ce{^{14}C} products.

To better understand how materials relevant to centrifugal mirrors are affected by fusion neutrons, a model was developed in OpenMC \citep{Romano2015}.

\subsubsection{Simulation Parameters}
\label{sec:Neutron Parameters}
All simulations were performed with the standard inputs listed in \cref{tab:openmc_settings_materials}. Several parameters were tracked in the target to quantify neutron-material interaction: \verb|flux|, \verb|absorption|, and \verb|damage-energy|. \Gls{dpa} is proportional to \verb|damage-energy| and is given by the standard Norgett, Robinson, and Torrens (NRT) formula \citep{Norgett1975}. It is inversely proportional to the threshold damage energy ($E_d$). The values of $E_d$ for \gls{hbn} from \citep{Kotakoski2010}, \ce{SiC} from \citep{Lucas2005} and \ce{W} from \citep{Banisalman2017} were used. For the 50/50 mixtures of \gls{hbn} and \ce{SiC}, the average of the two values of $E_d$ was used.

The materials modeled for neutron interactions parallel those discussed in \cref{sec:Plasma-Material Interaction Modeling}, and a comprehensive rationale for material selection is provided in \cref{sec:Material Selection}.  Neutrons from both the D-D and D-T reactions were simulated.

\begin{table}
	\begin{center}
		\begin{tblr}{colspec={rcc},
				hline{1}={2-3}{1pt},
				hline{2}={2-3}{0.5pt},
				hline{Z}={2-3}{1pt},}
			& Setting & Value \\
			\ldelim\{{4.2}{*}[Source] & Particles per batch & 10000 \\
			& Source angle & 0\degree \\
			& Source energy & {2.45 \si{\mega \electronvolt} (D-D) or\\14.1 \si{\mega \electronvolt} (D-T)} \\
			\ldelim\{{3.2}{*}[Target] & Material & Boron nitride \\
			& Side length & 120 \si{\centi \meter} \\
			& Cross-sections & ENDF/B-VIII.0 \citep{Brown2018} \\
			\ldelim\{{2}{*}[Simulation] & Run mode & Fixed source \\
			& Batches & 100 \\
		\end{tblr}
		\caption{Default settings for OpenMC materials damage model.}
		\label{tab:openmc_settings_materials}
	\end{center}
\end{table}

\subsubsection{Results}
\label{sec:Neutron Results}
\begin{figure}
	\centering
	\includegraphics[width=0.8\textwidth]{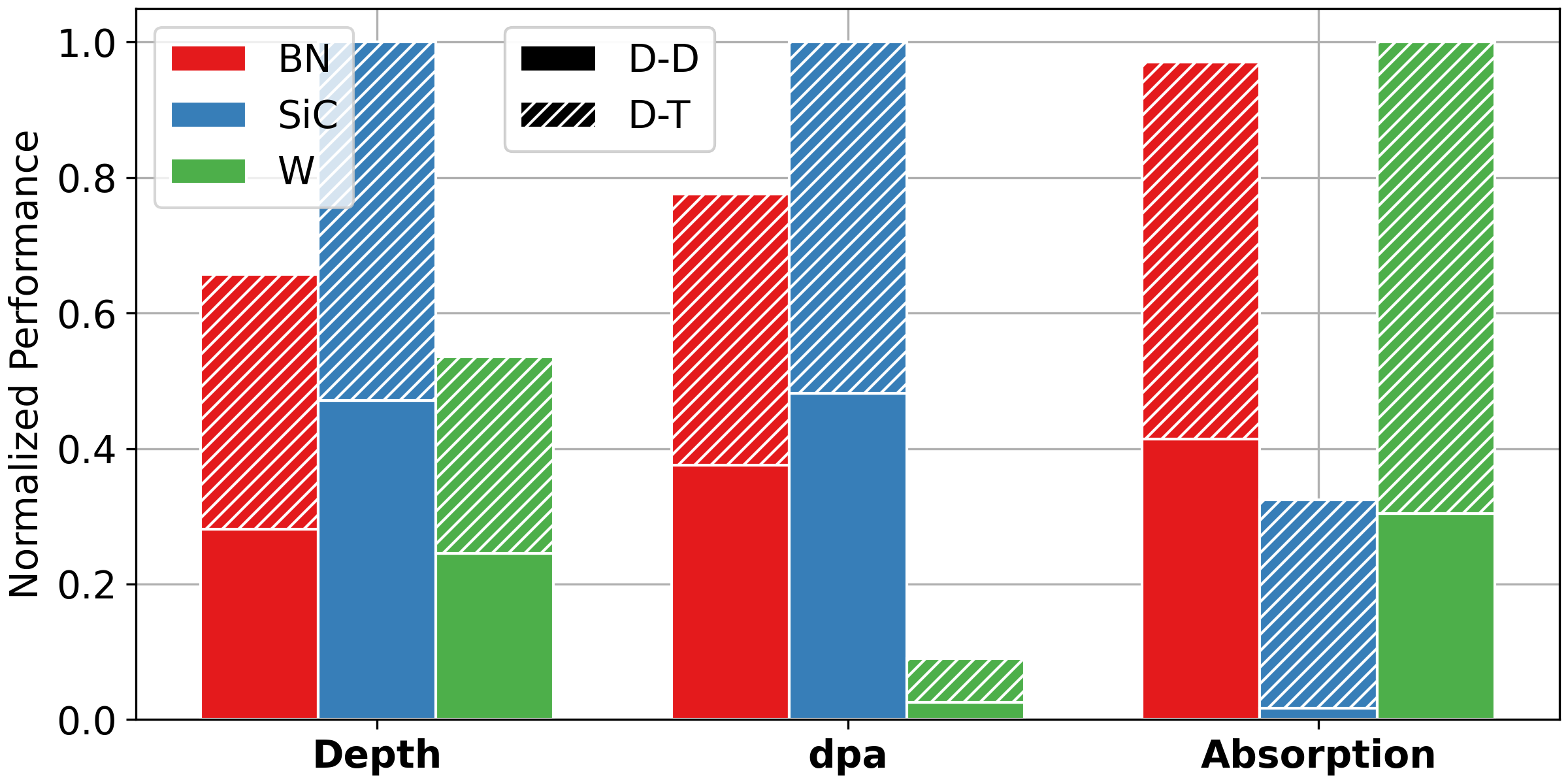}
	\caption{Summarized results of neutron-induced damage. The \gls{dpa} and absorption were integrated along the depth, and the `Depth' was chosen to be where the flux reached 1\% of its surface value. Note values in each bar segment are stacked progressively, such that the values are cumulative.}
	\label{fig:neutron_materials_damage_summary}
\end{figure}

\Cref{fig:neutron_materials_damage_summary} summarizes the integrated \gls{dpa}, integrated absorption, and penetration depth (defined as the depth where flux reaches 1\% of its surface value), normalized to the maximum value among the different materials for easy comparison between materials. \Cref{tab:openmc_damage_results} provides the numerical values for depth and absorption (\gls{dpa} was excluded because it does not carry physical meaning for these simulations). \ce{SiC} exhibits the highest neutron transparency, followed by \ce{BN}, then \ce{W}. For centrifugal mirror applications, \ce{SiC} may be preferable for neutron transparency, but \ce{BN} offers reasonable performance. However, while the penetration depth at these energies is $\sim$50-100 \si{\centi \meter}, the \gls{hbn} layer may only need to be $\sim$\si{\milli \meter}, based on the standoff needed for the applied voltage, allowing most neutrons to pass through.

\Gls{dpa} is an important metric to understand how the crystallinity of a \gls{pfm} may change. This is especially true for the insulators in centrifugal mirrors, as a degradation in their electrical properties can have drastic effects on device operation. However, prior studies for \ce{SiC} demonstrated that bulk properties do not scale linearly with \gls{dpa} and have a strong dependence on the temperature at which the material was exposed \citep{Katoh2003,Katoh2015}. Nevertheless, we see that \ce{SiC} sustains the most damage of all the materials, followed by \ce{BN} and then \ce{W}.

We also see that tungsten and boron nitride will absorb a significant portion of the incident neutrons in comparison to silicon carbide. The large \ce{^{10}B(n, $\alpha$)^7Li} cross-section at fast neutron energies leads to higher absorption. The transmutation products in boron nitride will lead to the production of $\alpha$ particles, potentially leading to the development of ``bubbles'' \citep{Kozlovskiy2021}, but this process has not been studied to date in boron nitride.

\begin{table}
	\begin{center}
		\begin{tblr}{colspec={rccc},
				hline{1}={2-4}{1pt},
				hline{2}={2-4}{0.5pt},
				hline{Z}={2-4}{1pt},}
			& Target & Depth (\si{\centi \meter}) & Absorption (per neutron) \\
			\ldelim\{{3.2}{*}[DD] & \gls{hbn} & 54 & \num{6.6e-2}  \\
			& \ce{SiC} & 90 & \num{2.6e-3} \\
			& \ce{W} & 47 & \num{4.8e-2} \\
			\ldelim\{{3.2}{*}[DT] & \gls{hbn} & 72 & \num{8.8e-2} \\
			& \ce{SiC} & 101 & \num{4.9e-2} \\
			& \ce{W} & 55 & 0.11  \\
		\end{tblr}
		\caption{Numerical values used to produced \cref{fig:neutron_materials_damage_summary}. \Gls{dpa} is not included because the quantitative value does not have physical meaning in this simulation.}
		\label{tab:openmc_damage_results}
	\end{center}
\end{table}

While computational models provide valuable predictions, experimental validation under fusion-relevant conditions is essential. Therefore, we developed a testing program using two complementary facilities: PISCES-A and \gls{cmfx}, which are described in the following section.

\section{Sample Exposure Materials and Methods}
\label{sec:Sample Exposure}
The overarching goal of this section is to experimentally investigate the behavior and changes induced in \gls{hbn} when exposed to the extreme conditions present in fusion-relevant plasmas.

While studies on the erosion of \ce{SiC} due to ion bombardment have been performed extensively \citep{Mohri1978,Plank1996,Balden2000,Balden2001,Koller2019,Sinclair2021}, prior research on \gls{hbn} degradation in fusion environments is limited. However, studies in related fields provide some insights. For instance, investigations into Hall thruster wall materials have examined the effects of \ce{Xe^+} ions on \gls{hbn} and its composites \citep{Duan2016_2,Satonik2014}. These studies identified two key mechanisms for surface erosion: (1) sputtering and (2) detachment of entire grains from the surface.
Each of these mechanisms plays a role in our analysis, so a brief description of the effects is as follows:
\begin{enumerate}
	\item \textbf{Sputtering} results in even surface wear, manifesting as grains with rounded edges.
	\item \textbf{Grain ejection} leads to a porous surface microstructure, where entire grains have been removed.
\end{enumerate} 
Building on this foundation, this section will explore the effects of fusion-relevant plasmas on \gls{hbn} in comparison to \ce{SiC}.

Two experimental facilities were utilized: PISCES-A at UC San Diego for controlled plasma exposure, and \gls{cmfx} for high-energy ion bombardment testing. Multiple analytical techniques were used to characterize the samples before and after exposure to examine surface morphology, composition, and erosion rate.

\subsection{Sample Materials}
\label{sec:Sample Materials}
Saint-Gobain Ceramics supplied AX05 grade \gls{hbn} samples, which have $>$99.5\% purity and the best thermal conductivity and dielectric strength of any grade provided. The coupons have a large porosity (19.3\%) and worse mechanical properties than some of the other composite grades. Nonetheless, the high purity is useful for establishing baseline performance, reducing interactions with impurities, and potentially guiding design of future composites. The \ce{SiC} samples, also supplied by Saint-Gobain \citep{saintgobain_sic}, were selected from the Hexoloy SA sintered grade. This choice was motivated by two factors: the grade's superior thermomechanical and electrical properties and its established use in other fusion devices \citep{Hinoki2005,Minami2007,Kishimoto2011}. The \gls{hbn} and \ce{SiC} coupons were manufactured 1" diameter and exhibited comparable purity levels. However, the \ce{SiC} samples have significantly lower porosity at $\sim$2\%.

\subsection{PISCES-A}
\label{sec:PISCES-A Exposure}
To test how \gls{hbn} is affected by plasma exposure, a collaboration was formed with \gls{ucsd} to expose samples on the PISCES-A device. The circular samples were 1" in diameter and 0.062" thick, and the mounting cap for the plasma exposure experiment was made of the same AX05 grade material to avoid impurity contamination.

The PISCES-A device is a linear plasma device that can operate for extended periods, has several plasma and in-situ materials diagnostics and has temperature control of the target \citep{Goebel1984}. Before exposure, the as-delivered samples were cleaned with isopropyl alcohol and outgassed in vacuum at temperatures of up to 650 \si{\degreeCelsius} to remove any absorbed water vapor. \cref{tab:pisces_parameters} provides the operating procedures for the exposure, where the only variable parameter was the ion energy. A control sample was left unexposed for comparison.

\begin{table}
	\begin{center}
		\begin{tblr}{Q[f,c]Q[f,c]Q[f,c]Q[f,c]Q[f,c]Q[f,c]Q[f,c]}
			\toprule
			Material & Gas & {Ion\\Energy (\si{\electronvolt})} & {Sample\\Temp. ($\si{\degreeCelsius}$)} & {Ion Flux\\(\si{\per \meter \squared \per \second})} & {Run\\time (\si{\minute})} & {Ion Fluence\\(\si{\per \meter \squared})} \\
			\midrule
			\gls{hbn} & Deuterium & 30, 70, 100 & 400 & \num{1.6e21} & 104 & \num{1e25} \\
			\bottomrule
		\end{tblr}
		\caption{Operational parameters for plasma exposure experiments on PISCES-A.}
		\label{tab:pisces_parameters}
	\end{center}
\end{table}

The surface analysis of PISCES-exposed samples provides valuable insights into the behavior of \gls{hbn} under plasma exposure. However, to fully evaluate \gls{hbn}'s potential as a \gls{pfm}, we subjected it to conditions similar to those in an actual fusion environment with higher ion energy in \gls{cmfx}.

\subsection{CMFX}
\label{sec:CMFX Exposure}
\Gls{cmfx} was constructed in part as a test-bed for characterizing insulating materials under high energy, steady-state ion flux. The high rotational velocity in \gls{cmfx} typically results in ion energies significantly larger than the ion temperature. This section discusses the design considerations, construction, and sample exposure methodology in \gls{cmfx}.

The samples were exposed at the midplane, where the particle density and temperature are highest. To estimate the ion energy and flux on the samples, we make several assumptions based on the $\bm{E} \times \bm{B}$ flow. The average rotational velocity $v$ is assumed to be entirely due to the applied electric field, giving $v = \infrac{E}{B}$, where $E = \infrac{\Phi}{2a}$, and $\Phi$ is the applied potential and $2a$ the radial plasma thickness. For an ion density $n_i$, the azimuthal (or toroidal) particle flux is calculated as $\varphi = n_i v = n_i \Phi / (2aB)$. We assume the particle energy is entirely due to its kinetic (and not thermal) energy; thus $\varepsilon_i = \frac{1}{2} m_i v^2$, where $m_i$ is the ion mass. The incident heat flux is then $\dot{W} = \varepsilon_i \varphi$. Lastly, the erosion rate is given by:
\begin{equation}
	\dot{z} = \frac{Y_s \varphi}{n_\mathrm{ins}},
\end{equation}
where $n_\mathrm{ins}$ is the insulators's particle density. \Cref{tab:sample_erosion_comparison} compares estimated parameters for \gls{cmfx} at midplane and a fusion power plant's insulators.

The ion flux and energy at the insulators in a reactor-relevant scenario were predicted using the \mctrans{} model \citep{Schwartz2024}. The parallel ion loss rate was calculated to be \num{2.8e20} \si{\per \second}, with ions lost at an energy of 170 \si{\kilo \electronvolt} (sum of kinetic and thermal energies). The particle flux was assumed to be distributed over the two insulators' area of 20 \si{\meter \squared}.

\begin{table}
	\begin{center}
		\begin{tblr}{lcc}
			\toprule
			Parameter & \Gls{cmfx} (at Midplane) & Reactor (at Insulator) \\
			\midrule
			$\Phi$ (\si{\kilo \volt}) & 15 & - \\
			$2a$ (\si{\centi \meter}) & 16  & - \\
			$B$ (\si{\tesla}) & 0.34 & - \\
			$n_i$ (\si{\per \meter \cubed}) & \num{1e18} & - \\
			$Y_s$ (-) & \num{2e-3} & \num{3e-4} \\
			$v$ (\si{\meter \per \second}) & \num{2.8e5} & - \\
			$\varphi$ (\si{\per \meter \squared \per \second}) & \num{2.8e23} & \num{1.4e19} \\
			$\varepsilon_i$ (\si{\kilo \electronvolt}) & 0.79 & 170 \\
			$\dot{W}$ (\si{\mega \watt \per \meter \squared}) & 35 & 0.38 \\
			$\dot{z}$ (\si{\nano \meter \per \second}) & 10 & \num{1.6e-4} \\
			\bottomrule
		\end{tblr}
		\caption{Comparison of plasma flux and erosion rate of \gls{hbn} samples exposed at the midplane of \gls{cmfx} and at the insulators at reactor-relevant scales. The numbers for the reactor scenario are estimated using the model developed in \citet{Schwartz2024}. The key takeaway is that one second of exposure in \gls{cmfx} results in erosion that is equivalent to many continuous days of runtime for an insulator in a reactor.}
		\label{tab:sample_erosion_comparison}
	\end{center}
\end{table}

In \gls{cmfx}, we can control the material erosion rate by changing the applied voltage, which affects both the flux and incident ion energy. While reactor ions have higher energies, the total power deposited ($\dot{W}$) at the midplane in \gls{cmfx} exceeds typical reactor insulator conditions due to much higher ion flux. However, the actual flux on the midplane sample is an overestimate because the sample creates a ``shadow'' that blocks the rotational paths of ions.

For the sample exposure, a design was developed incorporating the following key considerations: (i) rapid sample exchange capability, (ii) adjustable and repeatable sample positioning, and (iii) utilization of electrically insulating materials for components interacting with the plasma. The third requirement is crucial to prevent plasma shorting and high voltage arcs. While commercial sample transfer mechanisms exist, none satisfied all these criteria for \gls{cmfx}, and retrofitting them proved impractical and cost-prohibitive.

\begin{figure}
	\centering
	\includegraphics[width=0.9\textwidth]{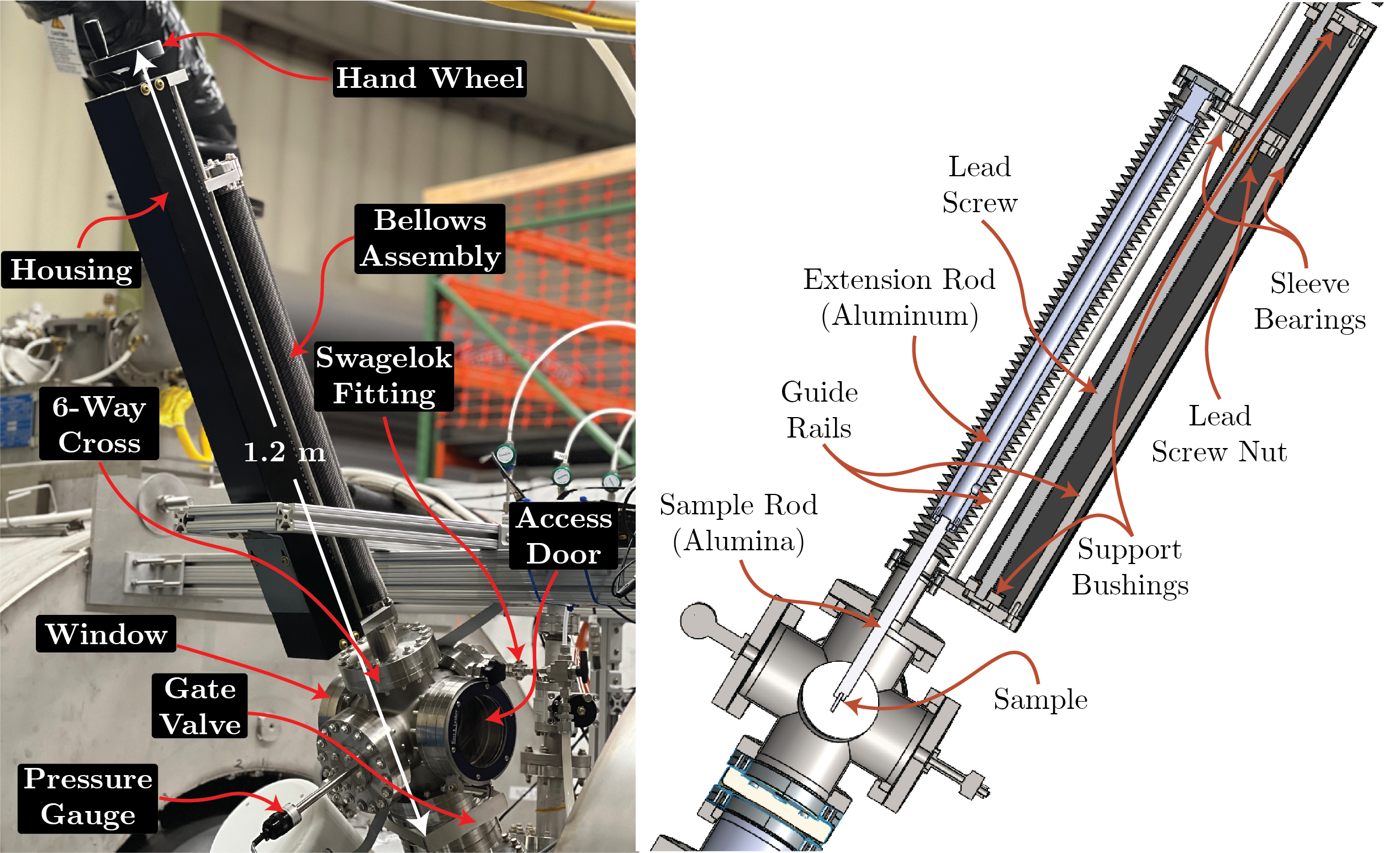}
	\caption{Pictures of sample feedthrough, including both an actual photo and cross-sectioned rendering.}
	\label{fig:feedthrough_labeled}
\end{figure}

\begin{figure}
	\centering
	\begin{subfigure}{0.49\textwidth}
		\centering
		\captionsetup{justification=centering,font=small}
		\includegraphics[width=\textwidth]{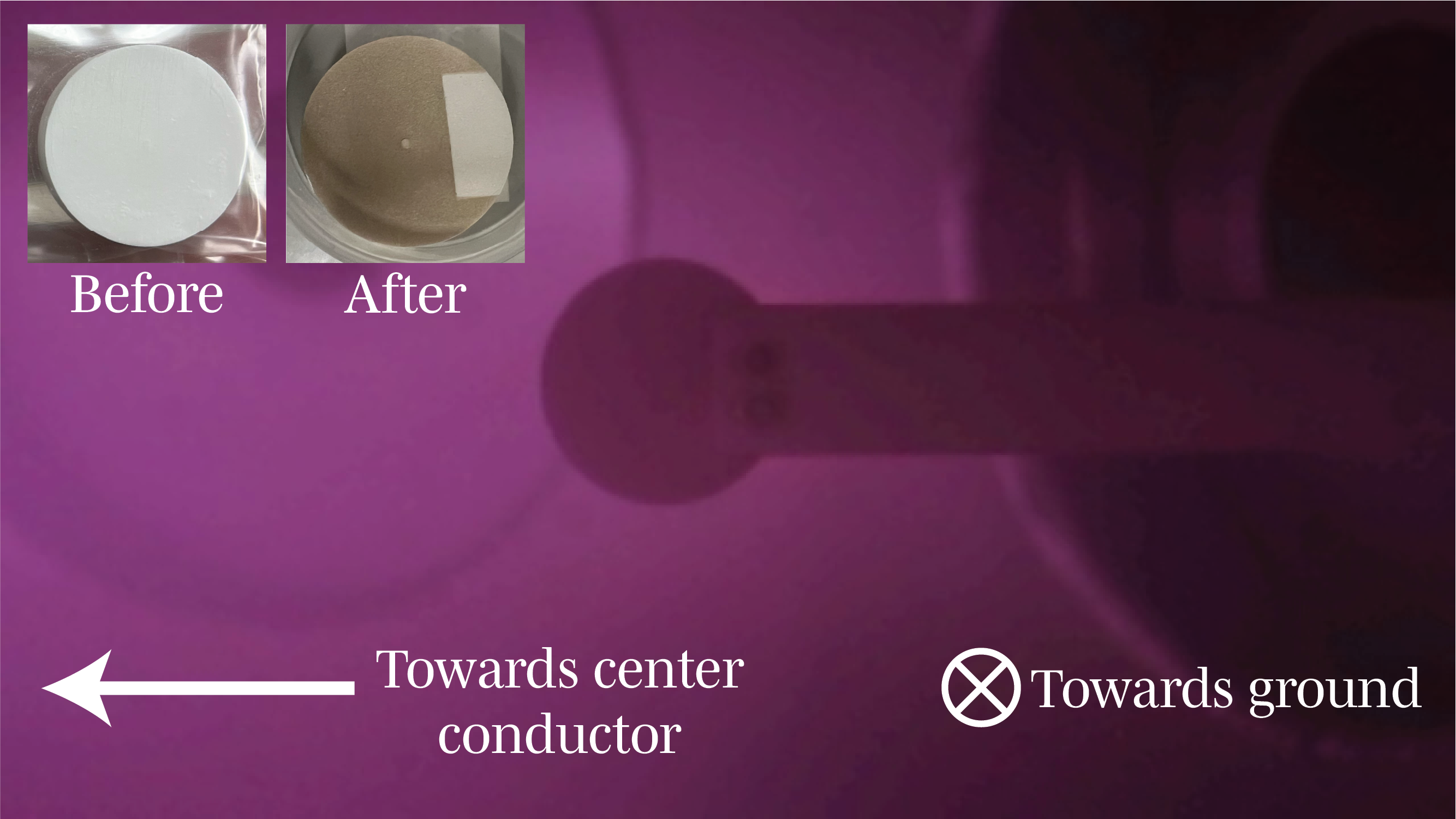}
		\caption{}
		\label{fig:hBN_plasma_back_annotated}
	\end{subfigure}
	\begin{subfigure}{0.49\textwidth}
		\centering
		\captionsetup{justification=centering,font=small}
		\includegraphics[width=\textwidth]{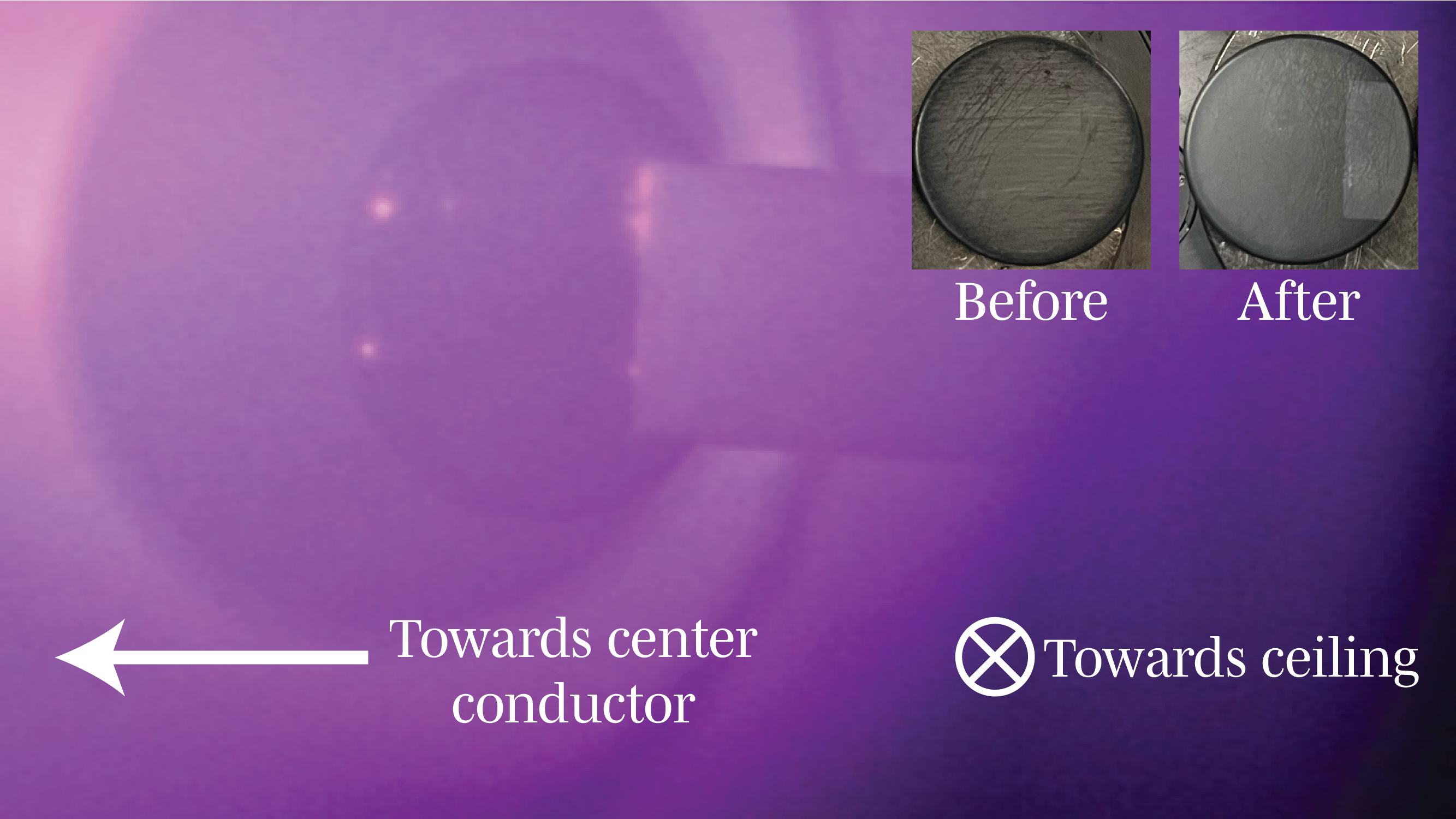}
		\caption{}
		\label{fig:SiC_plasma_front_annotated}
	\end{subfigure}
	\caption{Images of samples in \gls{cmfx} during plasma exposure. (a) is the backside of a \gls{hbn} sample (the side opposite the direction of rotation), and (b) is the frontside of a \ce{SiC} sample, which receives most of the energetic plasma bombardment. Both figures contain images of the samples taken before and after exposure. In the photos taken after exposure, the lighter, rectangular area is where the sample rod clamped onto the sample.}
	\label{fig:sample_in_plasma}
\end{figure}

The resulting assembly was installed on the west side of \gls{cmfx} at midplane (\cref{fig:feedthrough_labeled}). A ruler on the side of the housing is used to indicate sample position and for repeatable experiments. When inserted into \gls{cmfx}, the sample's interaction with the plasma was captured by two cameras (\cref{fig:sample_in_plasma}).

Samples were exposed to plasmas in \gls{cmfx} under the conditions provided in \cref{tab:cmfx_exposure}. An \gls{hbn} sample (exposed during discharges \#1326-1335) was inserted into the plasma with great care so as to not cause the sample rod to overheat. To establish a baseline for comparison with \gls{hbn}, a \ce{SiC} sample was exposed under identical conditions during discharges \#1360-1369, except at a larger radial distance of 21.5 \si{\centi \meter} to minimize the effects of hot spots that appear (\cref{fig:sample_in_plasma}).

\begin{table}
	\begin{center}
		\begin{tblr}{Q[l, m]Q[c, m]Q[c, m]Q[c, m]Q[c, m]Q[c, m]Q[c, m]}
			\toprule
			& \# Discharges & {Total\\exposure (\si{\second})} & {Voltage\\(\si{\kilo \volt})} & {Radial\\position (\si{\centi \meter})} & {Gas puff\\duration (\si{\milli \second})} & {Ion Fluence\\(\si{\per \meter \squared})} \\
			\midrule
			\gls{hbn} & 10 & 40 & 15 & 19 & 20 & \num{1e25} \\
			\ce{SiC} & 10 & 40 & 15 & 21.5 & 20 & \num{1e25} \\
			\bottomrule
		\end{tblr}
		\caption{Exposure conditions for samples in \gls{cmfx}.}
		\label{tab:cmfx_exposure}
	\end{center}
\end{table}

\subsection{Analytical Techniques}
\label{sec:Analytical Techniques}
To investigate the surface effects of plasma exposure, a suite of characterization techniques was employed. The surface morphology of the samples was analyzed using a Hitachi SU-70 Schottky field emission gun \gls{sem}. To mitigate surface charging, the samples were initially coated with a 3-6 nm layer of \ce{Au}/\ce{Pd} through sputter deposition. However, charging effects were still observed, particularly on the \gls{hbn} samples, likely due to their excellent insulation. Additionally, \gls{eds} was performed using a Bruker xFlash 6160 installed on the \gls{sem} to quantify changes to the surface composition.

The PISCES samples were subjected to Cu K-$\alpha$ \gls{xrd} analysis using a Bruker D8 diffractometer. This technique was employed to investigate potential changes in the crystalline structure of the samples following plasma exposure. The X-rays from this source have an attenuation length of approximately 1 \si{\milli \meter} in boron nitride \citep{Henke1993}. This penetration depth is significantly greater than the near-surface region where \gls{pmi} effects are expected to occur. Consequently, while this analysis method is capable of detecting changes in the bulk material, it was not expected to capture the localized effects of plasma interaction at the surface.

Surface profilometry was conducted to quantify variations in surface roughness between samples. Linear roughness ($R$) measurements were performed using the Tencor P-10 with a scan length of 400 \si{\micro \meter} at a speed of 20 \si{\micro \meter \per \second} and a 50 \si{\hertz} sampling rate. Large areal roughness ($S$) measurements were obtained with a Keyence VR-3200. Statistics for the areal measurements were found by creating 5 randomly placed 3.0 \si{\milli \meter}$\times$3.0 \si{\milli \meter} squares for profile analysis. Three key roughness parameters were selected for comparison: $R_a$ ($S_a$), the average deviation from the mean surface height; $R_q$ ($S_q$), the root-mean-squared deviation from the mean surface height; and $R_z$ ($S_z$), the mean separation between the five highest peaks and five lowest valleys. $R_a$ ($S_a$) and $R_q$ ($S_q$) provide overall measures of surface texture, with $R_q$ ($S_q$) being slightly more sensitive to outliers. $R_z$ ($S_z$) is particularly useful for assessing the presence of large features that may result from grain ejection.

\subsection{Erosion Rate Analysis}
\label{sec:Erosion Rate Analysis}
While ion sputtering contributes to surface degradation, the formation of cavities suggests an additional, more aggressive erosion mechanism: grain ejection \citep{Tielens1994,Jurac1998,Duan2016_2}. This process, where entire grains are dislodged after sufficient ion bombardment, is evidenced by the highly increased surface porosity.

\begin{figure}
	\centering
	\includegraphics[width=0.6\textwidth]{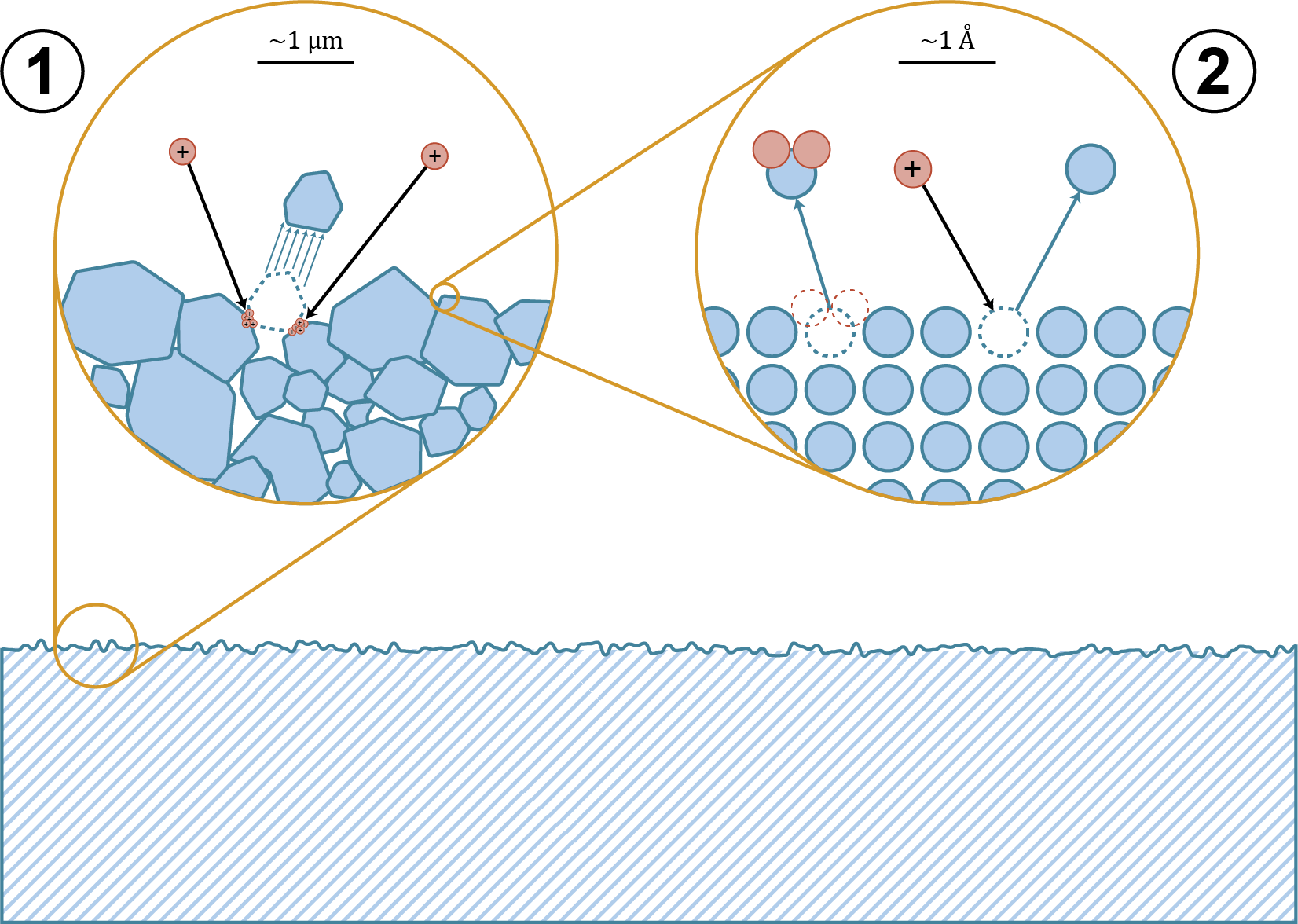}
	\caption{Diagram of surface erosion mechanisms, where (1) is grain ejection, by which an entire grain becomes detached from the surface, and (2) is sputtering, which removes individual atoms from the surface layers. Grain ejection can be attributed to a combination of electrostatic buildup, localized heating, or sputtering weakening intergranular bonds. Sputtering is a result of both chemical and physical sputtering processes.}
	\label{fig:erosion_mechanisms}
\end{figure}

Thus, two primary mechanisms of surface erosion have been identified (\cref{fig:erosion_mechanisms}):
\begin{enumerate}
	\item \textbf{Grain ejection}: This process leads to accelerated surface degradation through detachment of entire grains, and it seems to be heavily influenced by the relationship between ion energy and grain size. We use $R_z$ and $S_z$ as a measurement for the erosion caused by grain ejection. The observed surface degradation suggests that the current manufacturing method may result in suboptimal \gls{pmi} performance. This process has not been described mechanistically in prior literature, but it may be driven by three factors:
	\begin{itemize}
		\item Charge buildup at grain surfaces, creating a net repulsive electrostatic force between grains.
		\item Localized heating at grain edges, temporarily weakening bonds and facilitating grain removal.
		\item Physical weakening of intergranular bonds due to sputtering over time.
	\end{itemize}
	\item \textbf{Sputtering}: Sputtering manifests as even surface wear, smoothing and rounding grain edges. The amount of material removed by sputtering can be estimated by the radius of curvature of the grains. Combined with the known exposure time, this provides an approximate sputtering erosion rate. The erosion rate from sputtering was found to be significantly lower than that of grain ejection. However, the erosion rate of grain ejection can be decreased with lower energy incident ions. There are two processes which contribute to sputtering erosion:
	\begin{itemize}
		\item Physical sputtering occurs when an incident ion imparts enough energy to a near-surface atom to break the bond and remove it from the material.
		\item Incident ions can also form volatile compounds with material atoms, thereby weakening the bond strength, through a process known as chemical sputtering.
	\end{itemize}
\end{enumerate}

These methods provide order-of-magnitude estimates, but have limitations. For instance, the curvature of a grain may be affected by the movement of neighboring grains or perhaps by the elevated temperature during exposure. Despite these limitations, these measurements help elucidate general trends in surface erosion from exposure to deuterium plasma.

\section{Exposure Results}
\label{sec:Exposure Results}
Results from the exposure in both PISCES-A and \gls{cmfx} are presented in the following sections.

\subsection{PISCES-A}
\label{sec:PISCES-A Results}
We first inspected \gls{hbn} samples that were exposed in PISCES-A for any changes to the bulk crystalline structure. The \gls{xrd} results, as presented in \cref{tab:pisces_xrd}, indicate no significant changes after plasma exposure. This preservation of bulk crystallinity is a positive indicator for the potential use of \gls{hbn} in fusion applications, as it implies that the material's bulk properties are stable even when subjected to high ion and heat flux.

\begin{table}
	\begin{center}
		\begin{tblr}{lcc}
			\toprule
			Ion energy (\si{\electronvolt}) & $a$ (\si{\nano \meter}) & $c$ (\si{\nano \meter}) \\
			\midrule
			0 & 2.50499 & 6.65762 \\
			30 & 2.50411 & 6.65771 \\
			70 & 2.50442 & 6.65599 \\
			100 & 2.50427 & 6.65784 \\
			\bottomrule
		\end{tblr}
		\caption{Results of \gls{xrd} analysis on \gls{hbn} samples exposed to ion bombardment in PISCES. For \gls{hbn}, $a$ is the lattice constant that represents the distance between adjacent atoms in the hexagonal plane, and $c$ is the distance between hexagonal layers.}
		\label{tab:pisces_xrd}
	\end{center}
\end{table}

However, while the bulk material remains unaltered, the surface modifications observed through other analytical techniques are still significant.

\begin{figure}
	\centering
	\begin{subfigure}{0.75\textwidth}
		\centering
		\captionsetup{justification=centering,font=small}
		\includegraphics[width=\textwidth]{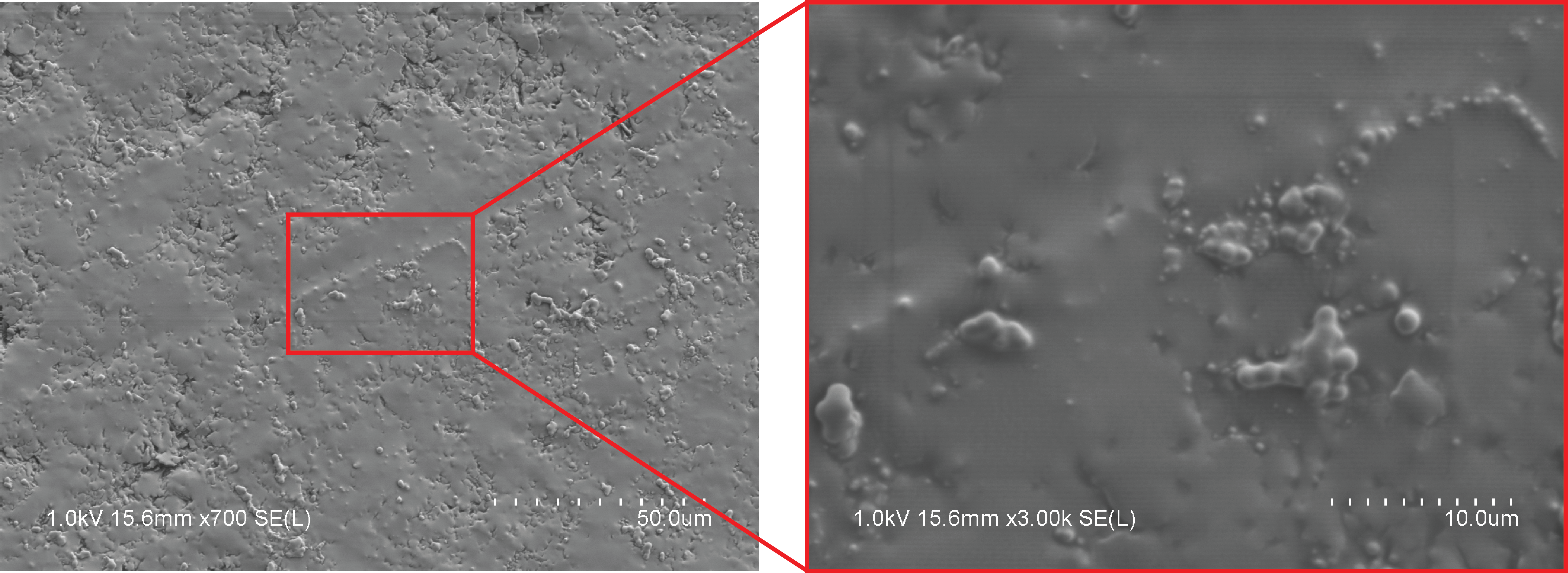}
		\caption{}
		\label{fig:pisces_sem_noplasma}
	\end{subfigure}
	\\
	\begin{subfigure}{0.75\textwidth}
		\centering
		\captionsetup{justification=centering,font=small}
		\includegraphics[width=\textwidth]{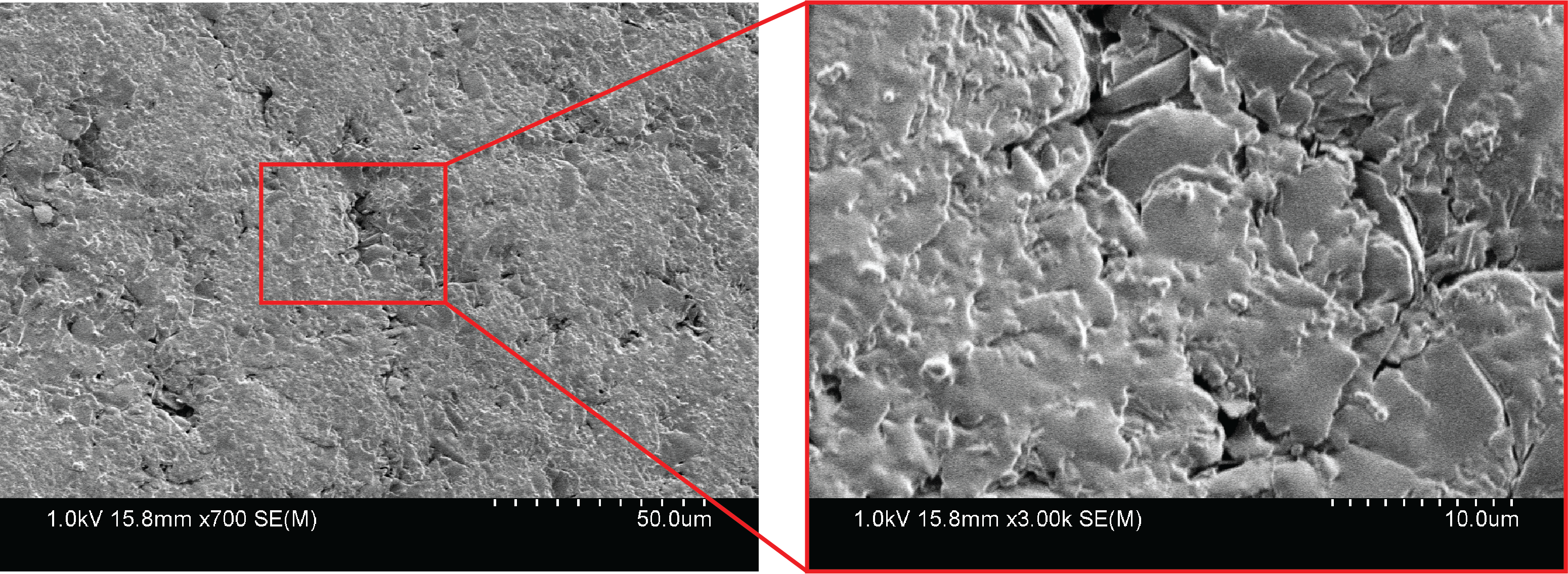}
		\caption{}
		\label{fig:pisces_sem_30eV}
	\end{subfigure}
	\\
	\begin{subfigure}{0.75\textwidth}
		\centering
		\captionsetup{justification=centering,font=small}
		\includegraphics[width=\textwidth]{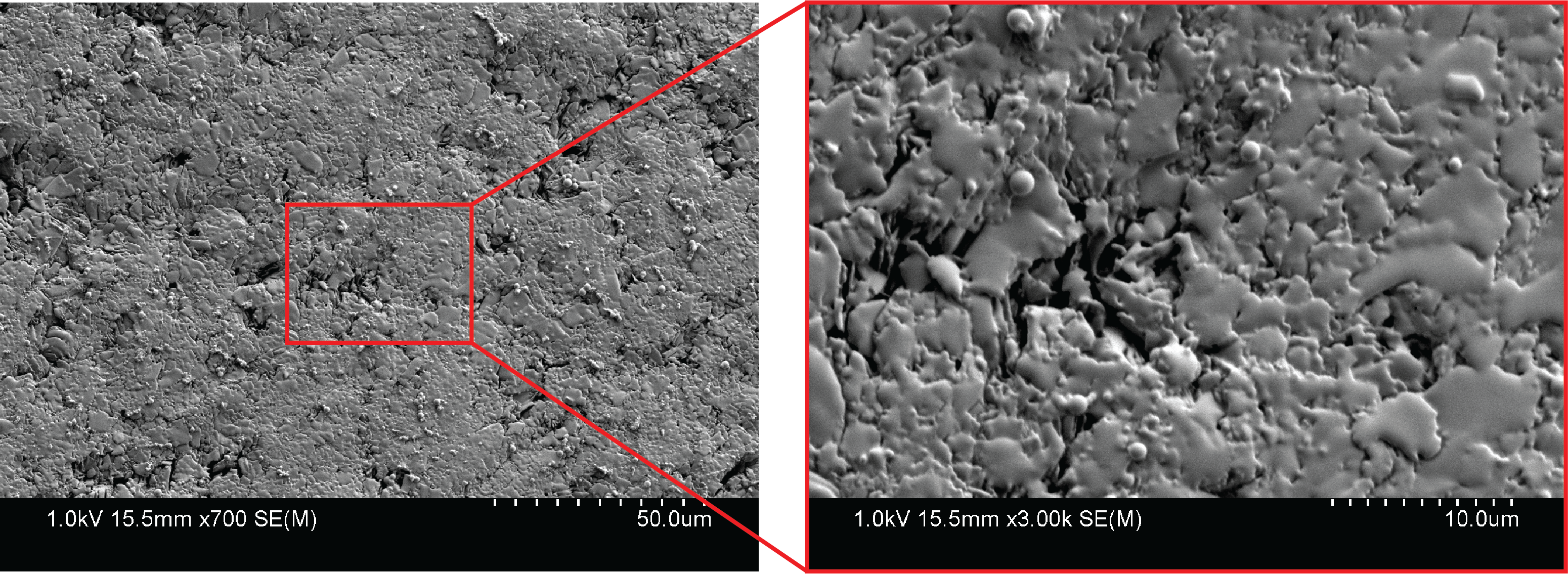}
		\caption{}
		\label{fig:pisces_sem_70eV}
	\end{subfigure}
	\\
	\begin{subfigure}{0.75\textwidth}
		\centering
		\captionsetup{justification=centering,font=small}
		\includegraphics[width=\textwidth]{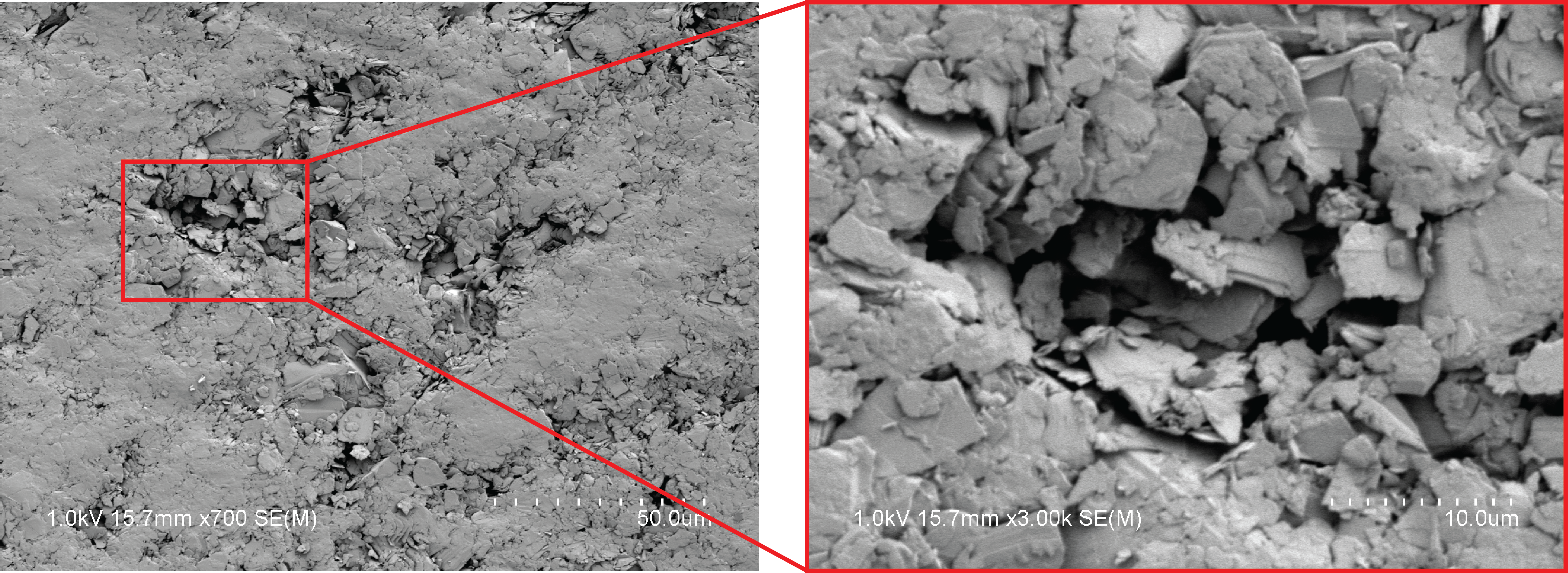}
		\caption{}
		\label{fig:pisces_sem_100eV}
	\end{subfigure}
	\caption{\gls{sem} images of \gls{hbn} samples, where (a) is the control sample without plasma exposure. Samples were exposed to deuterium ions at an energy of (b) 30 \si{\electronvolt}, (c) 70 \si{\electronvolt}, and (d) 100 \si{\electronvolt} in PISCES.}
	\label{fig:pisces_sem}
\end{figure}

\Gls{sem} analysis reveals striking differences between the control and ion-exposed \gls{hbn} samples (\cref{fig:pisces_sem}). The control sample exhibits a relatively flat surface with occasional protruding grains (1-10 \si{\micro\meter}) and minor cracks, likely resulting from the lathe machining process. In contrast, samples exposed to energetic ions (30-100 \si{\electronvolt}) display significantly increased surface roughness, prominent grain edges, and exposed subsurface porosity.

\begin{figure}
	\centering
	\includegraphics[width=0.7\textwidth]{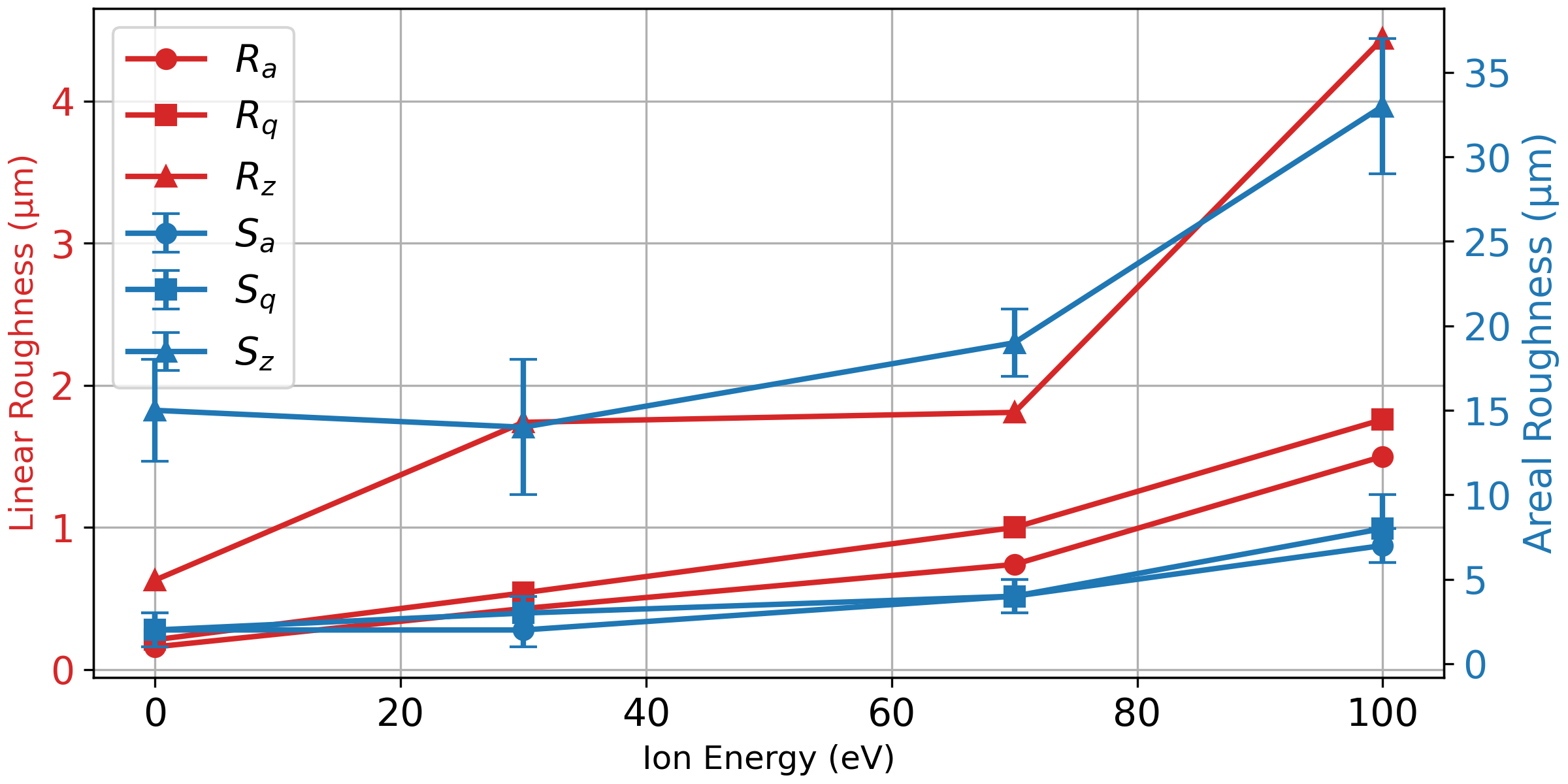}
	\caption{Surface roughness parameters measured by stylus (linear) and optical (areal) profilometer. Samples exposed in PISCES are compared to the control sample (0 \si{\electronvolt}).}
	\label{fig:PISCES_roughness}
\end{figure}

The surface roughness results for samples exposed in PISCES are presented in \cref{fig:PISCES_roughness}, revealing several important trends. A roughly linear increase in both $R_a$ and $R_q$ is observed as ion energy increases. This trend aligns qualitatively with the predicted increase in sputtering yield shown in \cref{fig:Ys_energy_BN}. The behavior of $R_z$ remains approximately constant between the 30 \si{\electronvolt} and 70 \si{\electronvolt} samples but then shows a rapid increase for the 100 \si{\electronvolt} sample. This abrupt change likely corresponds to grain ejection, suggesting the existence of an energy threshold between 70 and 100 \si{\electronvolt} for this erosion process. Areal measurements revealed similar trends, though with less pronounced $S_z$ variations. Notably, optical profilometry measurements were approximately an order of magnitude larger than stylus measurements. This can be attributed to the fact that the stylus tip has a diameter of 1 \si{\micro \meter}, and it will mechanically filter features of this size. Areal measurements will also tend to include the most extreme peaks and valleys because the sampling coverage is so much larger, and therefore offers a more accurate representation.

Note that the results only discuss physical sputtering for the samples exposed on PISCES-A. While chemical sputtering certainly would have a role and would lead to preferential sputtering of one element over the other, it was not surveyed in this current study, but it is discussed in the next section.

\subsection{CMFX}
\label{sec:CMFX Results}
The following sections detail morphological and compositional changes to the samples exposed in \gls{cmfx}.

\subsubsection{Boron Nitride}
\label{sec:Boron Nitride}
\begin{table}
	\begin{center}
			\begin{tblr}{colspec={Q[l, m]|Q[c, m]Q[c, m]Q[c, m]Q[c, m]|Q[c, m]|Q[c, m]|},
					vlines = {3-5}{solid}}
				\cline{2-10}
				& \SetCell[c=4]{c} Sample & & & & {Center\\conductor} & {Sample\\rod} \\
				\cline{2-10}
				& B & N & O & C & Fe & Al \\
				\midrule
				Control & {39.48\\(2.71)} & {39.46\\(2.60)} & {4.00\\(0.50)} & {16.87\\(1.60)} & - & {0.18\\(0.03)} \\
				{Exposed} & {44.81\\(2.82)} & {36.26\\(2.55)} & {10.47\\(1.10)} & {8.05\\(0.92)} & {0.33\\(0.07)} & {0.08\\(0.02)} \\
				\bottomrule
			\end{tblr}
		\caption{Surface composition of \gls{hbn} samples determined by \gls{eds}, expressed in \%. Each component is also grouped into its respective source. Control is compared to samples that have been exposed to plasma in \gls{cmfx}. One standard deviation, which is calculated based on the \gls{eds} spectrum, is provided in parentheses.}
		\label{tab:eds_hbn_quant}
	\end{center}
\end{table}

\begin{figure}
	\centering
	\includegraphics[width=0.9\textwidth]{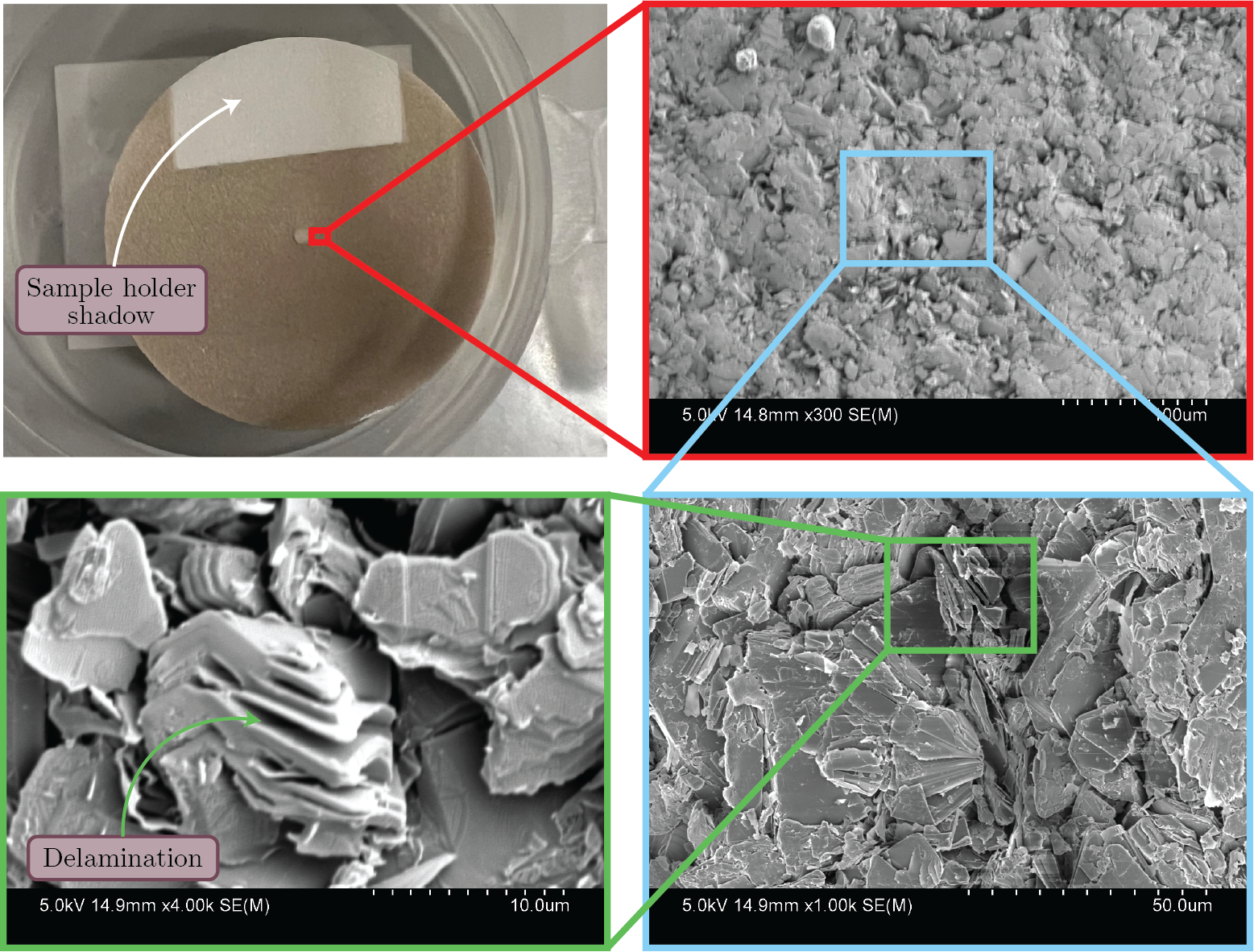}
	\caption{\Gls{sem} images of \gls{hbn} sample exposed to plasma during \gls{cmfx} discharges \#1326-1335.}
	\label{fig:sem_hBN2}
\end{figure}

\Gls{sem} images of the \gls{hbn} sample exposed in \gls{cmfx} during discharges 1326 to 1335 revealed increased porosity, grain separation, and delamination between \gls{hbn} layers in grains oriented perpendicular to the surface (\cref{fig:sem_hBN2}). The latter effect suggests that ion bombardment can act as a wedge, prying apart the weak van der Waals bonds between layers, if the grain is oriented perpendicular to the flux. Grains appeared more rounded, indicating even wear from physical sputtering.

\Gls{eds} results from these samples (\cref{tab:eds_hbn_quant}) showed negligible presence of the Macor sample rod elements, except for aluminum (which may also possibly be from chamber walls). Note that the control sample also contained trace aluminum, which may come the hot-pressing technique used to manufacture the samples in alumina dies. Moreover, elevated oxygen levels were observed, perhaps due to higher porosity allowing more binding spots during air exposure. It is also possible that impurity oxygen in the plasma were implanted during the plasma exposure.

\subsubsection{Silicon Carbide}
\label{sec:Silicon Carbide}
\begin{figure}
	\centering
	\begin{subfigure}{0.78\textwidth}
		\centering
		\captionsetup{justification=centering,font=small}
		\includegraphics[width=\textwidth]{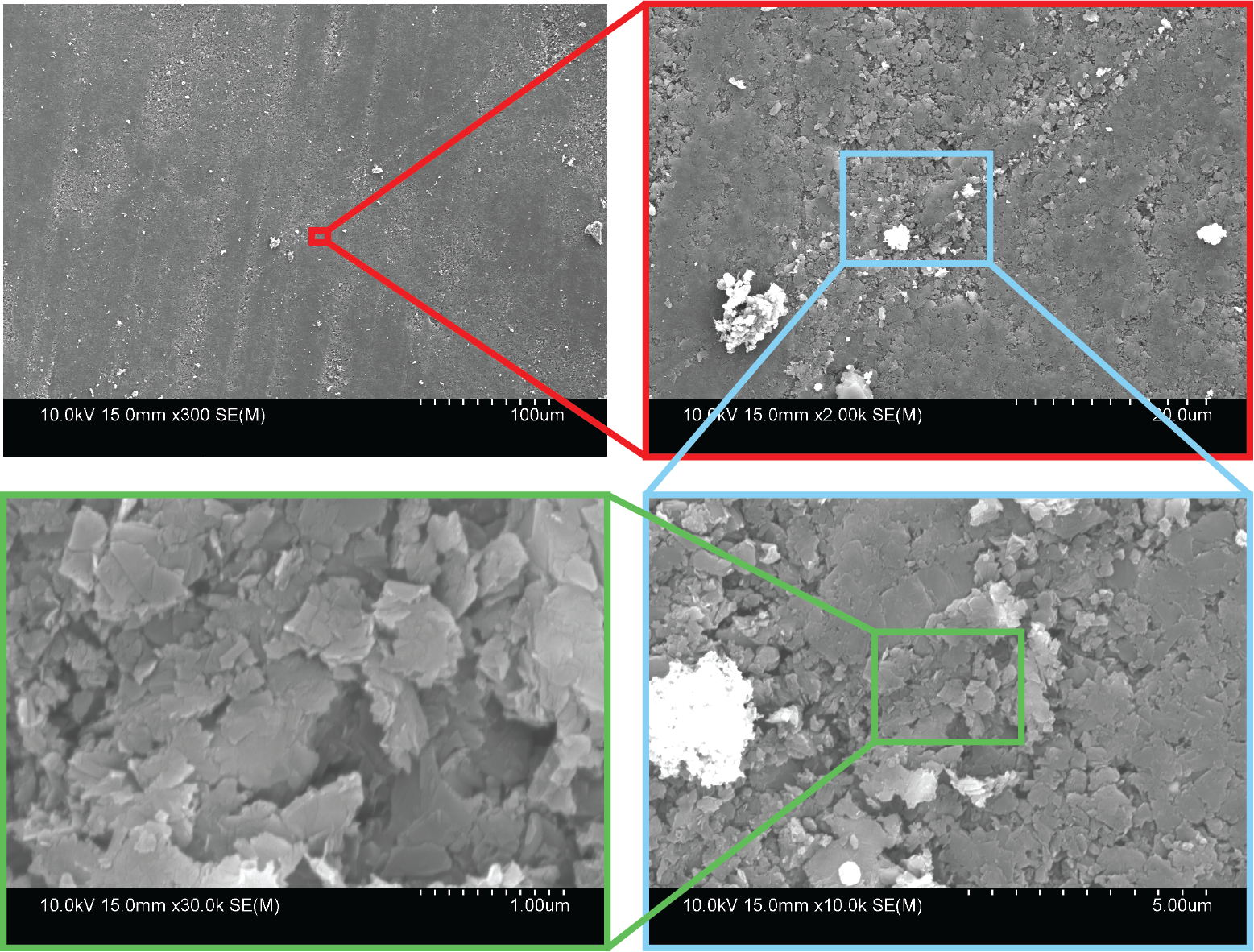}
		\caption{}
		\label{fig:sem_sic0}
	\end{subfigure}
	\\
	\begin{subfigure}{0.78\textwidth}
		\centering
		\captionsetup{justification=centering,font=small}
		\includegraphics[width=\textwidth]{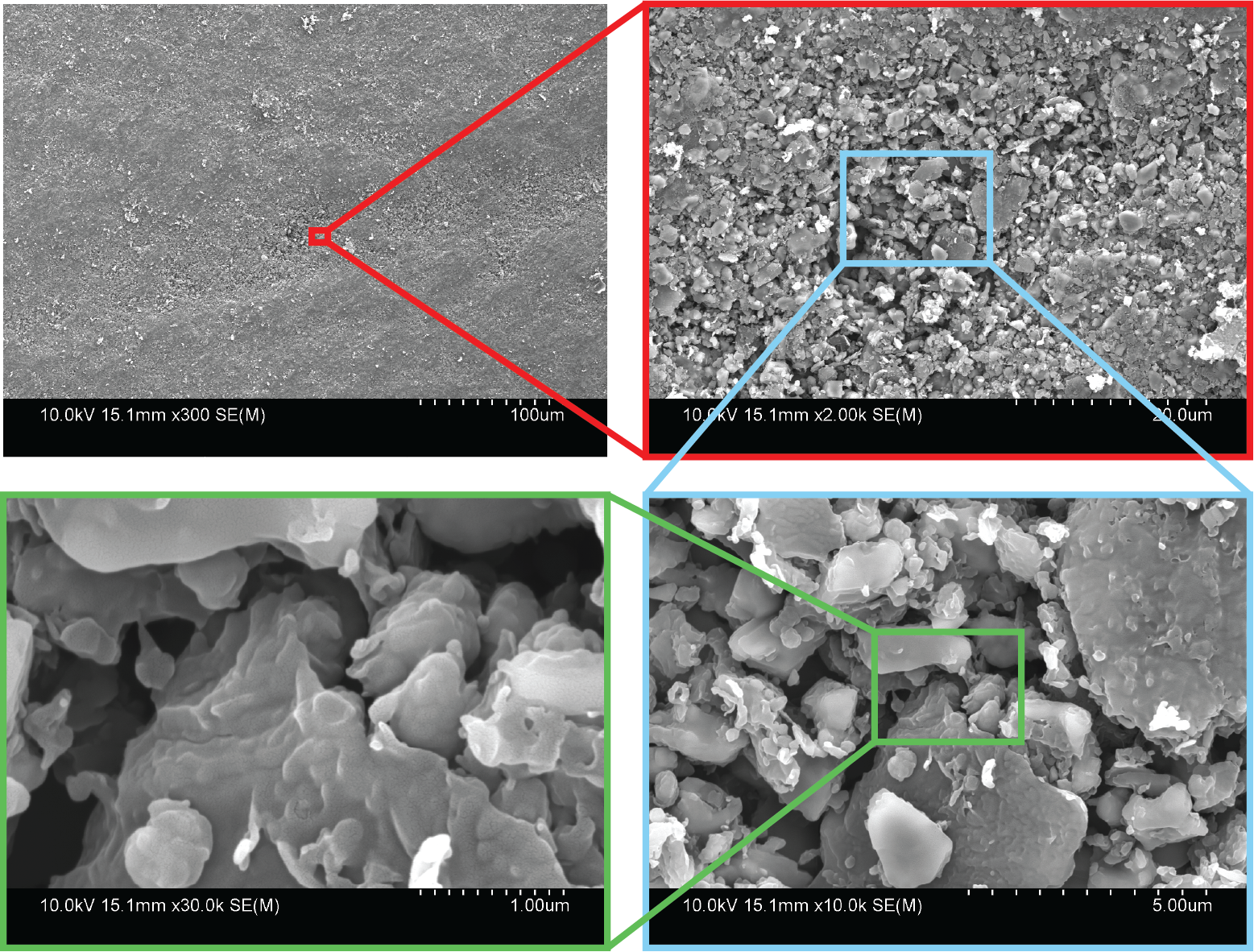}
		\caption{}
		\label{fig:sem_sic1}
	\end{subfigure}
	\caption{(a) \Gls{sem} images of \ce{SiC} control sample. (b) \Gls{sem} images of \ce{SiC} sample from discharges \#1360-1369, where the sample was placed at a radial distance of 21.5 \si{\centi \meter}.}
	\label{fig:cmfx_sem_sic}
\end{figure}

Higher-resolution \gls{sem} imaging (\cref{fig:cmfx_sem_sic}) was possible for \ce{SiC} due to its lower electrical resistivity, which minimized surface-charging effects. The control sample revealed a relatively flat surface, though rougher than the \gls{hbn} control (\cref{fig:pisces_sem_noplasma}), with grain sizes ranging from $\sim$100 \si{\nano \meter} to 1 \si{\micro \meter} and minimal surface porosity.

Post-irradiation analysis of the \ce{SiC} (discharges 1360 to 1369) sample showed marked changes. The surface exhibited significantly increased roughness and porosity, with extensive grain ejection surpassing that observed in \gls{hbn}. Intergranular layers were weakened due to ion bombardment, leading to grain detachment. Grain edges appeared very rounded and worn in comparison to the control sample, indicative of uniform erosion from sputtering.

\Gls{eds} analysis (\cref{tab:eds_sic_quant}) provided further insights. The control sample showed a highly enriched carbon layer on the surface, likely due to graphitization. This phenomenon occurs when silicon evaporates from the \ce{SiC} surface during annealing under high vacuum \citep{Forbeaux2000}, as in the pressureless sintering process used in Hexoloy production. After irradiation, a significant portion of this graphite layer was removed, and the silicon proportion increased nearly four-fold. These measurements reflect that there is real material loss, and not just surface modification. Notably, contaminants like iron and aluminum were missing. While these results alone were inconclusive regarding preferential sputtering of silicon or carbon, the model in \cref{fig:Ys_energy_SiC} suggests that we might expect preferential carbon removal.

\begin{table}
	\begin{center}
		\begin{tblr}{lccc}
			\toprule
			& Si & C & O \\
			\midrule
			Control & {6.96 (0.28)} & {90.74 (0.91)} & {2.30 (0.38)} \\
			Exposed & {23.24 (0.74)} & {73.96 (2.04)} & {2.80 (0.36)} \\
			\bottomrule
		\end{tblr}
		\caption{Surface composition of \ce{SiC} samples determined by \gls{eds}, expressed in \%. One standard deviation, which is calculated based on the \gls{eds} spectrum, is provided in parentheses.}
		\label{tab:eds_sic_quant}
	\end{center}
\end{table}

We acknowledge that the carbon-rich surface layer on \ce{SiC} would eventually erode in a fusion reactor, exposing more stoichiometric material beneath. This suggests a two-step erosion behavior: an initial period with potentially different erosion rates dominated by carbon-rich layer removal, followed by a steady-state phase once stoichiometric SiC is exposed. Our computational modeling with RustBCA already accounts for stoichiometric \ce{SiC} behavior, showing higher sputtering yields than \gls{hbn} (\cref{fig:sputtering}), which complements these experimental findings. Future work should include \ce{SiC} samples with the carbon-rich layer removed to directly measure erosion behavior of stoichiometric \ce{SiC}.

\subsubsection{Surface Roughness}
\label{sec:Surface Roughness}
\begin{figure}
	\centering
	\includegraphics[width=0.9\textwidth]{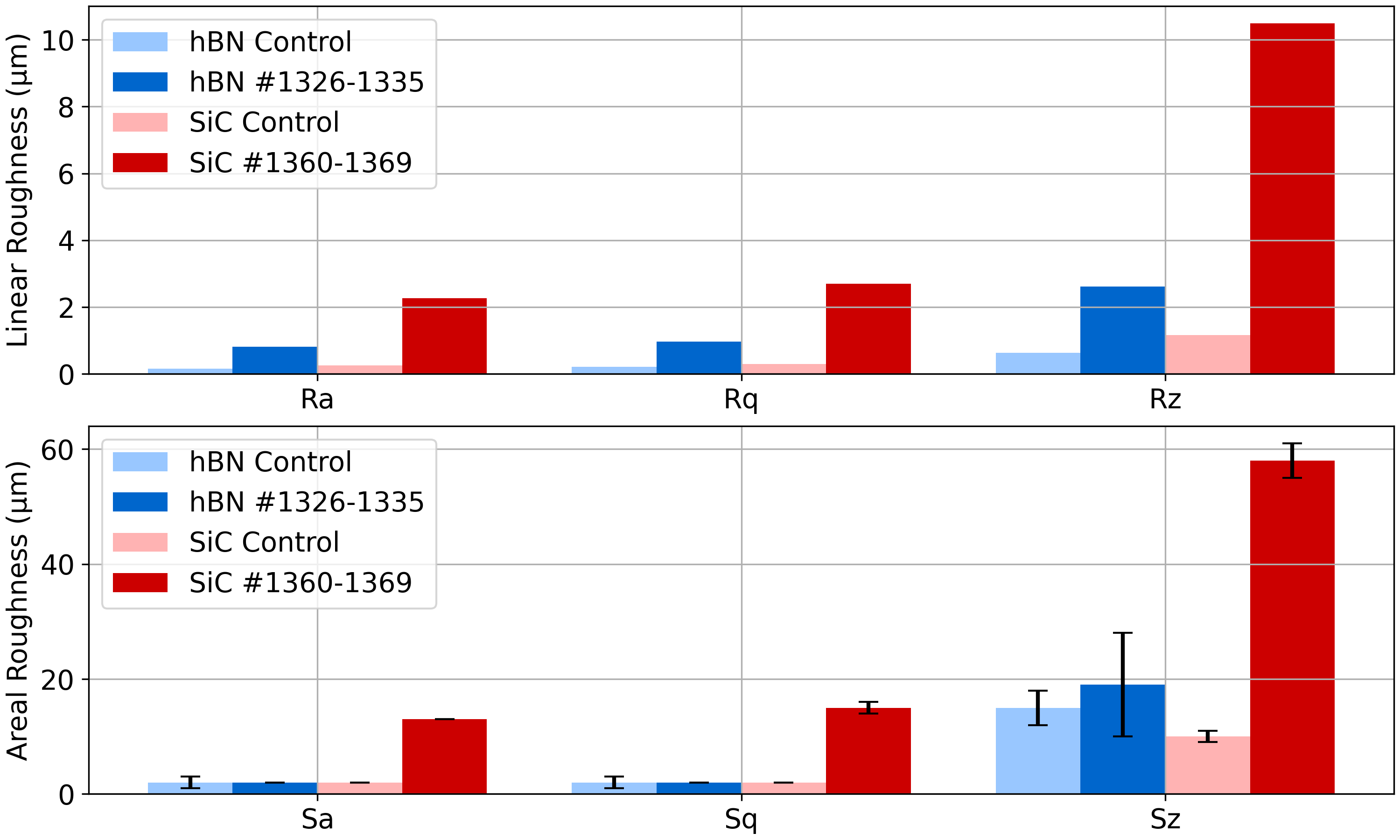}
	\caption{Surface roughness parameters measured by stylus (linear) and optical (areal) profilometer. Samples exposed in CMFX are compared to the control sample.}
	\label{fig:CMFX_roughness}
\end{figure}

Surface profilometry allowed for quantitative comparison of surface damage between \gls{hbn} and \ce{SiC} (\cref{fig:CMFX_roughness}). Most importantly, the \ce{SiC} sample (\#1360-1369) exhibited approximately twice the increase in surface roughness, if not more, across all $R$ parameters compared to \gls{hbn} (\#1326-1335). The areal parameters $S_a$ and $S_q$ approached the Keyence VR-3200's resolution limit (1 \si{\micro \meter}). Thus, $S_z$ provides the most reliable metric, confirming \ce{SiC}'s substantially greater roughness increase.

The combination of \gls{sem} and profilometry results lead to a crucial conclusion: under similar exposure conditions, the \ce{SiC} surface experienced more severe erosion than the \gls{hbn} surface. This finding suggests that \gls{hbn} may offer superior irradiation resistance in plasma-facing components.

\subsection{Erosion Rate}
\label{sec:Erosion Rate}
Estimates of the linear erosion rate normalized by incident fluence (see \cref{tab:pisces_xrd,tab:cmfx_exposure}) for both sputtering and grain ejection are provided in \cref{tab:erosion_rate}. These calculations incorporate several assumptions and should be interpreted primarily for qualitative comparison rather than as absolute measurements. The plasma density and velocity (and thus flux) estimates in \gls{cmfx} have significant uncertainty and could vary by an order of magnitude. Additionally, measurements of grain radii from \gls{sem} images show considerable variation across each sample's surface, leading to large uncertainties that warrant reporting only one significant figure.

\begin{table}
	\begin{center}
		\begin{tblr}{colspec={rccc|c},
				hline{1}={2-5}{1pt},
				hline{2}={2-5}{0.5pt},
				hline{Z}={2-5}{1pt},}
			& & {Radius\\(\si{\nano \meter})} & {Sputtering\\(\si{\nano \meter \cubed})} & {Grain Ejection\\(\si{\nano \meter \cubed})} \\
			\ldelim\{{3.3}{*}[PISCES] & \gls{hbn} (30 \si{\electronvolt}) & 200 & \num{2e-5} & - \\
			& \gls{hbn} (70 \si{\electronvolt}) & 300 & \num{3e-5} & \num{3.3e-4} \\
			& \gls{hbn} (100 \si{\electronvolt}) & 300 & \num{3e-5} & \num{1.5e-3} \\
			\ldelim\{{2}{*}[CMFX] & \gls{hbn} (\#1326-1335) & 700 & \num{6e-5} & \num{3.5e-4} \\
			& \ce{SiC} (\#1360-1369) & 900 & \num{8e-5} & \num{4.3e-3} \\
		\end{tblr}
		\caption{Estimates of erosion rate per unit flux for both sputtering and grain ejection. The total fluence of PISCES is found using the values in \cref{tab:pisces_parameters}, and that of CMFX is estimated with the value of flux in \cref{tab:cmfx_exposure}. The radius of the grain edges was found using \gls{sem} micrographs and is used as a proxy for sputtering erosion. The difference in $S_z$ between the control and the exposed sample is used for grain ejection.}
		\label{tab:erosion_rate}
	\end{center}
\end{table}

Regardless, we generally see that higher energy ions in PISCES leads to an increased sputtering rate. Additionally, there was no apparent grain ejection, as measured by changes in $S_z$, for the 30 \si{\electronvolt} exposure in PISCES, though we see a significant increase for the 100 \si{\electronvolt} case. Moreover, the sputtering rate in \gls{cmfx} for \ce{SiC} is slightly higher than that of \gls{hbn}, consistent with the results from \cref{sec:Ion Results}. Most importantly, the grain ejection erosion rate in \ce{SiC} is approximately an order of magnitude larger than in \gls{hbn}, suggesting that intergranular bonding strength plays a crucial role in material longevity under plasma exposure.

\section{Discussion and Conclusions}
\label{sec:Conclusion}

\subsection{Material Comparison}
A comparison of surface erosion mechanisms between \gls{hbn} and \ce{SiC} is provided in \cref{tab:material_comparison}. Notably, irradiated \ce{SiC} exhibited significantly higher porosity and smoother grain structure than irradiated \gls{hbn}. This qualitative observation was corroborated by quantitative measurements of surface roughness, which showed that the relative change in \ce{SiC} roughness was nearly twice that of \gls{hbn}.

\begin{table}
	\begin{center}
		\begin{tblr}{Q[l, m]|Q[c, m]Q[c, m]Q[c, m]}
			\toprule
			{Erosion\\process}& {Hexagonal\\boron nitride} & {Silicon\\carbide} & {Characterization\\technique} \\
			\midrule
			{Physical\\sputtering} & Moderate & Significant & {Predicted in \cref{sec:Ion Results},\\observed as even grain\\wear in \gls{sem}} \\
			{Chemical\\sputtering} & {Preferable N\\compound formation} & {Occurs for C} & {Deduced from \gls{eds} results} \\
			{Grain\\ejection} & Moderate & Significant & {Observed as surface porosity\\in \gls{sem}, quantified\\by surface profilometry} \\
			\bottomrule
		\end{tblr}
		\caption{Comparison between \gls{hbn} and \ce{SiC} for the various erosion processes that were observed after exposure to plasmas in \gls{cmfx}.}
		\label{tab:material_comparison}
	\end{center}
\end{table}

Contrary to our initial hypothesis, we did not observe preferential sputtering of boron leading to a nitrogen-rich surface. Instead, our results indicate an unexpected increase in the boron-to-nitrogen ratio following plasma exposure. This outcome may be attributed to reactive chemical sputtering, where nitrogen forms volatile compounds (such as ammonia in deuterium plasma) that are more readily removed from the surface. This process could lead to enhanced nitrogen loss, explaining the observed increase in boron concentration.

Based on the prior studies chemical removal of carbon by deuterium is significantly higher than chemical sputtering of silicon in \ce{SiC} \citep{Balden2000,Balden2001,Abrams2021,Sinclair2021}. While our \gls{eds} results show compositional changes after plasma exposure, they represent the combined effects of physical sputtering and potential chemical erosion. In either case, we would expect both higher physical and chemical sputtering of carbon.

\subsection{Conclusions}

In our search for suitable insulating \glspl{pfm}, \gls{hbn} emerged as a promising candidate, outperforming the baseline material, \ce{SiC}, in several key areas. However, \ce{SiC} does have the advantage of superior mechanical properties over \gls{hbn}, particularly in future power plants where there may be thermal stresses due to mismatches in \gls{cte} with a substrate. Using RustBCA for \gls{pmi} modeling, we found that \gls{hbn} exhibited superior performance in terms of sputtering yield and \gls{dpa} across all energies and angles of incidence compared to \ce{SiC}. Additionally, the penetration depth for the primary plasma ions remained below a few hundred nanometers, which is significantly less than the thickness required for the insulator ($\sim$millimeters), leaving the bulk unaffected. We note that these models do not account for crystalline effects or chemical sputtering.

Neutron-material interactions, modeled using OpenMC, showed that \gls{hbn} outperformed both \ce{SiC} and tungsten in terms of \gls{dpa} and depth of penetration, though it showed higher susceptibility to absorption damage due to the large neutron capture cross-section of \ce{^{10}B}. Moreover, \mctrans{} provides valuable outputs to quantify the ion and neutron flux at the insulators, and it can therefore be used in conjunction with RustBCA and OpenMC to predict damage to the insulators in future centrifugal mirrors.

Experimental exposure of \gls{hbn} samples to fusion-relevant plasmas in PISCES and \gls{cmfx} provided crucial insights into material behavior. \gls{sem} and surface profilometry analysis revealed two primary mechanisms of surface erosion: grain ejection and sputtering. Based off the PISCES results, grain ejection appears to be energy-dependent, with larger grains removed at higher ion energies (70-100 \si{\electronvolt}), and nearly no grains ejected at 30 \si{\electronvolt}, suggesting there is an energy threshold that depends on grain size. Drawing inspiration from other fusion devices, one potential strategy to mitigate high-energy ion bombardment is the implementation of radiative cooling near the insulators \citep{Gibson2007,Petrie2009,Zhang2016}. Comparative analysis of irradiated \gls{hbn} and \ce{SiC} samples from \gls{cmfx} showed a lower erosion rate for \gls{hbn}, corroborating our computational predictions.

The experiments conducted on \gls{cmfx} have demonstrated its usefulness as a test platform for exposing materials to fusion-relevant plasmas. Future enhancements to \gls{cmfx} aim to increase neutron yield, enabling neutron irradiation testing, and decrease impurities in the plasma. This addition would further solidify \gls{cmfx}'s role as a valuable tool for fusion materials research. Additionally, physical sputtering effects could be isolated from chemical interactions by puffing inert helium gas instead of hydrogen isotopes. These experiments would provide valuable distinction between the effects of physical and chemical sputtering.

While pure \gls{hbn} may not be ideally suited as a \gls{pfm} in future centrifugal mirror reactors due to its poor mechanical strength, its improved performance over pure \ce{SiC} in a fusion environment suggests that \gls{hbn} composites could be promising candidates for \glspl{pfm}. \gls{hbn} composites could be designed to enhance mechanical properties and mitigate the weak intergranular bonds that lead to grain ejection, thereby significantly reducing surface erosion. Future research should focus on developing and testing these composite materials under fusion-relevant conditions to identify optimal candidates for use in centrifugal mirror reactors.

\bibliographystyle{elsarticle-num-names}

\bibliography{references}

\begin{thebibliography}{88}
\expandafter\ifx\csname natexlab\endcsname\relax\def\natexlab#1{#1}\fi
\providecommand{\url}[1]{\texttt{#1}}
\providecommand{\href}[2]{#2}
\providecommand{\path}[1]{#1}
\providecommand{\DOIprefix}{doi:}
\providecommand{\ArXivprefix}{arXiv:}
\providecommand{\URLprefix}{URL: }
\providecommand{\Pubmedprefix}{pmid:}
\providecommand{\doi}[1]{\href{http://dx.doi.org/#1}{\path{#1}}}
\providecommand{\Pubmed}[1]{\href{pmid:#1}{\path{#1}}}
\providecommand{\bibinfo}[2]{#2}
\ifx\xfnm\relax \def\xfnm[#1]{\unskip,\space#1}\fi
\bibitem[{Schwartz et~al.(2024)Schwartz, Abel, Hassam, Kelly, and
  Romero-Talamás}]{Schwartz2024}
\bibinfo{author}{N.~R. Schwartz}, \bibinfo{author}{I.~G. Abel},
  \bibinfo{author}{A.~B. Hassam}, \bibinfo{author}{M.~Kelly},
  \bibinfo{author}{C.~A. Romero-Talamás},
\newblock \bibinfo{title}{Mctrans++: a 0-d model for centrifugal mirrors},
\newblock \bibinfo{journal}{Journal of Plasma Physics} \bibinfo{volume}{90}
  (\bibinfo{year}{2024}) \bibinfo{pages}{905900217}. \URLprefix
  \url{https://www.cambridge.org/core/product/identifier/S0022377824000424/type/journal\_article}.
  \DOIprefix\doi{10.1017/S0022377824000424}.
\bibitem[{Teodorescu et~al.(2010)Teodorescu, Young, Swan, Ellis, Hassam, and
  Romero-Talamas}]{Teodorescu2010}
\bibinfo{author}{C.~Teodorescu}, \bibinfo{author}{W.~C. Young},
  \bibinfo{author}{G.~W.~S. Swan}, \bibinfo{author}{R.~F. Ellis},
  \bibinfo{author}{A.~B. Hassam}, \bibinfo{author}{C.~A. Romero-Talamas},
\newblock \bibinfo{title}{Confinement of plasma along shaped open magnetic
  fields from the centrifugal force of supersonic plasma rotation},
\newblock \bibinfo{journal}{Physical Review Letters} \bibinfo{volume}{105}
  (\bibinfo{year}{2010}) \bibinfo{pages}{085003}. \URLprefix
  \url{https://link.aps.org/doi/10.1103/PhysRevLett.105.085003}.
  \DOIprefix\doi{10.1103/PhysRevLett.105.085003}.
\bibitem[{Huang and Hassam(2001)}]{Huang2001}
\bibinfo{author}{Y.-M. Huang}, \bibinfo{author}{A.~B. Hassam},
\newblock \bibinfo{title}{Velocity shear stabilization of centrifugally
  confined plasma},
\newblock \bibinfo{journal}{Physical Review Letters} \bibinfo{volume}{87}
  (\bibinfo{year}{2001}) \bibinfo{pages}{235002}. \URLprefix
  \url{https://link.aps.org/doi/10.1103/PhysRevLett.87.235002}.
  \DOIprefix\doi{10.1103/PhysRevLett.87.235002}.
\bibitem[{Uzun-Kaymak et~al.(2009)Uzun-Kaymak, Guzdar, Choi, Clary, Ellis,
  Hassam, and Teodorescu}]{Uzun2009}
\bibinfo{author}{I.~U. Uzun-Kaymak}, \bibinfo{author}{P.~N. Guzdar},
  \bibinfo{author}{S.~Choi}, \bibinfo{author}{M.~R. Clary},
  \bibinfo{author}{R.~F. Ellis}, \bibinfo{author}{A.~B. Hassam},
  \bibinfo{author}{C.~Teodorescu},
\newblock \bibinfo{title}{Nonlinear mode coupling and sheared flow in a
  rotating plasma},
\newblock \bibinfo{journal}{EPL (Europhysics Letters)} \bibinfo{volume}{85}
  (\bibinfo{year}{2009}) \bibinfo{pages}{15001}. \URLprefix
  \url{https://iopscience.iop.org/article/10.1209/0295-5075/85/15001}.
  \DOIprefix\doi{10.1209/0295-5075/85/15001}.
\bibitem[{Baker et~al.(1961)Baker, Hammel, and Ribe}]{Baker1961}
\bibinfo{author}{D.~A. Baker}, \bibinfo{author}{J.~E. Hammel},
  \bibinfo{author}{F.~L. Ribe},
\newblock \bibinfo{title}{Rotating plasma experiments. i. hydromagnetic
  properties},
\newblock \bibinfo{journal}{Physics of Fluids} \bibinfo{volume}{4}
  (\bibinfo{year}{1961}) \bibinfo{pages}{1534}. \URLprefix
  \url{https://aip.scitation.org/doi/10.1063/1.1706312}.
  \DOIprefix\doi{10.1063/1.1706312}.
\bibitem[{Baker and Hammel(1961)}]{Baker1961_2}
\bibinfo{author}{D.~A. Baker}, \bibinfo{author}{J.~E. Hammel},
\newblock \bibinfo{title}{Rotating plasma experiments. ii. energy measurements
  and the velocity limiting effect},
\newblock \bibinfo{journal}{Physics of Fluids} \bibinfo{volume}{4}
  (\bibinfo{year}{1961}) \bibinfo{pages}{1549}. \URLprefix
  \url{https://aip.scitation.org/doi/10.1063/1.1706313}.
  \DOIprefix\doi{10.1063/1.1706313}.
\bibitem[{Volosov(2009)}]{Volosov2009}
\bibinfo{author}{V.~I. Volosov},
\newblock \bibinfo{title}{Mhd stability of a hot rotating plasma: A brief
  review of psp-2 experiments},
\newblock \bibinfo{journal}{Plasma Physics Reports} \bibinfo{volume}{35}
  (\bibinfo{year}{2009}) \bibinfo{pages}{719--733}. \URLprefix
  \url{http://link.springer.com/10.1134/S1063780X09090025}.
  \DOIprefix\doi{10.1134/S1063780X09090025}.
\bibitem[{Ellis et~al.(2005)Ellis, Case, Elton, Ghosh, Griem, Hassam, Lunsford,
  Messer, and Teodorescu}]{Ellis2005}
\bibinfo{author}{R.~F. Ellis}, \bibinfo{author}{A.~Case},
  \bibinfo{author}{R.~Elton}, \bibinfo{author}{J.~Ghosh},
  \bibinfo{author}{H.~Griem}, \bibinfo{author}{A.~Hassam},
  \bibinfo{author}{R.~Lunsford}, \bibinfo{author}{S.~Messer},
  \bibinfo{author}{C.~Teodorescu},
\newblock \bibinfo{title}{Steady supersonically rotating plasmas in the
  maryland centrifugal experiment},
\newblock \bibinfo{journal}{Physics of Plasmas} \bibinfo{volume}{12}
  (\bibinfo{year}{2005}) \bibinfo{pages}{055704}. \URLprefix
  \url{http://aip.scitation.org/doi/10.1063/1.1896954}.
  \DOIprefix\doi{10.1063/1.1896954}.
\bibitem[{Lehnert(1971)}]{Lehnert1971}
\bibinfo{author}{B.~Lehnert},
\newblock \bibinfo{title}{Rotating plasmas},
\newblock \bibinfo{journal}{Nuclear Fusion} \bibinfo{volume}{11}
  (\bibinfo{year}{1971}) \bibinfo{pages}{485--533}. \URLprefix
  \url{https://iopscience.iop.org/article/10.1088/0029-5515/11/5/010}.
  \DOIprefix\doi{10.1088/0029-5515/11/5/010}.
\bibitem[{NAS(2021)}]{NASReport}
\bibinfo{title}{Bringing Fusion to the U.S. Grid}, \bibinfo{publisher}{National
  Academies Press}, \bibinfo{year}{2021}. \DOIprefix\doi{10.17226/25991}.
\bibitem[{Nishi et~al.(2006)Nishi, Yamanishi, and Hayashi}]{Nishi2006}
\bibinfo{author}{M.~Nishi}, \bibinfo{author}{T.~Yamanishi},
  \bibinfo{author}{T.~Hayashi},
\newblock \bibinfo{title}{Study on tritium accountancy in fusion demo plant at
  jaeri},
\newblock \bibinfo{journal}{Fusion Engineering and Design} \bibinfo{volume}{81}
  (\bibinfo{year}{2006}) \bibinfo{pages}{745--751}.
  \DOIprefix\doi{10.1016/J.FUSENGDES.2005.08.052}.
\bibitem[{Ebenhöch et~al.(2016)Ebenhöch, Niemes, Priester, and
  Röllig}]{Ebenhoch2016}
\bibinfo{author}{S.~Ebenhöch}, \bibinfo{author}{S.~Niemes},
  \bibinfo{author}{F.~Priester}, \bibinfo{author}{M.~Röllig},
\newblock \bibinfo{title}{Investigations of the applicability of a new
  accountancy tool in a closed tritium loop},
\newblock \bibinfo{journal}{Fusion Engineering and Design}
  \bibinfo{volume}{109-111} (\bibinfo{year}{2016}) \bibinfo{pages}{1376--1379}.
  \URLprefix
  \url{https://linkinghub.elsevier.com/retrieve/pii/S0920379615304026}.
  \DOIprefix\doi{10.1016/j.fusengdes.2015.12.018}.
\bibitem[{Ferry et~al.(2023)Ferry, Woller, Peterson, Sorensen, and
  Whyte}]{Ferry2023}
\bibinfo{author}{S.~E. Ferry}, \bibinfo{author}{K.~B. Woller},
  \bibinfo{author}{E.~E. Peterson}, \bibinfo{author}{C.~Sorensen},
  \bibinfo{author}{D.~G. Whyte},
\newblock \bibinfo{title}{The libra experiment: Investigating robust tritium
  accountancy in molten flibe exposed to a d-t fusion neutron spectrum},
\newblock \bibinfo{journal}{Fusion Science and Technology} \bibinfo{volume}{79}
  (\bibinfo{year}{2023}) \bibinfo{pages}{13--35}. \URLprefix
  \url{https://www.tandfonline.com/doi/abs/10.1080/15361055.2022.2078136}.
  \DOIprefix\doi{10.1080/15361055.2022.2078136}.
\bibitem[{Hopkins and Price(1985)}]{Hopkins1985}
\bibinfo{author}{G.~Hopkins}, \bibinfo{author}{R.~Price},
\newblock \bibinfo{title}{Fusion reactor design with ceramics},
\newblock \bibinfo{journal}{Nuclear Engineering and Design. Fusion}
  \bibinfo{volume}{2} (\bibinfo{year}{1985}) \bibinfo{pages}{111--143}.
  \URLprefix
  \url{https://linkinghub.elsevier.com/retrieve/pii/0167899X85900084}.
  \DOIprefix\doi{10.1016/0167-899X(85)90008-4}.
\bibitem[{Bolt(1993)}]{Bolt1993}
\bibinfo{author}{H.~Bolt}, \bibinfo{title}{Nonmetallic materials for plasma
  facing and structural applications in fusion reactors},
  \bibinfo{publisher}{Elsevier}, \bibinfo{year}{1993}, pp.
  \bibinfo{pages}{85--98}. \URLprefix
  \url{https://linkinghub.elsevier.com/retrieve/pii/B9780444899958500151}.
  \DOIprefix\doi{10.1016/B978-0-444-89995-8.50015-1}.
\bibitem[{Reinke et~al.(1995)Reinke, Kuhr, Kulisch, and Kassing}]{Reinke1995}
\bibinfo{author}{S.~Reinke}, \bibinfo{author}{M.~Kuhr},
  \bibinfo{author}{W.~Kulisch}, \bibinfo{author}{R.~Kassing},
\newblock \bibinfo{title}{Recent results in cubic boron nitride deposition in
  light of the sputter model},
\newblock \bibinfo{journal}{Diamond and Related Materials} \bibinfo{volume}{4}
  (\bibinfo{year}{1995}) \bibinfo{pages}{272--283}.
  \DOIprefix\doi{10.1016/0925-9635(94)05303-0}.
\bibitem[{Hu et~al.(2017)Hu, Wirth, and Maroudas}]{Hu2017}
\bibinfo{author}{L.~Hu}, \bibinfo{author}{B.~D. Wirth},
  \bibinfo{author}{D.~Maroudas},
\newblock \bibinfo{title}{Thermal conductivity of tungsten: Effects of
  plasma-related structural defects from molecular-dynamics simulations},
\newblock \bibinfo{journal}{Applied Physics Letters} \bibinfo{volume}{111}
  (\bibinfo{year}{2017}) \bibinfo{pages}{81902}. \URLprefix
  \url{https://pubs.aip.org/apl/article/111/8/081902/35076/Thermal-conductivity-of-tungsten-Effects-of-plasma}.
  \DOIprefix\doi{10.1063/1.4986956}.
\bibitem[{Han et~al.(2003)Han, Xu, Hu, Yu, Wang, Tian, and Huang}]{Han2003}
\bibinfo{author}{R.~Han}, \bibinfo{author}{X.~Xu}, \bibinfo{author}{X.~Hu},
  \bibinfo{author}{N.~Yu}, \bibinfo{author}{J.~Wang},
  \bibinfo{author}{Y.~Tian}, \bibinfo{author}{W.~Huang},
\newblock \bibinfo{title}{Development of bulk sic single crystal grown by
  physical vapor transport method},
\newblock \bibinfo{journal}{Optical Materials} \bibinfo{volume}{23}
  (\bibinfo{year}{2003}) \bibinfo{pages}{415--420}.
  \DOIprefix\doi{10.1016/S0925-3467(02)00330-0}.
\bibitem[{Yuan et~al.(2019)Yuan, Li, Lindsay, Cherns, Pomeroy, Liu, Edgar, and
  Kuball}]{Yuan2019}
\bibinfo{author}{C.~Yuan}, \bibinfo{author}{J.~Li},
  \bibinfo{author}{L.~Lindsay}, \bibinfo{author}{D.~Cherns},
  \bibinfo{author}{J.~W. Pomeroy}, \bibinfo{author}{S.~Liu},
  \bibinfo{author}{J.~H. Edgar}, \bibinfo{author}{M.~Kuball},
\newblock \bibinfo{title}{Modulating the thermal conductivity in hexagonal
  boron nitride via controlled boron isotope concentration},
\newblock \bibinfo{journal}{Communications Physics 2019 2:1}
  \bibinfo{volume}{2} (\bibinfo{year}{2019}) \bibinfo{pages}{1--8}. \URLprefix
  \url{https://www.nature.com/articles/s42005-019-0145-5}.
  \DOIprefix\doi{10.1038/s42005-019-0145-5}.
\bibitem[{Coe and Sussmann(2000)}]{Coe2000}
\bibinfo{author}{S.~E. Coe}, \bibinfo{author}{R.~S. Sussmann},
\newblock \bibinfo{title}{Optical, thermal and mechanical properties of cvd
  diamond},
\newblock \bibinfo{journal}{Diamond and Related Materials} \bibinfo{volume}{9}
  (\bibinfo{year}{2000}) \bibinfo{pages}{1726--1729}.
  \DOIprefix\doi{10.1016/S0925-9635(00)00298-3}.
\bibitem[{Hattori et~al.(2016)Hattori, Taniguchi, Watanabe, and
  Nagashio}]{Hattori2016}
\bibinfo{author}{Y.~Hattori}, \bibinfo{author}{T.~Taniguchi},
  \bibinfo{author}{K.~Watanabe}, \bibinfo{author}{K.~Nagashio},
\newblock \bibinfo{title}{Anisotropic dielectric breakdown strength of single
  crystal hexagonal boron nitride},
\newblock \bibinfo{journal}{ACS Applied Materials \& Interfaces}
  \bibinfo{volume}{8} (\bibinfo{year}{2016}) \bibinfo{pages}{27877--27884}.
  \URLprefix \url{https://pubs.acs.org/doi/10.1021/acsami.6b06425}.
  \DOIprefix\doi{10.1021/acsami.6b06425}.
\bibitem[{Boettger et~al.(1995)Boettger, Bluhm, Jiang, Schäfer, and
  Klages}]{Boettger1995}
\bibinfo{author}{E.~Boettger}, \bibinfo{author}{A.~Bluhm},
  \bibinfo{author}{X.~Jiang}, \bibinfo{author}{L.~Schäfer},
  \bibinfo{author}{C.-P. Klages},
\newblock \bibinfo{title}{Investigation of the high-field conductivity and
  dielectric strength of nitrogen containing polycrystalline diamond films},
\newblock \bibinfo{journal}{Journal of Applied Physics} \bibinfo{volume}{77}
  (\bibinfo{year}{1995}) \bibinfo{pages}{6332--6337}. \URLprefix
  \url{https://pubs.aip.org/jap/article/77/12/6332/528443/Investigation-of-the-high-field-conductivity-and}.
  \DOIprefix\doi{10.1063/1.359103}.
\bibitem[{Liu et~al.(2009)Liu, Zhang, Luo, and Lu}]{Liu2009}
\bibinfo{author}{Y.~L. Liu}, \bibinfo{author}{Y.~Zhang}, \bibinfo{author}{G.~N.
  Luo}, \bibinfo{author}{G.~H. Lu},
\newblock \bibinfo{title}{Structure, stability and diffusion of hydrogen in
  tungsten: A first-principles study},
\newblock \bibinfo{journal}{Journal of Nuclear Materials}
  \bibinfo{volume}{390-391} (\bibinfo{year}{2009}) \bibinfo{pages}{1032--1034}.
  \DOIprefix\doi{10.1016/J.JNUCMAT.2009.01.277}.
\bibitem[{Liu et~al.(2014)Liu, Wu, Yu, Li, Shu, and Lu}]{Liu2014}
\bibinfo{author}{Y.~N. Liu}, \bibinfo{author}{T.~Wu}, \bibinfo{author}{Y.~Yu},
  \bibinfo{author}{X.~C. Li}, \bibinfo{author}{X.~Shu}, \bibinfo{author}{G.~H.
  Lu},
\newblock \bibinfo{title}{Hydrogen diffusion in tungsten: A molecular dynamics
  study},
\newblock \bibinfo{journal}{Journal of Nuclear Materials} \bibinfo{volume}{455}
  (\bibinfo{year}{2014}) \bibinfo{pages}{676--680}.
  \DOIprefix\doi{10.1016/J.JNUCMAT.2014.09.003}.
\bibitem[{CAUSEY et~al.(1978)CAUSEY, FOWLER, RAVANBAKHT, ELLEMAN, and
  VERGHESE}]{CAUSEY1978}
\bibinfo{author}{R.~A. CAUSEY}, \bibinfo{author}{J.~D. FOWLER},
  \bibinfo{author}{C.~RAVANBAKHT}, \bibinfo{author}{T.~S. ELLEMAN},
  \bibinfo{author}{K.~VERGHESE},
\newblock \bibinfo{title}{Hydrogen diffusion and solubility in silicon
  carbide},
\newblock \bibinfo{journal}{Journal of the American Ceramic Society}
  \bibinfo{volume}{61} (\bibinfo{year}{1978}) \bibinfo{pages}{221--225}.
  \URLprefix
  \url{https://ceramics.onlinelibrary.wiley.com/doi/10.1111/j.1151-2916.1978.tb09284.x}.
  \DOIprefix\doi{10.1111/j.1151-2916.1978.tb09284.x}.
\bibitem[{Checchetto and Miotello(2000)}]{Checchetto2000}
\bibinfo{author}{R.~Checchetto}, \bibinfo{author}{A.~Miotello},
\newblock \bibinfo{title}{Deuterium diffusion through hexagonal boron nitride
  thin films},
\newblock \bibinfo{journal}{Journal of Applied Physics} \bibinfo{volume}{87}
  (\bibinfo{year}{2000}) \bibinfo{pages}{110--116}. \URLprefix
  \url{https://pubs.aip.org/jap/article/87/1/110/367998/Deuterium-diffusion-through-hexagonal-boron}.
  \DOIprefix\doi{10.1063/1.371831}.
\bibitem[{Cherniak et~al.(2018)Cherniak, Watson, Meunier, and
  Kharche}]{Cherniak2018}
\bibinfo{author}{D.~J. Cherniak}, \bibinfo{author}{E.~B. Watson},
  \bibinfo{author}{V.~Meunier}, \bibinfo{author}{N.~Kharche},
\newblock \bibinfo{title}{Diffusion of helium, hydrogen and deuterium in
  diamond: Experiment, theory and geochemical applications},
\newblock \bibinfo{journal}{Geochimica et Cosmochimica Acta}
  \bibinfo{volume}{232} (\bibinfo{year}{2018}) \bibinfo{pages}{206--224}.
  \DOIprefix\doi{10.1016/J.GCA.2018.04.029}.
\bibitem[{Eckstein and László(1991)}]{Eckstein1991}
\bibinfo{author}{W.~Eckstein}, \bibinfo{author}{J.~László},
\newblock \bibinfo{title}{Sputtering of tungsten and molybdenum},
\newblock \bibinfo{journal}{Journal of Nuclear Materials} \bibinfo{volume}{183}
  (\bibinfo{year}{1991}) \bibinfo{pages}{19--24}.
  \DOIprefix\doi{10.1016/0022-3115(91)90466-K}.
\bibitem[{Balden and Roth(2000)}]{Balden2000}
\bibinfo{author}{M.~Balden}, \bibinfo{author}{J.~Roth},
\newblock \bibinfo{title}{Comparison of the chemical erosion of si, c and sic
  under deuterium ion bombardment},
\newblock \bibinfo{journal}{Journal of Nuclear Materials} \bibinfo{volume}{279}
  (\bibinfo{year}{2000}) \bibinfo{pages}{351--355}.
  \DOIprefix\doi{10.1016/S0022-3115(00)00032-5}.
\bibitem[{Philipps(2011)}]{Philipps2011}
\bibinfo{author}{V.~Philipps},
\newblock \bibinfo{title}{Tungsten as material for plasma-facing components in
  fusion devices},
\newblock \bibinfo{journal}{Journal of Nuclear Materials} \bibinfo{volume}{415}
  (\bibinfo{year}{2011}) \bibinfo{pages}{S2--S9}.
  \DOIprefix\doi{10.1016/J.JNUCMAT.2011.01.110}.
\bibitem[{Wesson(2004)}]{Wesson2004}
\bibinfo{author}{J.~Wesson}, \bibinfo{title}{Confinement}, \bibinfo{edition}{3}
  ed., \bibinfo{publisher}{Oxford University Press}, \bibinfo{year}{2004}.
\bibitem[{Bortolon et~al.(2019)Bortolon, Rohde, Maingi, Wolfrum, Dux, Herrmann,
  Lunsford, McDermott, Nagy, Kallenbach, Mansfield, Nazikian, and
  Neu}]{Bortolon2019}
\bibinfo{author}{A.~Bortolon}, \bibinfo{author}{V.~Rohde},
  \bibinfo{author}{R.~Maingi}, \bibinfo{author}{E.~Wolfrum},
  \bibinfo{author}{R.~Dux}, \bibinfo{author}{A.~Herrmann},
  \bibinfo{author}{R.~Lunsford}, \bibinfo{author}{R.~M. McDermott},
  \bibinfo{author}{A.~Nagy}, \bibinfo{author}{A.~Kallenbach},
  \bibinfo{author}{D.~K. Mansfield}, \bibinfo{author}{R.~Nazikian},
  \bibinfo{author}{R.~Neu},
\newblock \bibinfo{title}{Real-time wall conditioning by controlled injection
  of boron and boron nitride powder in full tungsten wall asdex upgrade},
\newblock \bibinfo{journal}{Nuclear Materials and Energy} \bibinfo{volume}{19}
  (\bibinfo{year}{2019}) \bibinfo{pages}{384--389}.
  \DOIprefix\doi{10.1016/J.NME.2019.03.022}.
\bibitem[{Lunsford et~al.(2019)Lunsford, Rohde, Bortolon, Dux, Herrmann,
  Kallenbach, McDermott, David, Drenik, Laggner, Maingi, Mansfield, Nagy, Neu,
  and Wolfrum}]{Lunsford2019}
\bibinfo{author}{R.~Lunsford}, \bibinfo{author}{V.~Rohde},
  \bibinfo{author}{A.~Bortolon}, \bibinfo{author}{R.~Dux},
  \bibinfo{author}{A.~Herrmann}, \bibinfo{author}{A.~Kallenbach},
  \bibinfo{author}{R.~M. McDermott}, \bibinfo{author}{P.~David},
  \bibinfo{author}{A.~Drenik}, \bibinfo{author}{F.~Laggner},
  \bibinfo{author}{R.~Maingi}, \bibinfo{author}{D.~K. Mansfield},
  \bibinfo{author}{A.~Nagy}, \bibinfo{author}{R.~Neu},
  \bibinfo{author}{E.~Wolfrum},
\newblock \bibinfo{title}{Active conditioning of asdex upgrade tungsten
  plasma-facing components and discharge enhancement through boron and boron
  nitride particulate injection},
\newblock \bibinfo{journal}{Nuclear Fusion} \bibinfo{volume}{59}
  (\bibinfo{year}{2019}) \bibinfo{pages}{126034}. \URLprefix
  \url{https://iopscience.iop.org/article/10.1088/1741-4326/ab4095
  https://iopscience.iop.org/article/10.1088/1741-4326/ab4095/meta}.
  \DOIprefix\doi{10.1088/1741-4326/AB4095}.
\bibitem[{Gilson et~al.(2021)Gilson, Lee, Bortolon, Choe, Diallo, Hong, Lee,
  Lee, Maingi, Mansfield, Nagy, Park, Song, Song, Yun, Yoon, and
  Nazikian}]{Gilson2021}
\bibinfo{author}{E.~P. Gilson}, \bibinfo{author}{H.~H. Lee},
  \bibinfo{author}{A.~Bortolon}, \bibinfo{author}{W.~Choe},
  \bibinfo{author}{A.~Diallo}, \bibinfo{author}{S.~H. Hong},
  \bibinfo{author}{H.~M. Lee}, \bibinfo{author}{J.~Lee},
  \bibinfo{author}{R.~Maingi}, \bibinfo{author}{D.~K. Mansfield},
  \bibinfo{author}{A.~Nagy}, \bibinfo{author}{S.~H. Park},
  \bibinfo{author}{I.~W. Song}, \bibinfo{author}{J.~I. Song},
  \bibinfo{author}{S.~W. Yun}, \bibinfo{author}{S.~W. Yoon},
  \bibinfo{author}{R.~Nazikian},
\newblock \bibinfo{title}{Wall conditioning and elm mitigation with boron
  nitride powder injection in kstar},
\newblock \bibinfo{journal}{Nuclear Materials and Energy} \bibinfo{volume}{28}
  (\bibinfo{year}{2021}) \bibinfo{pages}{101043}.
  \DOIprefix\doi{10.1016/J.NME.2021.101043}.
\bibitem[{Lunsford et~al.(2022)Lunsford, Masuzaki, Nespoli, Ashikawa, Gilson,
  Gates, Ida, Kawamura, Morisaki, Nagy, Oishi, Shoji, Suzuki, and
  Yoshinuma}]{Lunsford2022}
\bibinfo{author}{R.~Lunsford}, \bibinfo{author}{S.~Masuzaki},
  \bibinfo{author}{F.~Nespoli}, \bibinfo{author}{N.~Ashikawa},
  \bibinfo{author}{E.~P. Gilson}, \bibinfo{author}{D.~A. Gates},
  \bibinfo{author}{K.~Ida}, \bibinfo{author}{G.~Kawamura},
  \bibinfo{author}{T.~Morisaki}, \bibinfo{author}{A.~Nagy},
  \bibinfo{author}{T.~Oishi}, \bibinfo{author}{M.~Shoji},
  \bibinfo{author}{C.~Suzuki}, \bibinfo{author}{M.~Yoshinuma},
\newblock \bibinfo{title}{Real-time wall conditioning and recycling
  modification utilizing boron and boron nitride powder injections into the
  large helical device},
\newblock \bibinfo{journal}{Nuclear Fusion} \bibinfo{volume}{62}
  (\bibinfo{year}{2022}) \bibinfo{pages}{086021}. \URLprefix
  \url{https://iopscience.iop.org/article/10.1088/1741-4326/ac6ff5
  https://iopscience.iop.org/article/10.1088/1741-4326/ac6ff5/meta}.
  \DOIprefix\doi{10.1088/1741-4326/AC6FF5}.
\bibitem[{Yu et~al.(2020)Yu, Tanigawa, Hamaguchi, and Nozawa}]{Yu2020}
\bibinfo{author}{J.~H. Yu}, \bibinfo{author}{H.~Tanigawa},
  \bibinfo{author}{D.~Hamaguchi}, \bibinfo{author}{T.~Nozawa},
\newblock \bibinfo{title}{Mechanical properties of three kinds of iter-grade
  pure tungsten with different manufacturing processes},
\newblock \bibinfo{journal}{Fusion Engineering and Design}
  \bibinfo{volume}{157} (\bibinfo{year}{2020}) \bibinfo{pages}{111679}.
  \DOIprefix\doi{10.1016/J.FUSENGDES.2020.111679}.
\bibitem[{Snead et~al.(2007)Snead, Nozawa, Katoh, Byun, Kondo, and
  Petti}]{Snead2007}
\bibinfo{author}{L.~L. Snead}, \bibinfo{author}{T.~Nozawa},
  \bibinfo{author}{Y.~Katoh}, \bibinfo{author}{T.~S. Byun},
  \bibinfo{author}{S.~Kondo}, \bibinfo{author}{D.~A. Petti},
\newblock \bibinfo{title}{Handbook of sic properties for fuel performance
  modeling},
\newblock \bibinfo{journal}{Journal of Nuclear Materials} \bibinfo{volume}{371}
  (\bibinfo{year}{2007}) \bibinfo{pages}{329--377}.
  \DOIprefix\doi{10.1016/J.JNUCMAT.2007.05.016}.
\bibitem[{Duan et~al.(2016)Duan, Jia, Wang, Cai, Tian, Yang, He, Wang, and
  Zhou}]{Duan2016}
\bibinfo{author}{X.~Duan}, \bibinfo{author}{D.~Jia}, \bibinfo{author}{Z.~Wang},
  \bibinfo{author}{D.~Cai}, \bibinfo{author}{Z.~Tian},
  \bibinfo{author}{Z.~Yang}, \bibinfo{author}{P.~He},
  \bibinfo{author}{S.~Wang}, \bibinfo{author}{Y.~Zhou},
\newblock \bibinfo{title}{Influence of hot-press sintering parameters on
  microstructures and mechanical properties of h-bn ceramics},
\newblock \bibinfo{journal}{Journal of Alloys and Compounds}
  \bibinfo{volume}{684} (\bibinfo{year}{2016}) \bibinfo{pages}{474--480}.
  \DOIprefix\doi{10.1016/J.JALLCOM.2016.05.153}.
\bibitem[{Sussmann et~al.(1994)Sussmann, Brandon, Scarsbrook, Sweeney,
  Valentine, Whitehead, and Wort}]{Sussmann1994}
\bibinfo{author}{R.~S. Sussmann}, \bibinfo{author}{J.~R. Brandon},
  \bibinfo{author}{G.~A. Scarsbrook}, \bibinfo{author}{C.~G. Sweeney},
  \bibinfo{author}{T.~J. Valentine}, \bibinfo{author}{A.~J. Whitehead},
  \bibinfo{author}{C.~J. Wort},
\newblock \bibinfo{title}{Properties of bulk polycrystalline cvd diamond},
\newblock \bibinfo{journal}{Diamond and Related Materials} \bibinfo{volume}{3}
  (\bibinfo{year}{1994}) \bibinfo{pages}{303--312}.
  \DOIprefix\doi{10.1016/0925-9635(94)90176-7}.
\bibitem[{Gludovatz et~al.(2010)Gludovatz, Wurster, Hoffmann, and
  Pippan}]{Gludovatz2010}
\bibinfo{author}{B.~Gludovatz}, \bibinfo{author}{S.~Wurster},
  \bibinfo{author}{A.~Hoffmann}, \bibinfo{author}{R.~Pippan},
\newblock \bibinfo{title}{Fracture toughness of polycrystalline tungsten
  alloys},
\newblock \bibinfo{journal}{International Journal of Refractory Metals and Hard
  Materials} \bibinfo{volume}{28} (\bibinfo{year}{2010})
  \bibinfo{pages}{674--678}. \DOIprefix\doi{10.1016/J.IJRMHM.2010.04.007}.
\bibitem[{Field(2012)}]{Field2012}
\bibinfo{author}{J.~E. Field},
\newblock \bibinfo{title}{The mechanical and strength properties of diamond},
\newblock \bibinfo{journal}{Reports on Progress in Physics}
  \bibinfo{volume}{75} (\bibinfo{year}{2012}) \bibinfo{pages}{126505}.
  \URLprefix
  \url{https://iopscience.iop.org/article/10.1088/0034-4885/75/12/126505}.
  \DOIprefix\doi{10.1088/0034-4885/75/12/126505}.
\bibitem[{Buzhinskij et~al.(1990)Buzhinskij, Opimach, Kabishev, Lopatin, and
  Surov}]{Buzhinskij1990}
\bibinfo{author}{O.~I. Buzhinskij}, \bibinfo{author}{I.~V. Opimach},
  \bibinfo{author}{A.~V. Kabishev}, \bibinfo{author}{V.~V. Lopatin},
  \bibinfo{author}{Y.~P. Surov},
\newblock \bibinfo{title}{Application of pyrolytic boron nitride in fusion
  devices},
\newblock \bibinfo{journal}{Journal of Nuclear Materials} \bibinfo{volume}{173}
  (\bibinfo{year}{1990}) \bibinfo{pages}{179--184}.
  \DOIprefix\doi{10.1016/0022-3115(90)90256-M}.
\bibitem[{Buzhinskij et~al.(1992)Buzhinskij, Lopatin, and
  Sharupin}]{Buzhinskij1992}
\bibinfo{author}{O.~I. Buzhinskij}, \bibinfo{author}{V.~V. Lopatin},
  \bibinfo{author}{B.~N. Sharupin},
\newblock \bibinfo{title}{Rhombohedral pyrolytic boron nitride as a protection
  material in fusion devices},
\newblock \bibinfo{journal}{Journal of Nuclear Materials}
  \bibinfo{volume}{196-198} (\bibinfo{year}{1992}) \bibinfo{pages}{1118--1120}.
  \DOIprefix\doi{10.1016/S0022-3115(06)80206-0}.
\bibitem[{Yamage et~al.(1992)Yamage, Ejima, Toyoda, and Sugai}]{Yamage1992}
\bibinfo{author}{M.~Yamage}, \bibinfo{author}{T.~Ejima},
  \bibinfo{author}{H.~Toyoda}, \bibinfo{author}{H.~Sugai},
\newblock \bibinfo{title}{In situ boron nitride coating and comparison with
  existing boronizations},
\newblock \bibinfo{journal}{Journal of Nuclear Materials}
  \bibinfo{volume}{196-198} (\bibinfo{year}{1992}) \bibinfo{pages}{618--621}.
  \DOIprefix\doi{10.1016/S0022-3115(06)80110-8}.
\bibitem[{Buzhinskij et~al.(1995)Buzhinskij, Opimach, Barsuk, Otrozhenko,
  Zhitlukhin, Trazhenkov, West, Trester, Valentine, Watson, Youchinson, Gahl,
  and Crawford}]{Buzhinskij1995}
\bibinfo{author}{O.~Buzhinskij}, \bibinfo{author}{I.~Opimach},
  \bibinfo{author}{V.~Barsuk}, \bibinfo{author}{V.~Otrozhenko},
  \bibinfo{author}{A.~Zhitlukhin}, \bibinfo{author}{A.~Trazhenkov},
  \bibinfo{author}{W.~West}, \bibinfo{author}{P.~Trester},
  \bibinfo{author}{P.~Valentine}, \bibinfo{author}{R.~Watson},
  \bibinfo{author}{D.~Youchinson}, \bibinfo{author}{J.~Gahl},
  \bibinfo{author}{J.~Crawford},
\newblock \bibinfo{title}{Performance of boron containing materials under
  disruption simulations and tokamak divertor plasma testing},
\newblock \bibinfo{journal}{Journal of Nuclear Materials}
  \bibinfo{volume}{220-222} (\bibinfo{year}{1995}) \bibinfo{pages}{922--925}.
  \URLprefix
  \url{https://linkinghub.elsevier.com/retrieve/pii/0022311594006121}.
  \DOIprefix\doi{10.1016/0022-3115(94)00612-1}.
\bibitem[{Okuno(1966)}]{Okuno1966}
\bibinfo{author}{A.~F. Okuno}, \bibinfo{title}{Ablative and insulating
  properties of outgassed boron nitride and boron nitride composite},
  \bibinfo{type}{Technical Report}, NASA Ames Research Center,
  \bibinfo{year}{1966}. \URLprefix
  \url{https://ntrs.nasa.gov/api/citations/19660028127/downloads/19660028127.pdf}.
\bibitem[{Naujoks(2006)}]{Naujoks2006}
\bibinfo{author}{D.~Naujoks}, \bibinfo{title}{Plasma-Material Interaction in
  Controlled Fusion}, \bibinfo{publisher}{Springer}, \bibinfo{year}{2006}.
\bibitem[{Mil(2024)}]{MilliporeSigma}
\bibinfo{title}{Millipore sigma}, \bibinfo{year}{2024}. \URLprefix
  \url{www.sigmaaldrich.com}.
\bibitem[{Drobny and Curreli(2021)}]{Drobny2021}
\bibinfo{author}{J.~T. Drobny}, \bibinfo{author}{D.~Curreli},
\newblock \bibinfo{title}{Rustbca: A high-performance
  binary-collision-approximation code for ion-material interactions},
\newblock \bibinfo{journal}{Journal of Open Source Software}
  \bibinfo{volume}{6} (\bibinfo{year}{2021}) \bibinfo{pages}{3298}. \URLprefix
  \url{https://joss.theoj.org/papers/10.21105/joss.03298}.
  \DOIprefix\doi{10.21105/JOSS.03298}.
\bibitem[{Lipp et~al.(1989)Lipp, Schwetz, and Hunold}]{Lipp1989}
\bibinfo{author}{A.~Lipp}, \bibinfo{author}{K.~A. Schwetz},
  \bibinfo{author}{K.~Hunold},
\newblock \bibinfo{title}{Hexagonal boron nitride: Fabrication, properties and
  applications},
\newblock \bibinfo{journal}{Journal of the European Ceramic Society}
  \bibinfo{volume}{5} (\bibinfo{year}{1989}) \bibinfo{pages}{3--9}.
  \DOIprefix\doi{10.1016/0955-2219(89)90003-4}.
\bibitem[{Hvizdoš and Vencl(2021)}]{Hvisdos2021}
\bibinfo{author}{P.~Hvizdoš}, \bibinfo{author}{A.~Vencl},
  \bibinfo{title}{Ceramic Matrix Composites With Carbon Nanophases:
  Development, Structure, Mechanical and Tribological Properties and Electrical
  Conductivity}, volume~\bibinfo{volume}{2}, \bibinfo{publisher}{Elsevier},
  \bibinfo{year}{2021}, pp. \bibinfo{pages}{116--133}. \URLprefix
  \url{https://linkinghub.elsevier.com/retrieve/pii/B9780128035818118582}.
  \DOIprefix\doi{10.1016/B978-0-12-803581-8.11858-2}.
\bibitem[{Ooi et~al.(2005)Ooi, Rairkar, Lindsley, and Adams}]{Ooi2005}
\bibinfo{author}{N.~Ooi}, \bibinfo{author}{A.~Rairkar},
  \bibinfo{author}{L.~Lindsley}, \bibinfo{author}{J.~B. Adams},
\newblock \bibinfo{title}{Electronic structure and bonding in hexagonal boron
  nitride},
\newblock \bibinfo{journal}{Journal of Physics: Condensed Matter}
  \bibinfo{volume}{18} (\bibinfo{year}{2005}) \bibinfo{pages}{97}. \URLprefix
  \url{https://iopscience-iop-org.proxy-um.researchport.umd.edu/article/10.1088/0953-8984/18/1/007
  https://iopscience-iop-org.proxy-um.researchport.umd.edu/article/10.1088/0953-8984/18/1/007/meta}.
  \DOIprefix\doi{10.1088/0953-8984/18/1/007}.
\bibitem[{Heera and Skorupa(1996)}]{Heera1996}
\bibinfo{author}{V.~Heera}, \bibinfo{author}{W.~Skorupa},
\newblock \bibinfo{title}{Ion implantation and annealing effects in silicon
  carbide},
\newblock \bibinfo{journal}{Materials Research Society Symposium - Proceedings}
  \bibinfo{volume}{438} (\bibinfo{year}{1996}) \bibinfo{pages}{241--252}.
  \URLprefix
  \url{https://link-springer-com.proxy-um.researchport.umd.edu/article/10.1557/PROC-438-241}.
  \DOIprefix\doi{10.1557/PROC-438-241/METRICS}.
\bibitem[{Kittel et~al.(2005)Kittel, McEuen, and Sons}]{kittel2005introduction}
\bibinfo{author}{C.~Kittel}, \bibinfo{author}{P.~McEuen},
  \bibinfo{author}{J.~W.~. Sons}, \bibinfo{title}{Introduction to Solid State
  Physics}, \bibinfo{publisher}{John Wiley \& Sons}, \bibinfo{year}{2005}.
  \URLprefix \url{https://books.google.com/books?id=rAMujwEACAAJ}.
\bibitem[{Smith and Boyd(2016)}]{Smith2016}
\bibinfo{author}{B.~D. Smith}, \bibinfo{author}{I.~D. Boyd},
\newblock \bibinfo{title}{Molecular dynamics investigation of hexagonal boron
  nitride sputtering and sputtered particle characteristics},
\newblock \bibinfo{journal}{Journal of Applied Physics} \bibinfo{volume}{120}
  (\bibinfo{year}{2016}) \bibinfo{pages}{55}. \URLprefix
  \url{https://pubs.aip.org/jap/article/120/5/053301/344729/Molecular-dynamics-investigation-of-hexagonal}.
  \DOIprefix\doi{10.1063/1.4958869}.
\bibitem[{Yang and Hassanein(2014)}]{Yang2014}
\bibinfo{author}{X.~Yang}, \bibinfo{author}{A.~Hassanein},
\newblock \bibinfo{title}{Atomic scale calculations of tungsten surface binding
  energy and beryllium-induced tungsten sputtering},
\newblock \bibinfo{journal}{Applied Surface Science} \bibinfo{volume}{293}
  (\bibinfo{year}{2014}) \bibinfo{pages}{187--190}.
  \DOIprefix\doi{10.1016/J.APSUSC.2013.12.129}.
\bibitem[{Bauer et~al.(1979)Bauer, Wilson, Bisson, Haggmark, and
  Goldston}]{Bauer_1979}
\bibinfo{author}{W.~Bauer}, \bibinfo{author}{K.~Wilson},
  \bibinfo{author}{C.~Bisson}, \bibinfo{author}{L.~Haggmark},
  \bibinfo{author}{R.~Goldston},
\newblock \bibinfo{title}{Alpha transport and blistering in tokamaks},
\newblock \bibinfo{journal}{Nuclear Fusion} \bibinfo{volume}{19}
  (\bibinfo{year}{1979}) \bibinfo{pages}{93}. \URLprefix
  \url{https://dx.doi.org/10.1088/0029-5515/19/1/009}.
  \DOIprefix\doi{10.1088/0029-5515/19/1/009}.
\bibitem[{Jung and Li(2024)}]{Jung2024}
\bibinfo{author}{Y.~Jung}, \bibinfo{author}{J.~Li},
\newblock \bibinfo{title}{Boron-10 stimulated helium production and accelerated
  radiation displacements for rapid development of fusion structural
  materials},
\newblock \bibinfo{journal}{Journal of Materiomics} \bibinfo{volume}{10}
  (\bibinfo{year}{2024}) \bibinfo{pages}{377--385}.
  \DOIprefix\doi{10.1016/J.JMAT.2023.06.009}.
\bibitem[{Brown et~al.(2018)Brown, Chadwick, Capote, Kahler, Trkov
  et~al.}]{Brown2018}
\bibinfo{author}{D.~A. Brown}, \bibinfo{author}{M.~B. Chadwick},
  \bibinfo{author}{R.~Capote}, \bibinfo{author}{A.~C. Kahler},
  \bibinfo{author}{A.~Trkov}, et~al.,
\newblock \bibinfo{title}{Endf/b-viii.0: The 8th major release of the nuclear
  reaction data library with cielo-project cross sections, new standards and
  thermal scattering data},
\newblock \bibinfo{journal}{Nuclear Data Sheets} \bibinfo{volume}{148}
  (\bibinfo{year}{2018}) \bibinfo{pages}{1--142}.
  \DOIprefix\doi{10.1016/J.NDS.2018.02.001}.
\bibitem[{Kozlovskiy and Zdorovets(2021)}]{Kozlovskiy2021}
\bibinfo{author}{A.~L. Kozlovskiy}, \bibinfo{author}{M.~V. Zdorovets},
\newblock \bibinfo{title}{Study of the radiation disordering mechanisms of aln
  ceramic structure as a result of helium swelling},
\newblock \bibinfo{journal}{Journal of Materials Science: Materials in
  Electronics} \bibinfo{volume}{32} (\bibinfo{year}{2021})
  \bibinfo{pages}{21658--21669}. \URLprefix
  \url{https://link.springer.com/article/10.1007/s10854-021-06684-x}.
  \DOIprefix\doi{10.1007/S10854-021-06684-X/FIGURES/6}.
\bibitem[{Hammond(2017)}]{Hammond2017}
\bibinfo{author}{K.~D. Hammond},
\newblock \bibinfo{title}{Helium, hydrogen, and fuzz in plasma-facing
  materials},
\newblock \bibinfo{journal}{Materials Research Express} \bibinfo{volume}{4}
  (\bibinfo{year}{2017}) \bibinfo{pages}{104002}. \URLprefix
  \url{https://iopscience.iop.org/article/10.1088/2053-1591/aa8c22}.
  \DOIprefix\doi{10.1088/2053-1591/aa8c22}.
\bibitem[{Romano et~al.(2015)Romano, Horelik, Herman, Nelson, Forget, and
  Smith}]{Romano2015}
\bibinfo{author}{P.~K. Romano}, \bibinfo{author}{N.~E. Horelik},
  \bibinfo{author}{B.~R. Herman}, \bibinfo{author}{A.~G. Nelson},
  \bibinfo{author}{B.~Forget}, \bibinfo{author}{K.~Smith},
\newblock \bibinfo{title}{Openmc: A state-of-the-art monte carlo code for
  research and development},
\newblock \bibinfo{journal}{Annals of Nuclear Energy} \bibinfo{volume}{82}
  (\bibinfo{year}{2015}) \bibinfo{pages}{90--97}.
  \DOIprefix\doi{10.1016/J.ANUCENE.2014.07.048}.
\bibitem[{Norgett et~al.(1975)Norgett, Robinson, and Torrens}]{Norgett1975}
\bibinfo{author}{M.~J. Norgett}, \bibinfo{author}{M.~T. Robinson},
  \bibinfo{author}{I.~M. Torrens},
\newblock \bibinfo{title}{A proposed method of calculating displacement dose
  rates},
\newblock \bibinfo{journal}{Nuclear Engineering and Design}
  \bibinfo{volume}{33} (\bibinfo{year}{1975}) \bibinfo{pages}{50--54}.
  \DOIprefix\doi{10.1016/0029-5493(75)90035-7}.
\bibitem[{Kotakoski et~al.(2010)Kotakoski, Jin, Lehtinen, Suenaga, and
  Krasheninnikov}]{Kotakoski2010}
\bibinfo{author}{J.~Kotakoski}, \bibinfo{author}{C.~H. Jin},
  \bibinfo{author}{O.~Lehtinen}, \bibinfo{author}{K.~Suenaga},
  \bibinfo{author}{A.~V. Krasheninnikov},
\newblock \bibinfo{title}{Electron knock-on damage in hexagonal boron nitride
  monolayers},
\newblock \bibinfo{journal}{Physical Review B - Condensed Matter and Materials
  Physics} \bibinfo{volume}{82} (\bibinfo{year}{2010}).
  \DOIprefix\doi{10.1103/PHYSREVB.82.113404}.
\bibitem[{Lucas and Pizzagalli(2005)}]{Lucas2005}
\bibinfo{author}{G.~Lucas}, \bibinfo{author}{L.~Pizzagalli},
\newblock \bibinfo{title}{Ab initio molecular dynamics calculations of
  threshold displacement energies in silicon carbide},
\newblock \bibinfo{journal}{Physical Review B - Condensed Matter and Materials
  Physics} \bibinfo{volume}{72} (\bibinfo{year}{2005}) \bibinfo{pages}{161202}.
  \URLprefix
  \url{https://journals.aps.org/prb/abstract/10.1103/PhysRevB.72.161202}.
  \DOIprefix\doi{10.1103/PHYSREVB.72.161202/FIGURES/3/THUMBNAIL}.
\bibitem[{Banisalman et~al.(2017)Banisalman, Park, and Oda}]{Banisalman2017}
\bibinfo{author}{M.~J. Banisalman}, \bibinfo{author}{S.~Park},
  \bibinfo{author}{T.~Oda},
\newblock \bibinfo{title}{Evaluation of the threshold displacement energy in
  tungsten by molecular dynamics calculations},
\newblock \bibinfo{journal}{Journal of Nuclear Materials} \bibinfo{volume}{495}
  (\bibinfo{year}{2017}) \bibinfo{pages}{277--284}.
  \DOIprefix\doi{10.1016/J.JNUCMAT.2017.08.019}.
\bibitem[{Katoh et~al.(2003)Katoh, Kohyama, Hinoki, and Snead}]{Katoh2003}
\bibinfo{author}{Y.~Katoh}, \bibinfo{author}{A.~Kohyama},
  \bibinfo{author}{T.~Hinoki}, \bibinfo{author}{L.~L. Snead},
\newblock \bibinfo{title}{Progress in sic-based ceramic composites for fusion
  applications},
\newblock \bibinfo{journal}{Fusion Science and Technology} \bibinfo{volume}{44}
  (\bibinfo{year}{2003}) \bibinfo{pages}{155--162}. \URLprefix
  \url{https://www.tandfonline.com/doi/abs/10.13182/FST03-A326}.
  \DOIprefix\doi{10.13182/FST03-A326}.
\bibitem[{Katoh et~al.(2015)Katoh, Nozawa, Shih, Ozawa, Koyanagi, Porter, and
  Snead}]{Katoh2015}
\bibinfo{author}{Y.~Katoh}, \bibinfo{author}{T.~Nozawa},
  \bibinfo{author}{C.~Shih}, \bibinfo{author}{K.~Ozawa},
  \bibinfo{author}{T.~Koyanagi}, \bibinfo{author}{W.~Porter},
  \bibinfo{author}{L.~L. Snead},
\newblock \bibinfo{title}{High-dose neutron irradiation of hi-nicalon type s
  silicon carbide composites. part 2: Mechanical and physical properties},
\newblock \bibinfo{journal}{Journal of Nuclear Materials} \bibinfo{volume}{462}
  (\bibinfo{year}{2015}) \bibinfo{pages}{450--457}.
  \DOIprefix\doi{10.1016/J.JNUCMAT.2014.12.121}.
\bibitem[{Mohri et~al.(1978)Mohri, Watanabe, and Yamashina}]{Mohri1978}
\bibinfo{author}{M.~Mohri}, \bibinfo{author}{K.~Watanabe},
  \bibinfo{author}{T.~Yamashina},
\newblock \bibinfo{title}{Sputtering process of a silicon carbide surface with
  energetic ions by means of an aes-sims-fds combined system},
\newblock \bibinfo{journal}{Journal of Nuclear Materials} \bibinfo{volume}{75}
  (\bibinfo{year}{1978}) \bibinfo{pages}{7--13}.
  \DOIprefix\doi{10.1016/0022-3115(78)90023-5}.
\bibitem[{Plank et~al.(1996)Plank, Schwörer, and Roth}]{Plank1996}
\bibinfo{author}{H.~Plank}, \bibinfo{author}{R.~Schwörer},
  \bibinfo{author}{J.~Roth},
\newblock \bibinfo{title}{Erosion behaviour and surface composition
  modifications of sic under d+ ion bombardment},
\newblock \bibinfo{journal}{Nuclear Instruments and Methods in Physics Research
  Section B: Beam Interactions with Materials and Atoms} \bibinfo{volume}{111}
  (\bibinfo{year}{1996}) \bibinfo{pages}{63--69}.
  \DOIprefix\doi{10.1016/0168-583X(95)01258-3}.
\bibitem[{Balden et~al.(2001)Balden, Picarle, and Roth}]{Balden2001}
\bibinfo{author}{M.~Balden}, \bibinfo{author}{S.~Picarle},
  \bibinfo{author}{J.~Roth},
\newblock \bibinfo{title}{Mechanism of the chemical erosion of sic under
  hydrogen irradiation},
\newblock \bibinfo{journal}{Journal of Nuclear Materials}
  \bibinfo{volume}{290-293} (\bibinfo{year}{2001}) \bibinfo{pages}{47--51}.
  \DOIprefix\doi{10.1016/S0022-3115(00)00505-5}.
\bibitem[{Koller et~al.(2019)Koller, Davis, Goodland, Abrams, Gonderman,
  Herdrich, Frieß, and Zuber}]{Koller2019}
\bibinfo{author}{M.~T. Koller}, \bibinfo{author}{J.~W. Davis},
  \bibinfo{author}{M.~E. Goodland}, \bibinfo{author}{T.~Abrams},
  \bibinfo{author}{S.~Gonderman}, \bibinfo{author}{G.~Herdrich},
  \bibinfo{author}{M.~Frieß}, \bibinfo{author}{C.~Zuber},
\newblock \bibinfo{title}{Deuterium retention in silicon carbide, sic ceramic
  matrix composites, and sic coated graphite},
\newblock \bibinfo{journal}{Nuclear Materials and Energy} \bibinfo{volume}{20}
  (\bibinfo{year}{2019}) \bibinfo{pages}{100704}.
  \DOIprefix\doi{10.1016/J.NME.2019.100704}.
\bibitem[{Sinclair et~al.(2021)Sinclair, Abrams, Bringuier, Thomas, Holland,
  Gonderman, Yu, and Doerner}]{Sinclair2021}
\bibinfo{author}{G.~Sinclair}, \bibinfo{author}{T.~Abrams},
  \bibinfo{author}{S.~Bringuier}, \bibinfo{author}{D.~M. Thomas},
  \bibinfo{author}{L.~Holland}, \bibinfo{author}{S.~Gonderman},
  \bibinfo{author}{J.~H. Yu}, \bibinfo{author}{R.~P. Doerner},
\newblock \bibinfo{title}{Quantifying erosion and retention of silicon carbide
  due to d plasma irradiation in a high-flux linear plasma device},
\newblock \bibinfo{journal}{Nuclear Materials and Energy} \bibinfo{volume}{26}
  (\bibinfo{year}{2021}) \bibinfo{pages}{100939}.
  \DOIprefix\doi{10.1016/J.NME.2021.100939}.
\bibitem[{Duan et~al.(2016)Duan, Yang, Chen, Tian, Cai, Wang, Jia, and
  Zhou}]{Duan2016_2}
\bibinfo{author}{X.~Duan}, \bibinfo{author}{Z.~Yang},
  \bibinfo{author}{L.~Chen}, \bibinfo{author}{Z.~Tian},
  \bibinfo{author}{D.~Cai}, \bibinfo{author}{Y.~Wang},
  \bibinfo{author}{D.~Jia}, \bibinfo{author}{Y.~Zhou},
\newblock \bibinfo{title}{Review on the properties of hexagonal boron nitride
  matrix composite ceramics},
\newblock \bibinfo{journal}{Journal of the European Ceramic Society}
  \bibinfo{volume}{36} (\bibinfo{year}{2016}) \bibinfo{pages}{3725--3737}.
  \DOIprefix\doi{10.1016/J.JEURCERAMSOC.2016.05.007}.
\bibitem[{Satonik et~al.(2014)Satonik, Rovey, and Hilmas}]{Satonik2014}
\bibinfo{author}{A.~J. Satonik}, \bibinfo{author}{J.~L. Rovey},
  \bibinfo{author}{G.~Hilmas},
\newblock \bibinfo{title}{Effects of plasma exposure on boron nitride ceramic
  insulators for hall-effect thrusters},
\newblock \bibinfo{journal}{https://doi.org/10.2514/1.B34877}
  \bibinfo{volume}{30} (\bibinfo{year}{2014}) \bibinfo{pages}{656--663}.
  \URLprefix \url{https://arc.aiaa.org/doi/10.2514/1.B34877}.
  \DOIprefix\doi{10.2514/1.B34877}.
\bibitem[{sai(2024)}]{saintgobain_sic}
\bibinfo{title}{Hexoloy sa sintered silicon carbide material},
  \bibinfo{year}{2024}. \URLprefix
  \url{https://www.ceramicsrefractories.saint-gobain.com/materials/silicon-carbide-sic/hexoloy-silicon-carbide-material/hexoloy-sa-sic-material}.
\bibitem[{Hinoki et~al.(2005)Hinoki, Snead, and Blue}]{Hinoki2005}
\bibinfo{author}{T.~Hinoki}, \bibinfo{author}{L.~L. Snead},
  \bibinfo{author}{C.~A. Blue},
\newblock \bibinfo{title}{Development of refractory armored silicon carbide by
  infrared transient liquid phase processing},
\newblock \bibinfo{journal}{Journal of Nuclear Materials} \bibinfo{volume}{347}
  (\bibinfo{year}{2005}) \bibinfo{pages}{207--216}.
  \DOIprefix\doi{10.1016/J.JNUCMAT.2005.08.020}.
\bibitem[{Minami et~al.(2007)Minami, Niigawa, Ueno, Hinoki, Yamamoto, and
  Konishi}]{Minami2007}
\bibinfo{author}{T.~Minami}, \bibinfo{author}{S.~Niigawa},
  \bibinfo{author}{Y.~Ueno}, \bibinfo{author}{T.~Hinoki},
  \bibinfo{author}{Y.~Yamamoto}, \bibinfo{author}{S.~Konishi},
\newblock \bibinfo{title}{Hydrogen isotopes permeation evaluation in the
  advanced materialfor nuclearfusion blanket use},
\newblock \bibinfo{journal}{Proceedings - Symposium on Fusion Engineering}
  (\bibinfo{year}{2007}). \DOIprefix\doi{10.1109/FUSION.2007.4337867}.
\bibitem[{Kishimoto et~al.(2011)Kishimoto, Shibayama, Shimoda, Kobayashi, and
  Kohyama}]{Kishimoto2011}
\bibinfo{author}{H.~Kishimoto}, \bibinfo{author}{T.~Shibayama},
  \bibinfo{author}{K.~Shimoda}, \bibinfo{author}{T.~Kobayashi},
  \bibinfo{author}{A.~Kohyama},
\newblock \bibinfo{title}{Microstructural and mechanical characterization of
  w/sic bonding for structural material in fusion},
\newblock \bibinfo{journal}{Journal of Nuclear Materials} \bibinfo{volume}{417}
  (\bibinfo{year}{2011}) \bibinfo{pages}{387--390}.
  \DOIprefix\doi{10.1016/J.JNUCMAT.2010.12.079}.
\bibitem[{Goebel et~al.(1984)Goebel, Campbell, and Conn}]{Goebel1984}
\bibinfo{author}{D.~M. Goebel}, \bibinfo{author}{G.~Campbell},
  \bibinfo{author}{R.~W. Conn},
\newblock \bibinfo{title}{Plasma surface interaction experimental facility
  (pisces) for materials and edge physics studies},
\newblock \bibinfo{journal}{Journal of Nuclear Materials} \bibinfo{volume}{121}
  (\bibinfo{year}{1984}) \bibinfo{pages}{277--282}.
  \DOIprefix\doi{10.1016/0022-3115(84)90135-1}.
\bibitem[{Henke et~al.(1993)Henke, Gullikson, and Davis}]{Henke1993}
\bibinfo{author}{B.~L. Henke}, \bibinfo{author}{E.~M. Gullikson},
  \bibinfo{author}{J.~C. Davis},
\newblock \bibinfo{title}{X-ray interactions: Photoabsorption, scattering,
  transmission, and reflection at e = 50-30,000 ev, z = 1-92},
\newblock \bibinfo{journal}{Atomic Data and Nuclear Data Tables}
  \bibinfo{volume}{54} (\bibinfo{year}{1993}) \bibinfo{pages}{181--342}.
  \DOIprefix\doi{10.1006/ADND.1993.1013}.
\bibitem[{Forbeaux et~al.(2000)Forbeaux, Themlin, Charrier, Thibaudau, and
  Debever}]{Forbeaux2000}
\bibinfo{author}{I.~Forbeaux}, \bibinfo{author}{J.~M. Themlin},
  \bibinfo{author}{A.~Charrier}, \bibinfo{author}{F.~Thibaudau},
  \bibinfo{author}{J.~M. Debever},
\newblock \bibinfo{title}{Solid-state graphitization mechanisms of silicon
  carbide 6h–sic polar faces},
\newblock \bibinfo{journal}{Applied Surface Science} \bibinfo{volume}{162-163}
  (\bibinfo{year}{2000}) \bibinfo{pages}{406--412}.
  \DOIprefix\doi{10.1016/S0169-4332(00)00224-5}.
\bibitem[{Tielens et~al.(1994)Tielens, McKee, Seab, Hollenbach, Tielens, McKee,
  Seab, and Hollenbach}]{Tielens1994}
\bibinfo{author}{A.~G. G.~M. Tielens}, \bibinfo{author}{C.~F. McKee},
  \bibinfo{author}{C.~G. Seab}, \bibinfo{author}{D.~J. Hollenbach},
  \bibinfo{author}{A.~G. G.~M. Tielens}, \bibinfo{author}{C.~F. McKee},
  \bibinfo{author}{C.~G. Seab}, \bibinfo{author}{D.~J. Hollenbach},
\newblock \bibinfo{title}{The physics of grain-grain collisions and gas-grain
  sputtering in interstellar shocks},
\newblock \bibinfo{journal}{ApJ} \bibinfo{volume}{431} (\bibinfo{year}{1994})
  \bibinfo{pages}{321}. \URLprefix
  \url{https://ui.adsabs.harvard.edu/abs/1994ApJ...431..321T/abstract}.
  \DOIprefix\doi{10.1086/174488}.
\bibitem[{Jurac et~al.(1998)Jurac, Johnson, and Donn}]{Jurac1998}
\bibinfo{author}{S.~Jurac}, \bibinfo{author}{R.~E. Johnson},
  \bibinfo{author}{B.~Donn},
\newblock \bibinfo{title}{Monte carlo calculations of the sputtering of grains:
  Enhanced sputtering of small grains},
\newblock \bibinfo{journal}{The Astrophysical Journal} \bibinfo{volume}{503}
  (\bibinfo{year}{1998}) \bibinfo{pages}{247--252}. \URLprefix
  \url{https://iopscience.iop.org/article/10.1086/305994
  https://iopscience.iop.org/article/10.1086/305994/meta}.
  \DOIprefix\doi{10.1086/305994/FULLTEXT/}.
\bibitem[{Abrams et~al.(2021)Abrams, Bringuier, Thomas, Sinclair, Gonderman,
  Holland, Rudakov, Wilcox, Unterberg, and Scotti}]{Abrams2021}
\bibinfo{author}{T.~Abrams}, \bibinfo{author}{S.~Bringuier},
  \bibinfo{author}{D.~M. Thomas}, \bibinfo{author}{G.~Sinclair},
  \bibinfo{author}{S.~Gonderman}, \bibinfo{author}{L.~Holland},
  \bibinfo{author}{D.~L. Rudakov}, \bibinfo{author}{R.~S. Wilcox},
  \bibinfo{author}{E.~A. Unterberg}, \bibinfo{author}{F.~Scotti},
\newblock \bibinfo{title}{Evaluation of silicon carbide as a divertor armor
  material in diii-d h-mode discharges},
\newblock \bibinfo{journal}{Nuclear Fusion} \bibinfo{volume}{61}
  (\bibinfo{year}{2021}) \bibinfo{pages}{066005}. \URLprefix
  \url{https://iopscience.iop.org/article/10.1088/1741-4326/abecee
  https://iopscience.iop.org/article/10.1088/1741-4326/abecee/meta}.
  \DOIprefix\doi{10.1088/1741-4326/ABECEE}.
\bibitem[{Gibson et~al.(2007)Gibson, van Well, Hodgkinson, al, Koomson,
  Ebenezer, Lee, Fatkiyah, Persada, Andayati, Kukushkin, Pacher, Kotov, Reiter,
  Coster, and Pacher}]{Gibson2007}
\bibinfo{author}{G.~Gibson}, \bibinfo{author}{B.~van Well},
  \bibinfo{author}{J.~Hodgkinson}, \bibinfo{author}{al},
  \bibinfo{author}{S.~Koomson}, \bibinfo{author}{A.~Ebenezer},
  \bibinfo{author}{C.~Lee}, \bibinfo{author}{E.~Fatkiyah},
  \bibinfo{author}{D.~Persada}, \bibinfo{author}{D.~Andayati},
  \bibinfo{author}{A.~Kukushkin}, \bibinfo{author}{H.~Pacher},
  \bibinfo{author}{V.~Kotov}, \bibinfo{author}{D.~Reiter},
  \bibinfo{author}{D.~Coster}, \bibinfo{author}{G.~Pacher},
\newblock \bibinfo{title}{Effect of conditions for gas recirculation on
  divertor operation in iter},
\newblock \bibinfo{journal}{Nuclear Fusion} \bibinfo{volume}{47}
  (\bibinfo{year}{2007}) \bibinfo{pages}{698}. \URLprefix
  \url{https://iopscience.iop.org/article/10.1088/0029-5515/47/7/021
  https://iopscience.iop.org/article/10.1088/0029-5515/47/7/021/meta}.
  \DOIprefix\doi{10.1088/0029-5515/47/7/021}.
\bibitem[{Petrie et~al.(2009)Petrie, Porter, Brooks, Fenstermacher, Ferron,
  Groth, Hyatt, Haye, Lasnier, Leonard, Luce, Politzer, Rensink, Schaffer,
  Wade, Watkins, and West}]{Petrie2009}
\bibinfo{author}{T.~W. Petrie}, \bibinfo{author}{G.~D. Porter},
  \bibinfo{author}{N.~H. Brooks}, \bibinfo{author}{M.~E. Fenstermacher},
  \bibinfo{author}{J.~R. Ferron}, \bibinfo{author}{M.~Groth},
  \bibinfo{author}{A.~W. Hyatt}, \bibinfo{author}{R.~J.~L. Haye},
  \bibinfo{author}{C.~J. Lasnier}, \bibinfo{author}{A.~W. Leonard},
  \bibinfo{author}{T.~C. Luce}, \bibinfo{author}{P.~A. Politzer},
  \bibinfo{author}{M.~E. Rensink}, \bibinfo{author}{M.~J. Schaffer},
  \bibinfo{author}{M.~R. Wade}, \bibinfo{author}{J.~G. Watkins},
  \bibinfo{author}{W.~P. West},
\newblock \bibinfo{title}{Impurity behaviour under puff-and-pump radiating
  divertor conditions},
\newblock \bibinfo{journal}{Nuclear Fusion} \bibinfo{volume}{49}
  (\bibinfo{year}{2009}) \bibinfo{pages}{065013}. \URLprefix
  \url{https://iopscience.iop.org/article/10.1088/0029-5515/49/6/065013
  https://iopscience.iop.org/article/10.1088/0029-5515/49/6/065013/meta}.
  \DOIprefix\doi{10.1088/0029-5515/49/6/065013}.
\bibitem[{Zhang et~al.(2016)Zhang, Bobkov, Lunt, Noterdaeme, Coster, Bilato,
  Jacquet, Brida, Feng, Wolfrum, and Guimarais}]{Zhang2016}
\bibinfo{author}{W.~Zhang}, \bibinfo{author}{V.~Bobkov},
  \bibinfo{author}{T.~Lunt}, \bibinfo{author}{J.~M. Noterdaeme},
  \bibinfo{author}{D.~Coster}, \bibinfo{author}{R.~Bilato},
  \bibinfo{author}{P.~Jacquet}, \bibinfo{author}{D.~Brida},
  \bibinfo{author}{Y.~Feng}, \bibinfo{author}{E.~Wolfrum},
  \bibinfo{author}{L.~Guimarais},
\newblock \bibinfo{title}{3d simulations of gas puff effects on edge density
  and icrf coupling in asdex upgrade},
\newblock \bibinfo{journal}{Nuclear Fusion} \bibinfo{volume}{56}
  (\bibinfo{year}{2016}) \bibinfo{pages}{036007}. \URLprefix
  \url{https://iopscience.iop.org/article/10.1088/0029-5515/56/3/036007
  https://iopscience.iop.org/article/10.1088/0029-5515/56/3/036007/meta}.
  \DOIprefix\doi{10.1088/0029-5515/56/3/036007}.

\end{thebibliography}
	
\end{document}